\documentclass[fleqn,usenatbib]{mnras}

\usepackage{amssymb}	

\usepackage{newtxtext,newtxmath}

\usepackage[T1]{fontenc}

\usepackage{graphicx}	
\usepackage{amsmath}	
\usepackage{subcaption}
\usepackage{graphicx} 
\usepackage{ulem}

\title[The nature of $z \gtrsim 10$ galaxies]{Exploring the nature of UV-bright $z \gtrsim 10$ galaxies detected by JWST: \\ 
star formation, black hole accretion, or a non-universal IMF? }

\author[Trinca et al.]{Alessandro Trinca$^{1,2,3}$\thanks{E-mail: alessandro.trinca@inaf.it},
Raffaella Schneider$^{1,2,3,4}$,
Rosa Valiante$^{1,3}$, Luca Graziani$^{2,3}$, Arianna Ferrotti$^{2}$, 
\newauthor
Kazuyuki Omukai$^{5}$ and Sunmyon Chon$^{5,6}$ \\
\\
$^{1}$INAF/Osservatorio Astronomico di Roma, Via Frascati 33, 00040 Monte Porzio Catone, Italy \\
$^{2}$Dipartimento di Fisica, ``Sapienza'' Universit$\grave{a}$ di Roma, Piazzale Aldo Moro 2, 00185 Roma, Italy \\
$^{3}$INFN, Sezione Roma1, Dipartimento di Fisica, ``Sapienza'' Universit$\grave{a}$ di Roma, Piazzale Aldo Moro 2, 00185, Roma, Italy \\
$^{4}$Sapienza School for Advanced Studies, Viale Regina Elena 291, 00161 Roma, Italy\\
$^{5}$Astronomical Institute, Graduate School of Science, Tohoku University, Aoba, Sendai 980-8578, Japan\\
$^{6}$Max-Planck-Institut f$\ddot{u}$r Astrophysik, Karl-Schwarzschild-Str. 1, D-85741 Garching, Germany\\ 
}
\date{Accepted XXX. Received XXX; in original form XXX}

\pubyear{2024}

\begin{document}
\label{firstpage}
\pagerange{\pageref{firstpage}--\pageref{lastpage}}
\maketitle

\begin{abstract}
We use the Cosmic Archaeology Tool (CAT) semi-analytical model to explore the contribution of Population (Pop) III/II stars and active galactic nuclei (AGNs) to the galaxy UV luminosity function (LF) evolution at $4 \leq z \leq 20$.  We compare in particular with recent JWST data in order to explore the apparent tension between observations and theoretical models in the number density of bright galaxies at $z \gtrsim 10$. The model predicts a star formation history dominated by UV faint ($M_{\rm UV} > - 18$) galaxies, with a Pop III contribution of $\lesssim 10\%$ ($\lesssim 0.5\%$) at $z \simeq 20$ ($z \simeq 10$). Stars are the primary sources of cosmic reionization, with $5 - 10 \%$ of ionizing photons escaping into the intergalatic medium at $5 \leq z \leq 10$, while the contribution of unobscured AGNs becomes dominant only at $z \lesssim 5$. The predicted stellar and AGN UV LFs reproduce the observational data at $5 \lesssim z \lesssim 9 - 10$. At higher redshift, CAT predicts a steeper evolution in the faint-end slope ($M_{\rm UV} > - 18$), and a number density of bright galaxies ($M_{\rm UV} \simeq -20$) consistent with data at $z \sim 10 - 11$, but smaller by 0.8 dex at $z \sim 12 - 13$, and 1.2 dex at $z \sim 14 - 16$, when compared to the values estimated by recent studies. Including the AGN emission does not affect the above findings, as AGNs contribute at most to $\lesssim 10 \%$ of the total UV luminosity at $M_{\rm UV} < - 19$ and $z \gtrsim 10$. Interestingly, considering a gradual transition in the stellar IMF, modulated by metallicity and redshift as suggested by recent simulations, the model agrees with JWST data at $z \sim 12 - 13$, and the disagreement at $z \sim 14 - 16$ is reduced to 0.5 dex.
\end{abstract}




\section{Introduction}
\label{sec:introduction}

The launch of {\it James Webb Space Telescope} (JWST) represents a major breakthrough in our understanding of the high-redshift Universe. The early release observations enabled the detection of several galaxy candidates with photometric redshifts $z \gtrsim 10$ \citep{Castellano2022, Naidu2022, Labbe2022, Bradley2022,  Harikane2023photo,  Tacchella2022, williams2023, Adams2023, Atek2023, Donnan2023, Tacchella2023}, some of which with spectroscopic confirmation
\citep{Schaerer2022, curti2023, Wang2022,  Heintz2022a, Heintz2022b, Bunker2023, ArrabalHaro2023, Harikane2023spectro}, including the currently most distant galaxy at $z = 13.2$ \citep{Curtislake2022}. This provides an unprecedented opportunity to explore the properties of the galaxy population in the first few hundreds million years of cosmic history, and to constrain their evolution. In particular, deep JWST observations are starting to constrain the low-mass end of the black hole (BH) mass function at $z \sim 4 - 7$ \citep{Ubler2023, Kocevski2023, Harikane2023bh, Maiolino2023AGN}, observe accreting supermassive black holes (SMBHs) in the nuclei of galaxies at $z \sim 9 - 11$ \citep{Larson2023, Maiolino2023BHGNz11}, and detect potential signatures of the first stellar populations \citep{Wang2022, Welch2022, Vanzella2023, Maiolino2023popIII}.    

The wealth of early JWST observations is already starting to challenge theoretical models. In particular, the number density of bright galaxies and its apparent lack of evolution between $z \sim 9$ and $z \sim 13 - 17$ are in tension with standard model predictions \citep{Finkelstein2022}.
Several physical causes have already been proposed to relieve this tension. A first possibility is that the observed UV bright galaxies arise as outliers of the general population, due to variations in halo formation histories, which lead to young ages ($\sim 10$ Myr) and high star formation rates \citep{Mason2023}. Alternatively, the observed UV luminosity function at $z \gtrsim 10$ can be reproduced if the interstellar dust is evacuated by radiatively driven outflows during the earliest phase of galaxy build-up \citep{ferrara2022, fiore2022, ziparo2023}. An overabundance of bright objects could also be explained if galaxies produce stars or UV photons more efficiently than expected due to a lack of star formation suppression at pre-reionization epochs \citep{Harikane2023photo}, feedback-free starbursts \citep{Dekel2023}, less efficient stellar-driven winds \citep{yung2023}, or to a more top-heavy stellar initial mass function (IMF) characterizing stellar populations at high redshift \citep{Harikane2023photo, Harikane2023spectro, Finkelstein2022,inayoshi2022, yung2023}. Indeed, the fraction of massive stars in high redshift stellar populations 
is expected to increase as a result of both the low metallicity \citep{omukai2005, hirano2014, hirano2015, chon2021} and the higher temperature of the Cosmic Microwave Background (CMB; \citealt{schneider2010, chon2022}). A considerable evolution of the stellar IMF might therefore be expected at $z \gtrsim 10$, resulting into a reduced mass-to-light ratio in early galaxies. To test this hypothesis, spectroscopic signatures associated to the presence of massive stars, such as strong nebular line emission and significant SN-driven galactic winds, will be tested by forthcoming observations \citep{Nakajima2022, Katz2022, Trussler2022}. 
By looking at the inferred properties of JWST photometric sources (stellar masses, ages, and UV slopes), \citet{Mirocha2023} argue that a combination of increased star formation efficiency, short-term temporal variations in the star formation rate, and dust attenuation is required to match the observations. When the comparison is made with spectroscopically confirmed sources \citep{Curtislake2022, robertson2023, ArrabalHaro2023}, \citet{keller2023} and \citet{mccaffrey2023} show that that existing cosmological simulations with varying resolution can generally reproduce the observations in terms of galaxy stellar masses, star formation rates, and number density of galaxies at $z > 10$. Similarly, \citet{Prada2023}
show that forecasts of a standard cosmological galaxy formation model\footnote{They combine Uchuu N-body simulations \citep{Ishiyama2021} with the Universe Machine galaxy formation algorithm \citep{Behroozi2019}.} are consistent with the abundance of photometrically selected JWST/HST galaxies at $z = 8, 9,$ and 10, and with the properties of spectroscopically confirmed galaxies at the same redshift; an exception is represented by the sample of red massive photometrically selected galaxies identified by \citet{Labbe2022} at $z \sim 8$, whose stellar masses exceeding $10^{10} M_\odot$ may be affected by systematic errors, but if real would require a revision of the standard galaxy formation model (see e.g. \citealt{Menci2022}). 

The above quoted studies show the complexity of interpreting early JWST measurements. One additional possibility is that at least a fraction of the observed UV luminosity densities at $z \gtrsim 10$ are produced by active galactic nucleus (AGN) activity \citep{Pacucci2022}. Indeed, 
ongoing JWST surveys, such as JADES Medium/Deep, CEERS, and PRIMER, are expected to be sensitive enough to detect tens of accreting BHs with masses $M_{\rm BH} = 10^6 - 10^8 M_\odot$ at $7 \leq z \leq 10$, with JADES Deep having the sensitivity to detect growing BHs with
masses $M_{\rm BH} = 10^4 - 10^6 M_\odot$ at $z \gtrsim 10$ \citep{Trinca2023}. While most of the $z \sim 12 - 16$ JWST candidates show extended morphologies \citep{Harikane2023photo}, JWST NIRSpec spectroscopy have detected broad emission lines from low-luminosity AGNs at $z \sim 4 - 7$ \citep{Kocevski2023, Ubler2023, Harikane2023bh, Maiolino2023AGN} and from sources at $z = 9 - 11$ \citep{Larson2023, Maiolino2023BHGNz11}, suggesting that AGN activity may contribute to the UV luminosity of at least a fraction of these sources. The estimated BH masses are $\sim 10^7 - 10^8 M_\odot$ and the $M_{\rm BH}/M_{\rm *}$ ratio is higher than the empirical relationship measured for nearby broad-line AGNs with comparable BH masses \citep{reines2015}. These findings are consistent with the expectations of theoretical models \citep{schneider2023}, which show that JWST surveys have the sensitivity to detect early accreting black holes out to $z \sim 10$ \citep{Trinca2023}, provided that these systems are overmassive with respect to their host galaxy stellar mass \citep{volonteri2022}. Interestingly, none of the systems is detected in X-rays, and their position on the BPT \citep{baldwin1981}, or the OHNO \citep{backhaus2022} diagrams is similar to that of star-forming galaxies observed at similar redshift. This means that the identification of these systems must be aided by properly designed color-selection techniques \citep{natarajan2017unveiling, valiante2018observability, zhang2021, Goulding2022, inayoshi2022}, or spectral diagnostics based on high-ionization lines, such as HeII and NeV \citep[][]{Nakajima2022, cleri2023}, but may ultimately rely on the detection of their broad emission lines (\citealt{Kocevski2023, Ubler2023, Larson2023, Harikane2023bh, Maiolino2023AGN}).

In this work, we aim to assess the impact of AGN emission and Population (Pop) III/II stars on the high-redshift galaxy UV luminosity function. We base our predictions on the Cosmic Archaeology Tool (CAT, \citealt{trinca2022}), and we focus, in particular, on the redshift range probed by JWST, from $z \gtrsim 4$ out to $z \sim 16 -18$ \citep{Harikane2023photo, Bouwens2023, perezgonzalez2023, Harikane2023spectro}.

The paper is organized as follows. In section \ref{sec:model} we summarize the mean features of CAT and the improvements made in the modeling of the first sources. In section \ref{sec:global} and \ref{sec:UVlum} we present our results. In particular, we first illustrate how the model complies with global constraints on the star formation (sec. \ref{sec:sfh}) and cosmic reionization histories (sec. \ref{sec:reionization}), and how our predictions compare with independent models. Then, we present the predicted redshift evolution of the galaxy luminosity function (sec. \ref{sec:galaxyUVlum}), tracing the contribution of AGNs (sec. \ref{sec:AGNcont}) and Pop III stars (sec. \ref{sec:PopIII_LF}) in different luminosity bins, and the potential effects of a gradual change in the stellar IMF with redshift and metallicity  (sec. \ref{sec:CompIMF}). Finally, in section \ref{sec:conclusion} we summarize our findings and draw our main conclusions.

\section{Modeling the first sources with CAT}
\label{sec:model}
Here we briefly recall the main features of the CAT model, presented in detail in \citet{trinca2022}, and the improved modeling that we have made for the present study (see sections \ref{sec:PopIIIstars} and \ref{sec:ModelReionization}). 

\textsc{CAT} is a semi-analyical model which has been developed to follow the co-evolution of the first galaxies and their nuclear black holes through cosmic times. A large sample of dark matter (DM) hierarchical merger histories, representative of the evolution of the entire galaxy population, is generated from $z = 4$ up to $z = 24$ using the \textsc{galform} galaxy formation model \citep{parkinson2008}, which is based on the extended Press-Schechter formalism. To describe star formation occurring in molecular and atomic-cooling halos, corresponding to virial temperatures $1200 {\rm K} \leq T_{\rm vir} < 10^4 {\rm K}$ and $T_{\rm vir} \geq 10^4 {\rm K}$, respectively, 
we adopt a mass resolution that corresponds to a virial temperature of $T_{\rm vir} = 1200 {\rm K}$. Hence, the minimum resolved halo mass ranges from $9.4 \times 10^5 \, \rm M_\odot$ at $z \sim 24$ to $1.0 \times 10^7 \, \rm M_\odot$ at $z \sim 4$.
Once each DM halo virializes and collapses, the gas is accreted and, depending on the virial temperature, redshift and composition, it cools down and forms stars. The baryonic evolution is followed by means of a set of physically motivated prescriptions, that we briefly summarize below, with a set of free parameters: the star formation efficiency $\epsilon_{\rm SF}$, the BH accretion parameter $\alpha$, the SN and AGN wind efficiencies $\epsilon_{\rm w,SN}$, $\epsilon_{\rm w,AGN}$ (see Eq.\ref{eq:SFR}, \ref{eq:BHLacc}, \ref{eq:SNfbk}, and \ref{eq:AGNfbk}). These are calibrated to reproduce the observed stellar mass density and star formation rate density at $4 \leq z \leq 6$, and the SMBHs masses and luminosities inferred for the $z \sim 6$ quasar population (for further details see section 3 in \citealt{trinca2022}).

\subsection{Star formation}
\label{sec:starformation}
\begin{figure}
\centering
\includegraphics[width=0.8\linewidth]{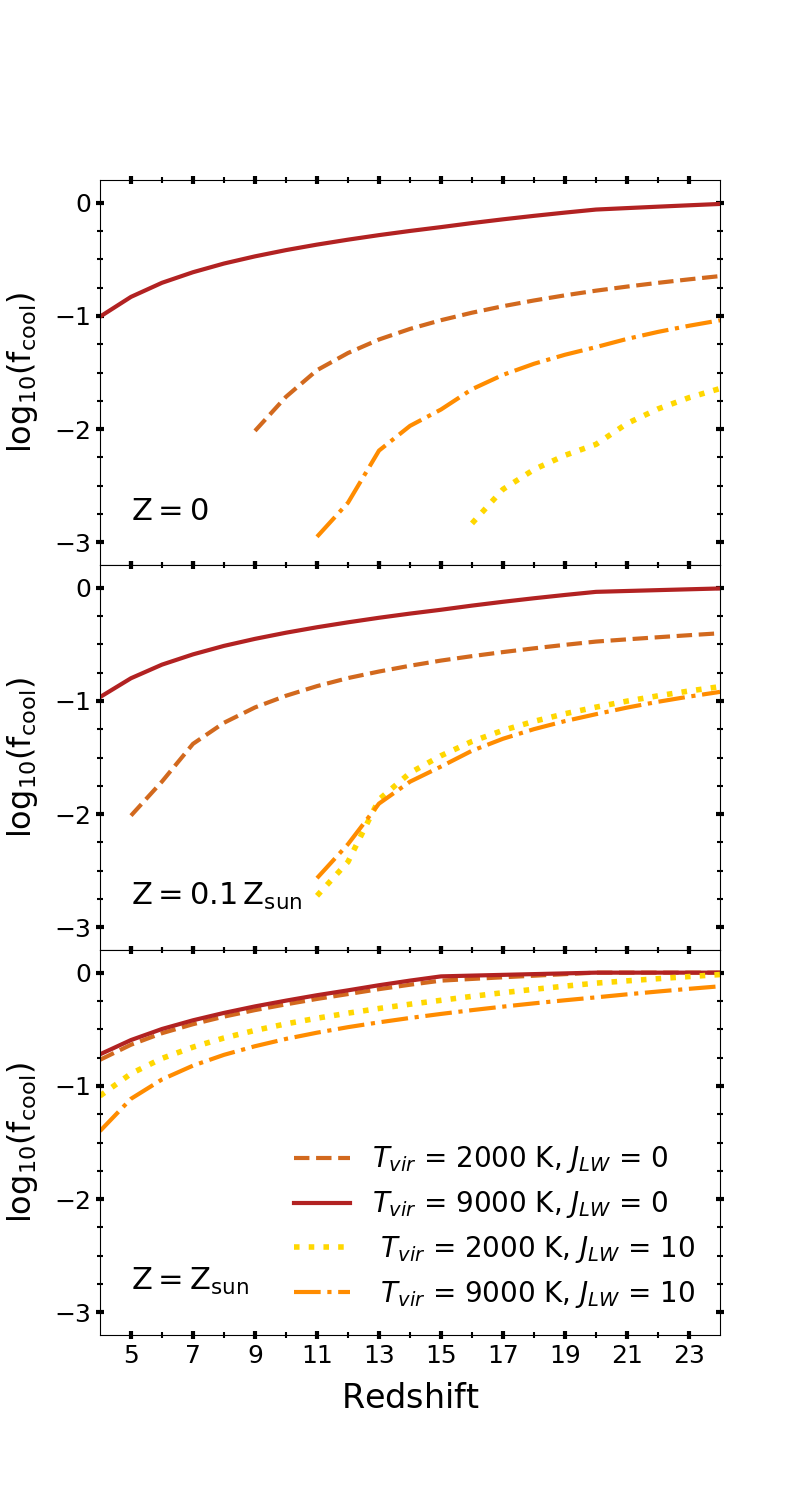}
\caption{The mass fraction of gas that is able to cool in one free-fall time, $f_{\rm cool}$, as a function of redshift for
two halo virial temperatures, $T_{\rm vir} =$ 2000 K, 9000 K, which enclose the mini-halo mass range, and considering two 
different values of the illuminating LW flux, $J_{\rm LW} =$ 0, 10. We show the evolution assuming three different values of 
gas metallicity, $Z = 0$ (upper panel), $Z = 0.1 \, Z_\odot$ (central panel) and $Z = Z_\odot$ (lower panel).}
\label{fig:fcool_evo}
\end{figure}

The fraction of gas mass available for star formation depends on the balance between cooling and dynamical timescales. The star formation rate (SFR) is computed in each galaxy as:
\begin{equation}
{\rm SFR} = f_{\rm cool} \, M_{\rm gas} \, \epsilon_{\rm SF} / \tau_{\rm dyn},
\label{eq:SFR}
\end{equation}
where $M_{\rm gas}$ is the available gas mass reservoir, $\tau_{\rm dyn} = [R_{\rm vir}^3 / (G \, M_{\rm halo}) ]^{1/2}$ is the dynamical time of the system, and $\epsilon_{\rm SF} = 0.05$ is the star formation efficiency, which has been calibrated to match the star formation rate density and stellar mass density evolution at $4 \leq z \leq 8$ (see section 3 in \citealt{trinca2022}). The parameter $f_{\rm cool}$ quantifies the reduced cooling efficiency inside molecular cooling halos (also referred to as \textit{minihalos}) with respect to more massive atomic-cooling halos. In fact, in minihalos, where $T_{\rm vir} < 10^4 \, \rm{K}$, the only efficient cooling channel in the primordial gas is provided by the roto-vibrational emission of molecular hydrogen ($\rm{H_2}$). If the star forming regions are exposed to a sufficiently large flux in the \textit{Lyman-Werner} energy band, $[11.2-13.6] \rm \, eV$, emitted by nearby galaxies, the formation of $\rm{H_2}$ can be strongly suppressed, decreasing the cooling efficiency \citep{Fialkov2013, sugimura2014critical, valiante2016, wolcottgreen2017}. Once the gas in the galaxy inter-stellar medium (ISM) starts to become metal-enriched due to stellar evolution, metals and dust enhance the cooling efficiency in minihalos. Hence, the parameter $f_{\rm cool}$ in Eq. \ref{eq:SFR} depends on the halo virial temperature, redshift, gas metallicity, and intensity of the illuminating LW radiation. 
A thorough discussion of these physical dependencies of $f_{\rm cool}$ can be found in the Appendix A of \citet{valiante2016}. Here we provide a more synthetic description, and we show in Fig. \ref{fig:fcool_evo} the value of $f_{\rm cool}$ as a function of redshift for minihalos with $T_{\rm vir} = 2000 \, \rm{K}$ and $9000 \, \rm{K}$, assuming two different values for the intensity of the illuminating LW flux ( $J_{\rm LW} = 0$ and $10$), and three different gas metallicities: $Z = 0$ (primordial composition), $Z = 0.1 \, Z_\odot$ and $Z = Z_\odot$ (see also Fig.1 in \citealt{sassano2021}). When $J_{\rm LW} = 0$, the value of $f_{\rm cool}$ is close to unity for the most massive ($T_{\rm vir} = 9000 \, \rm{K}$) mini-halos at high redshift, but it decreases with increasing $J_{\rm LW}$ and with decreasing metallicity. Concurrently, the cooling efficiency is reduced for smaller mini-halos ($T_{\rm vir} = 2000 \, \rm{K}$) and it decreases toward lower redshift. Even moderate values of $J_{\rm LW}$ ($J_{\rm LW} = 10$) lead to a significant drop in the cooling efficiency, particularly for metal-free and metal-poor gas compositions. Conversely, in atomic cooling halos we set $f_{\rm cool} = 1$.

\subsection{Improved treatment of Population III stars}
\label{sec:PopIIIstars}
Relying on recent state-of-the-art hydrodynamical simulations, which follow in detail the formation of stellar systems in very metal-poor environments \citep{chon2021,chon2022}, we imposed a further requirement for the onset of star formation inside pristine halos. Since simulations suggest that cold gas clouds need to accumulate a sufficient amount of gas before undergoing collapse and fragmentation, we assumed that the cold gas mass inside each halo must be larger than a minimum value of $M_{\rm cold, min} = 10^3 \, M_\odot$ to form stars. We also adopted an enhanced SF efficiency during Pop III star formation, assuming $\epsilon_{\rm SF, \, PopIII} = 0.15$, while maintaining the standard efficiency $\epsilon_{\rm SF} = 0.05$ \citep[see][]{trinca2022} for subsequent stellar populations. This improved modeling sets the minimum total stellar mass formed in each Pop III star formation episodes to $\sim 150 \, M_\odot$, in agreement with simulation results \citep{chon2022}. 

Theoretical studies suggest that at extremely low metallicities, such as the ones characterizing the first star forming regions, very massive stars are preferentially formed \citep{omukai1998formation, bromm2001,omukai2003formation,yoshida2008,hosokawa2011protostellar}. 
The initial mass function (IMF) of the first generation of stars, also referred to as PopIII stars, is therefore supposed to be top-heavy, with typical masses ranging from few 10s up to 100s of solar \citep{hirano2014,hirano2015,susa2014,hosokawa2016,sugimura2020}. 
Here we assume, for Pop III stars, a Larson IMF:
\begin{equation}
\Phi (m_*) \propto m_*^{\alpha -1} \, e^{-m_*/m_{\rm ch}}
\label{eq:imf}
\end{equation}
where $m_{\rm ch} = 20 \, \rm M_{\odot}$ is the characteristic mass, $\alpha = -1.35$ and the mass ranges between $10 \, \rm M_{\odot} \, \leq m_*  \, \leq 300 \, \rm M_{\odot}$.
Our choice is motivated by stellar archaeology studies and appears to best match the observed Galactic halo metallicity distribution function and the properties of C-enhanced and C-normal stars at [Fe/H] $< -3$ \citep{debennassuti2014, debennassuti2017, Fraser_2017, Magg_2022, Aguado_2023a}.

Under the environmental conditions where Pop III stars form (pristine galaxies hosted in DM minihalos), the total stellar mass formed in a dynamical timescale might be too small to fully sample the IMF of Pop III stars.\footnote{In fact, a total stellar mass of $M_{\rm *, tot} \gtrsim 10^6 \, M_{\odot}$ \citep{valiante2016} would be required to fully sample the Pop III IMF following a single star formation episode.} Therefore, we stochastically sample the IMF during each episode of Pop III star formation, building up the population star by star until we saturate the total stellar mass formed. In addition, Pop III stars are often assumed to die instantaneously due to the rapid evolutionary timescales of very massive stars. However, for less massive stars, with $m_* \sim 10 \, M_{\odot}$ and lifetimes $\sim 20 \rm Myr$, this assumption might lead to a significant underestimation of their total radiative output, as well as of the characteristic timescale of metal enrichment following their supernova (SN) explosions. For this reason, following each star formation episode, we define the lifetime of the stellar population $\tau_{\rm life, PopIII}$ as the lifetime of the most massive star formed, which represents the characteristic timescale for metal enrichment inside the host halo. As a result, multiple Pop III star formation episodes can occur within the same host halo before the most massive stars explode as SN, enriching the gas above the critical threshold of metallicity $\rm Z_{\rm crit}$, and suppressing star formation due to mechanical feedback (see section \ref{sec:feedback}). 

\begin{figure}
    \includegraphics[width=0.9\linewidth]{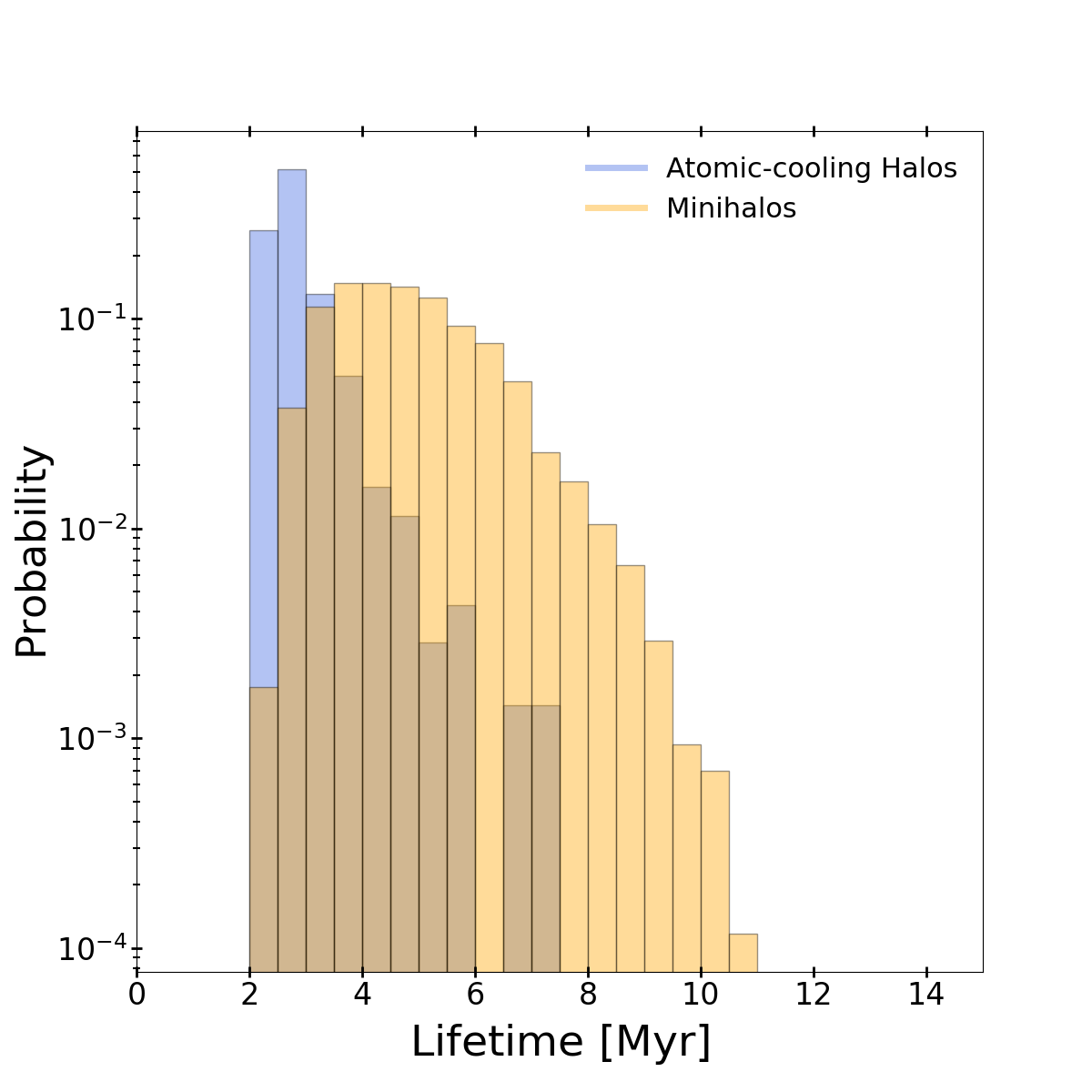}
	\caption{Distribution of lifetimes of Pop III stellar populations formed inside minihalos (orange) and atomic-cooling halos (blue). We show the two probability distributions assuming that the cold gas mass inside each halo must be larger than a minimum value of $M_{\rm cold,min} = 10^3 \rm \,  M_\odot$ to trigger Pop III star formation. The two distributions have a different shape, reflecting the reduced amount of gas that is available for star formation in minihalos, that leads to the preferential formation of Pop III stars with masses comparable to the characteristic mass of the IMF, $m_{\rm ch} = 20 \, \rm M_{\odot}$, and to an undersampling of the high-mass tail of the IMF (see text).}
    \label{fig:PopIII_lt}
\end{figure}

Figure \ref{fig:PopIII_lt} compares the resulting distribution of stellar lifetimes for Pop III stars formed in minihalos and atomic cooling halos. It is evident how the distribution changes depending on the halo properties. Inside minihalos, Pop III stars are characterized by longer lifetimes since fewer and less massive stars form. In fact, atomic cooling halos are more massive and characterized by more efficient cooling, providing a reservoir of cold gas large enough to sample the high mass end of the IMF ($m_* > m_{\rm ch}$), leading to shorter Pop III lifetimes. In this case, the host halo will be more rapidly enriched, promptly transitioning towards Pop II star formation. 
At higher metallicities, emission through metal-fine structure lines and the presence of dust increase the cooling efficiency. This leads to a transition toward lower characteristic masses. Therefore, above a critical metallicity of $Z_{\rm crit}=10^{-3.8} \, \rm M_\odot$, we assume that the formation of Pop II stars follows a Larson IMF, given in Eq.\ref{eq:imf}, with $m_{\rm ch} = 0.35 \rm M_{\odot}$ in the mass range [0.1, 100] $\rm M_{\odot}$.

\subsection{Black hole seeds formation and growth}
\label{sec:BHseeds}
At the end of each Pop III star formation episode, we assume that the heaviest among the newly formed BH remnants forms a light BH seed. Inside atomic-cooling halos (where $T_{\rm vir} \geq 10^4 \, \rm{K}$), if metal and dust cooling are still inefficient ($\rm{Z} \leq \rm{Z_{cr}}$) and molecular cooling is suppressed by a strong illuminating LW flux\footnote{This condition is usually expressed as $\rm{J_{LW}} \geq \rm{J_{cr}}$, where $\rm{J_{LW}}$ is the cumulative flux into the LW energy band in units of $10^{-21} \, \rm{ erg \, s^{-1} \, cm^{-2} \, Hz^{-1} \,sr^{-1}}$. 
For consistency with our previous studies \citep{trinca2022, Trinca2023}, here we adopt a threshold value of $\rm{J_{cr}} = 300$. 
A thorough discussion on the value of $\rm{J_{cr}}$ for heavy BH seed formation can be found in \citet{woods2019} and \citet{inayoshi2020}.}, the gas collapses almost isothermally with no fragmentation. This leads to the formation of a single supermassive star that becomes unstable, due to nuclear exhaustion or GR instabilities, forming a heavy BH seed with a mass of $10^5 \, \rm{M_\odot}$ \citep{hosokawa2012, inayoshi2014, latif2013, ferrara2014, becerra2015, latif2016a, becerra2018}. Note that we do not consider intermediate mass BH seeds, which are supposed to from runaway mergers in dense stellar clusters (see \citealt{sassano2021} for a recent investigation that considers all the three BH seeds populations).

Once formed, BH seeds are assumed to settle at the center of the host galaxy, where they can accrete gas and grow in mass. High resolution zoom-in simulations show that if the BH seed mass is less than $10^5 M_\odot$, its dynamical evolution is very perturbed by the irregular gas and stellar distribution in high-redshift galaxies \citep{pfister2019, sassano2023}. 
This effect will further suppress the growth of light BH seeds, as discussed by \citet{trinca2022}, but have a smaller impact on the observable population of accreting BHs, which largely descend from heavy BH seeds \citep{Trinca2023}.

The gas accretion rate onto BHs is described by the Bondi-Hoyle-Lyttleton (BHL) formula \citep{hoyle1941, bondi1952}:
\begin{equation}
    \dot{M}_{\rm BHL} = \alpha \frac{4 \pi G^2 M^2_{\rm BH}\rho_{\rm gas}(r_{\rm A})}{c^3_{\rm s}},
\label{eq:BHLacc}
\end{equation}
\noindent
where $c_{\rm s}$ is the sound speed, $\rho_{\rm gas}(r_{\rm A})$ is the gas
density evaluated at the radius of gravitational influence of the BH and $r_{\rm A} = 2 G M_{\rm BH}/c_{\rm s}^2$. The boost factor $\alpha = 90$ is included to take into account the density enhancement at small scales around the BH and is one of the free parameters of the model, which has been calibrated to reproduce the range of BH mass and bolometric luminosity of the observed quasar population at $z > 5$ \citep{trinca2022}. 

In our reference model, the gas accretion rate, $\dot{M}_{\rm accr}$, cannot exceed the Eddington limit, so that:
\begin{equation}
\dot{M}_{\rm accr} = {\rm min} (\dot{M}_{\rm BHL}, \dot{M}_{\rm Edd}),
\end{equation}
\noindent
and the BH mass growth rate is computed as:
\begin{equation}
\dot{M}_{\rm BH} = (1 - \epsilon_{\rm r}) \dot{M}_{\rm accr}. 
\end{equation}
\noindent
where, $\dot{M}_{\rm Edd} = L_{\rm Edd}/(\epsilon_{\rm r} c^2)$, $\epsilon_{\rm r} = 0.1$ is the adopted radiative efficiency, and $L_{\rm Edd} = 4 \pi c G M_{\rm BH} m_{\rm p}/\sigma_{\rm T}$ is the Eddington luminosity ($c$ is the speed of light, 
$m_{\rm p}$ is the proton mass and $\sigma_{\rm T}$ is the Thomson scattering cross section).

During galaxy mergers, the two nuclear BHs may sink to the center of the newly formed galaxy, form a binary system and merge. 
The timescale of this process can be considerably longer than halo sinking timescales \citep{tremmel2018}, but the formation of
coalescing BH pairs may be facilitated if the BHs have masses $\gtrsim 10^5 M_\odot$ and are hosted in galaxies with high central stellar and gas densities \citep{volonteri2020}. Here we take a very simplified approach, and assume that two BHs coalesce during major mergers, i.e. if the mass ratio of their interacting host DM halos is $\mu > 1/10$ \citep{tanaka2009, valiante2011}. Conversely, in minor mergers ($\mu < 1/10$), only the most massive among the two nuclear BHs is assumed to migrate to the center of the newly formed galaxy. We note here that this oversimplification has a relatively small impact on BH mass growth, which is largely dominated by gas accretion \citep{dubois2014,valiante2016,pacucci2020}.

\subsection{Mechanical and radiative feedback}
\label{sec:feedback}
The abundance of gas inside each galaxy is affected by mechanical feedback due to galaxy-scale outflows driven by the energy 
released by SN explosions and BH accretion, 
\begin{equation}
\dot{M}_{\rm{ej}} = \dot{M}_{\rm{ej, SN}} + \dot{M}_{\rm{ej, AGN}} 
\label{eq:fbk}
\end{equation}
where $\dot{M}_{\rm ej,SN}$ and $\dot{M}_{\rm ej, AGN}$ are the SN- and AGN-driven outflow rates. The first term is defined as:
\begin{equation}
\dot{M}_{\rm{ej, SN}} = \frac{2 E_{\rm SN} \epsilon_{\rm w,SN} R_{\rm SN} (t)}{v_{\rm e}^2} ,
\label{eq:SNfbk}
\end{equation}
where $E_{\rm SN}$ is the explosion energy per SN, $v_{\rm e} = (2 G M/R_{\rm vir} )^{1/2}$ is the escape velocity of the galaxy, $\epsilon_{\rm w,SN} = 1.6 \times 10^{-3}$ is a free parameter representing the SN-driven wind efficiency, and $R_{\rm SN} (t)$ is the SN explosion rate. The latter quantity depends on the star formation history and on the nature of the stellar populations hosted by each galaxy: for Pop III stars,  $E_{\rm SN}$ is assumed to be $2.7\times 10^{52}$\, erg, while for Pop II/I stars, $E_{\rm SN} = 1.2\times 10^{51}$ erg.

The second term in Eq. \ref{eq:fbk} is computed as,
\begin{equation}
\dot{M}_{\rm{ej, AGN}} = 2 \, \epsilon_{\rm{w, AGN}} \, \epsilon_r \, \dot{M}_{\rm accr} \, \biggl( \frac{c}{v_{\rm e}} \biggl)^2.
\label{eq:AGNfbk}
\end{equation}
\noindent
where $\epsilon_{\rm w,AGN}$ is the AGN-driven wind efficiency. Following \citet{trinca2022}, in our reference model, we assume that $\epsilon_{\rm w,AGN} = 2.5 \times 10^{-3}$.
While quasars accreting at (or above) the Eddington limit are more likely to show strong outflows that contribute to evacuate the gas from the inner galactic regions, broad, blushifted absorption lines tracing outflowing gas have also been detected in AGNs accreting at more moderately pace, down to few percent Eddington (see e.g. \citealt{Ganguly2007}). Following \citet{trinca2022,Trinca2023}, we parametrize BH feedback as described in Eq. \ref{eq:AGNfbk} even when BHs do not exceed the Eddington limit, as in the model that we discuss in the present work (see also \citealt{weinberger2017,negri2017,tremmel2019,piana2022} for similar descriptions of BH feedback).

In addition to the radiative feedback induced by LW photons, described in section \ref{sec:starformation}, during the process of cosmic reionization the photo-heating due to the increased gas temperature in photo-ionized regions can suppress star formation in haloes with virial temperatures below the temperature of the intergalactic medium (IGM, \citealt{valiante2016}). We consider $T_{\rm IGM} = Q_{\rm HII} \, T_{\rm reio} + (1 - Q_{\rm HII}) \, T_{\rm HI}$, where $T_{\rm reio} = 2 \times 10^4 \rm \, K$, $T_{\rm HI} = 0.017 (1+z)^2$ and the filling factor of HII regions, $Q_{\rm HII}$, is computed as described below.

\subsection{Metal and dust enrichment}
\label{sec:metal}
Following \citet{valiante2014} and \citet{debennassuti2014}, CAT follows the metal and dust enrichment in each galaxy adopting a two-phase ISM model, with a cold atomic/molecular phase, where star formation occurs and where dust grains can grow in mass by accreting gas-phase metals, and a hot/warm diffuse phase where dust can be destroyed by SN
shocks. Following DM halo virialization, the gas is initially accreted into the diffuse phase, then it condenses into the cold/molecular phase, where star formation occurs. Stars evolve and return gas, metal and dust into the diffuse phase. Finally, mechanical feedback due to SN explosions and AGN eject gas from the diffuse and condensed phases, following the description provided in section \ref{sec:feedback}. 
Metal and dust enrichment in the two-phase ISM is described by a system of differential equations, which relies on mass- and metallicity-dependent metal and dust yields and follows the release of nucleosynthetic products on the stellar characteristic lifetimes. We refer the interested readers to \citet{valiante2014} and \citet{debennassuti2014} for a thorough description of the chemical evolution model implemented in CAT.

\subsection{Photo-ionizing emission and reionization}
\label{sec:ModelReionization}
CAT follows the formation of the first stars and BHs across cosmic epochs in our galaxy sample. Therefore, we can investigate the relative contribution of different classes of sources to cosmic reionization.
In particular, at different redshifts, we can compute the photo-ionizing emissivities from Pop II stars, Pop III stars and early accreting BHs evolving in each galaxy of our sample.

For Pop III stars we compute the photo-ionizing emission rate $\dot{n}_\gamma$ from the mass-dependent emissivities tabulated by \citet{schaerer2002} for zero metallicity stars with no mass loss. For Pop II stars, we adopt the metallicity- and age-dependent intrinsic emissivities computed using \citet{bruzual2003} population synthesis model.
To compute the emission rate of ionizing photons 
from early accreting BHs, we model their spectral energy distribution (SED) as a multicolor-disk spectrum up to energies of $k T_{\rm max} \sim 1 {\rm keV} (M_{BH}/M_\odot)^{-1/4}$ plus a non-thermal power-law component $L_\nu \propto \nu^{-\alpha}$ with $\alpha \simeq 2$ at higher energies \citep[][]{shakura1973}.

Two additional effects need to be considered in order to model cosmic reionization: {\it (i)} only a fraction of the ionizing photons emitted will escape the galaxy and reach the outer medium, and {\it (ii)} the IGM density increases its inhomogeneity with time, leading to higher gas opacity to ionizing photons, and to a slower reionization process.
These effects are usually modeled with two parameters that are still poorly constrained by theoretical models and observations; the escape fraction, $f_{\rm esc}$, i.e. the fraction of ionizing photons that are able to escape the galaxy, and the clumping factor, $C$, which quantifies the increased clumpiness of the IGM.

Starting from the source emissivities, these two additional parameters allow us to predict the redshift evolution of the volume filling factor of ionized hydrogen, $Q_{\rm HII}$.
Several works showed that a decreasing trend with redshift of $f_{\rm esc}$ is required to simultaneously accommodate the production rate of ionizing photons associated with star formation and the available constraints on the IGM electron scattering optical depth $\tau_e$.
Therefore, following \citet{dayal2020}, we assume a redshift dependent escape fraction for ionizing photons emitted by Pop II and Pop III stars:
\begin{equation}
    f_{\rm esc, *} (z) = f_0 \, [(1+z)/5]^\beta \,\,\, ,
    \label{eq:fesc}
\end{equation}
where we choose $f_0 = 0.03$ and $\beta = 1.5$, such that $f_{\rm esc}$ varies between $\sim 3 - 35 \%$ for $z = 4 - 24$. The range of values assumed for the galaxy escape fraction is in broad agreement with empirical constraints obtained from early JWST observations. Recent results by \citet{mascia2023}, based on a sample of 24 lensed galaxies at $4.5 < z < 8$, suggest typical mean values of $f_{\rm esc} \sim 0.10$, while \citet{Schaerer2022} find smaller values, $f_{\rm esc} \sim 0.03 - 0.08$, for three systems at $z \sim 8$.
For the AGN ionizing emission, instead, we make the assumption that the fraction of unobscured AGNs can be used as a tracer of the escape fraction \citep[][]{ricci2017, dayal2020}. Therefore, following \citet{ueda2014}, we adopt a luminosity-dependent parametrization:
\begin{equation}
    f_{\rm esc, AGN} (L_{\rm X})= {\rm min} \bigl[ f_{\rm max}, {\rm max}[f_0 - \beta(\log(L_{\rm X} - 43.75)),f_{\rm min}] \bigr] 
    \label{eq:fobsUeda}
\end{equation}
with $f_{\rm max} = 0.84$, $f_{\rm min} = 0.20$, $f_{\rm 0} = 0.73$, $\beta = 0.24$ and $L_X$ is the AGN X-ray luminosity in the $[2 - 10] \, \rm keV$ energy band \citep[for further details see ][]{trinca2022}.

For the IGM clumping, we rely on the parametrization proposed by \citet{Iliev2005}, and we adopt the following redshift dependent
clumping factor:
\begin{equation}
    C(z) = 17.6 \, e^{-0.10 \,z + 0.0011 \, z^2} .
    \label{eq:clumping}
\end{equation}
The time evolution of the volume filling factor for ionized hydrogen, $Q_{\rm HII}$, can be written as \citep{Barkana2001}:

\begin{equation}
\dot{Q}_{\rm HII} = f_{\rm esc} \, \dot{n}_\gamma/n_{\rm H} \, - \alpha_{\rm B} \, C \, n_{\rm H} \, (1+z)^3 \, Q_{\rm HII} 
\label{eq:QHII}
\end{equation}
where $\dot{n}_\gamma$ is the total emission rate of ionizing photons per unit volume computed summing over all the available sources, $n_{\rm H} = X_{\rm H} \, n_{\rm IGM}$ is the number density of the hydrogen gas in the IGM, $X_{\rm H}$ is the hydrogen mass fraction, $n_{\rm IGM}$ is the IGM gas number density and $\alpha_{\rm B}= 2.6 \times 10^{-13} \, {\rm cm^3 \, s^{-1}}$ is case-B hydrogen recombination rate. 
From $Q_{\rm HII} (z)$, we can compute the IGM optical depth to electron scattering $\tau_{\rm e} (z)$ as:
\begin{equation}
    \tau_{\rm e} (z) = \int_0^z n_{\rm e} (z') \sigma_{\rm T} \, c \,    \biggl| \frac{dt}{dz'} \biggl| \, dz' 
	\label{eq:tau_e}
\end{equation}
where $\sigma_{\rm T} = 6.65 \times 10^{-25} \, \rm cm^2$ is the Thomson cross section, $c$ is the speed of light and $n_{\rm e} (z')$ is the mean electron number density at $z'$, which can be written as:
\begin{equation}
    n_{\rm e} (z') = \, Q_{\rm HII} (z') \, n_{\rm 0, B} \, X_{\rm H} (1+z)^3 \, 
	\label{eq:n_e}
\end{equation}
where $n_{\rm 0, B} = 2.51 \times 10^{-7} \, \rm cm^{-3}$ is the mean baryon number density at $z=0$.

\section{Global observational constraints}
\label{sec:global}
In this section, we first describe the redshift evolution of the comoving star formation rate density predicted by the model. Then, we show the predicted redshift evolution of the hydrogen ionizing emissivity and neutral hydrogen fraction, and how these compare with available observational data. 

\subsection{Star Formation History}
\label{sec:sfh}
\begin{figure*}
\centering
\includegraphics[width=12cm]{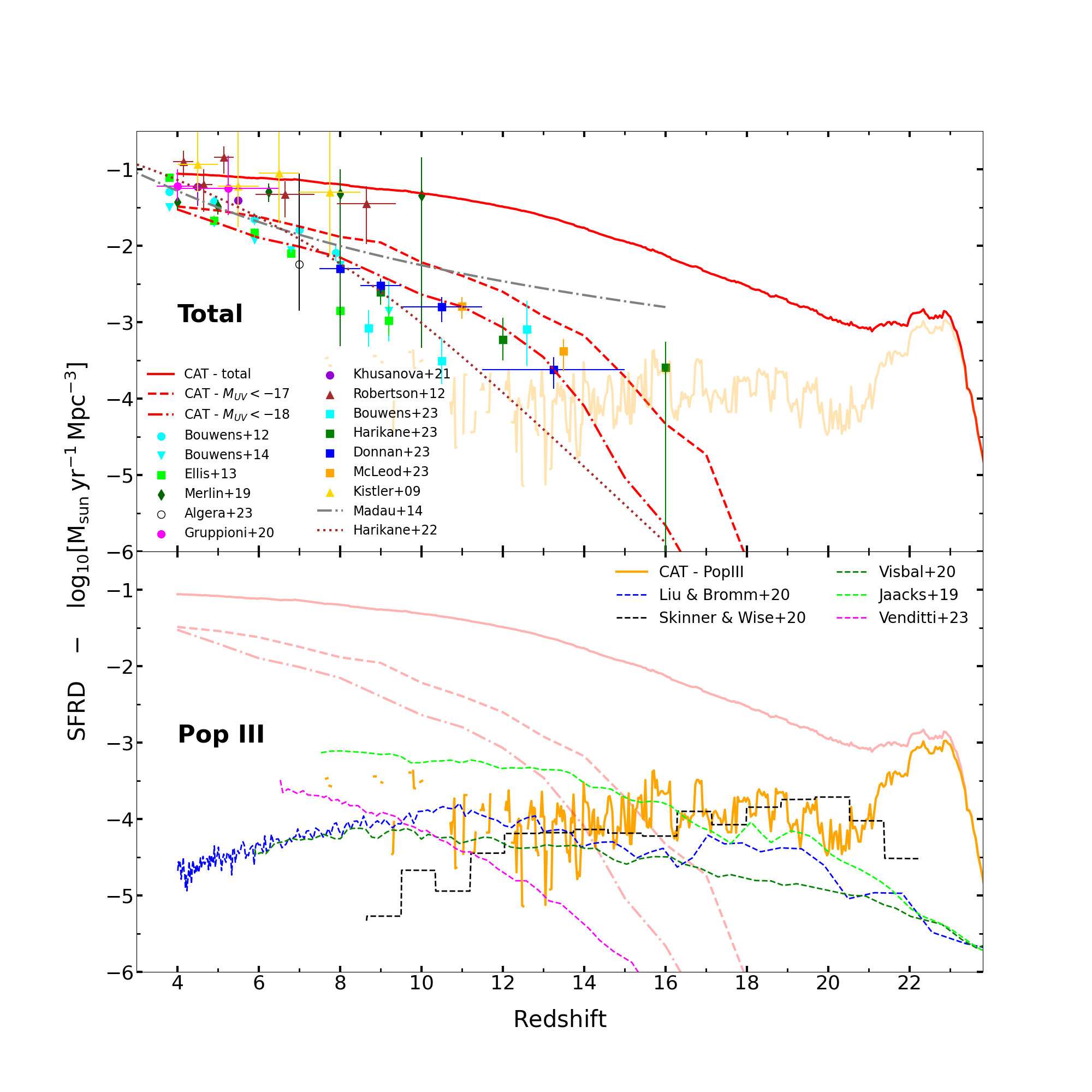}
\caption{Evolution of the global (PopIII + PopII, upper panel) and PopIII-only (lower panel) SFR density as a function of redshift. In the upper panel, CAT predictions illustrate the total SFRD (solid red lines) and the SFRD of UV-bright sources with $\rm M_{UV} < -17$ (red dashed line) and $\rm M_{UV} < -18$ (red dash-dotted line). The SFRD inferred from observations sampling the rest-frame UV luminosity are taken from \citet{bouwens2012,bouwens2014,ellis2013,schenker2013} and from recent JWST data \citep{Bouwens2023,Harikane2023photo,Donnan2023,McLeod2023}.
In addition, we also show the SFRD derived by \citep{merlin2019} from the SFRHs of passive galaxies during their active phase, by ALMA large surveys \citep{khusanova2020, gruppioni2020, algera2023}, and by gamma-ray bursts observations \citep{kistler2009, robertson2012}, which are sensitive to both obscured and unobscured star formation. Finally, we also show the extrapolation of the empirical models by \citet{madau2014a} and the constant star formation efficiency model by \citet{Harikane2023photo}.
In the lower panel, we compare the Pop III SFRD predicted by CAT with independent theoretical models by \citet{SkinnerWise2020, LiuBromm2020,Jaacks2019,Visbal2020,Venditti2023}. For illustrative purposes, we also report CAT predictions for the global and PopIII-only SFR density  with fainter colors in the lower and upper panel, respectively.}
\label{fig:SFR_PopII_PopIII}
\end{figure*}

In the upper panel of Figure \ref{fig:SFR_PopII_PopIII} we show the star formation rate density (SFRD) evolution predicted by CAT at $z>4$. The red solid line represents the total SFRD, while the dashed and dash-dotted red lines indicate the SFRD for galaxies with intrinsic UV magnitude $M_{\rm UV}$ smaller, i.e., brighter, than $-17.0$ and $-18.0$, respectively (see also \citealt{trinca2022}). We compare our results with observational constraints by \citet{bouwens2012,bouwens2014,ellis2013,merlin2019}, and with recent JWST data at $z \gtrsim 10$ \citep{Bouwens2023,Harikane2023photo,Donnan2023,McLeod2023}. We also show the empirical relations proposed by \citet{madau2014a} extrapolated to $z > 6$ (gray dashed-dot line) and the constant star formation efficiency model by \citet{Harikane2023photo} (maroon dotted line).
In the redshift range $4 < z < 10$, CAT predictions for the total SFRD are in good agreement with data from \citet{merlin2019}, who estimated the contribution of high-redshift passive galaxies to the global SFRD during their phase of activity. It is also consistent with the SFRD estimated from ALMA large surveys, in particular from ALPINE data by \citet{khusanova2020} and \citet{gruppioni2020}, and from REBELS data by \citet{algera2023}, and it is in very good agreement with the SFRD inferred from gamma-ray burst observations, which are sensitive to both obscured and unobscured star formation \citep{kistler2009, robertson2012}. Conversely, the SFRD derived from the rest-frame UV luminosity are better reproduced by CAT when only the contribution of the sources brightest than $M_{\rm UV} \lesssim -17.0$ is considered \citep{trinca2022}. We stress that the total SFRD computed with CAT accounts for both intrinsically faint objects and for obscured sources, which are better traced by rest-frame FIR observations. 

\subsection{Pop III Star Formation History}
\label{sec:pop3sfh}
In the lower panel of Figure \ref{fig:SFR_PopII_PopIII} we also show the predicted SFRD for Pop III stars (solid orange line). The Pop III SFRD predicted by CAT is characterised by an initial steep rise at $z \gtrsim 22$, and then declines to follow a relatively flat evolution. Indeed, despite the scatter due to the intrinsic burstiness of Pop III star formation, we find an almost constant SFRD of $\sim 10^{-4} \, M_\odot \, \rm yr^{-1} \, Mpc^{-1}$ over the redshift range 
$10 \lesssim z \lesssim 20$, below which the ISM metal enrichment of galaxies causes Pop III star formation to become progressively rarer. 
In Figure \ref{fig:SFR_PopII_PopIII}, we also show the Pop III SFRD predicted by high-resolution \citep{Jaacks2019, LiuBromm2020, SkinnerWise2020} and large-scale \citep{Venditti2023} hydrodynamical simulations, and the results of the semi-analytical model by \citet{Visbal2020}. Orders of magnitude differences are found between different studies. In particular, the mass resolution and scale of the simulations significantly affect the predicted SFRD at early times, where SAM models and small scale hydrodynamical simulations better resolve the large population of star forming \textit{minihalos} \citep{Visbal2020,SkinnerWise2020}. Further discrepancies can be ascribed to the different treatment of radiative feedback, which can be implemented sub-grid or properly accounted for by adopting a full radiative transfer scheme \citep{Maio2016,xu2016}. Likewise, the implementation of chemical evolution can vary among different studies, with only few models adopting mass and metallicity-dependent stellar yields and accounting for the presence of cosmic dust.
Finally, distinct assumptions are made in different simulations regarding the Pop III stellar IMF and the critical metallicity threshold for Pop II star formation \citep[see e.g.][for a detailed discussion]{Venditti2023}. 
CAT results appear to be in broad agreement with high-resolution simulations \citep{Jaacks2019,SkinnerWise2020}, which tend to predict larger SFRDs, especially in the redshift range $10 < z < 20$. This might be related to the ability of resolving Pop III star formation in small DM minihalos, which are generally below the resolution threshold of large-scale simulations \citep{sarmento2022,Venditti2023}.
Interestingly, the Pop III SFRD predicted by CAT is in very good agreement with the predictions of \citet{SkinnerWise2020} down to $z \sim 12$, below which their SFRD quickly decreases, due to the strong impact of LW radiation inside the simulated $\rm 1 Mpc^{3}$ comoving box.
The high Pop III SFRD predicted by CAT at $z>22$ is a consequence of the lack of self-regulation in the early star-forming regions hosted by the first mini-halos. The description of Pop III star formation in these systems implemented in CAT accounts for the stochasticity in their IMF, and the population is evolved over the typical lifetimes of the more massive stars formed; these masses are generally smaller in mini-halos than in Lyman-alpha cooling halos (see Figure \ref{fig:PopIII_lt}), as a result of the reduced gas mass available for star formation. Because of this, mechanical feedback acts on longer timescales. Once feedback becomes efficient, star formation begins to self-regulate, as indicated by the descending trend of the SFRD at $z \lesssim 22$, only $\sim 20$ Myr after the first star-forming episodes at $z = 24$.
Different assumptions are made in other semi-analytical models. For instance, in the work by \citet{Visbal2020}, a constant time delay of $10 \rm \, Myrs$ is assumed after the first episode of Pop III SF before subsequent metal-enriched SF can occur, to account for the gas recovery time after SN explosions.
This different treatment of Pop III SF can account for the large discrepancy in the predicted SFRD at very early times, before stellar feedback starts to efficiently self-regulate the SF process, leading to much closer trends at $z \lesssim 20$.

\begin{figure*}
\centering
  \includegraphics[width=12cm]{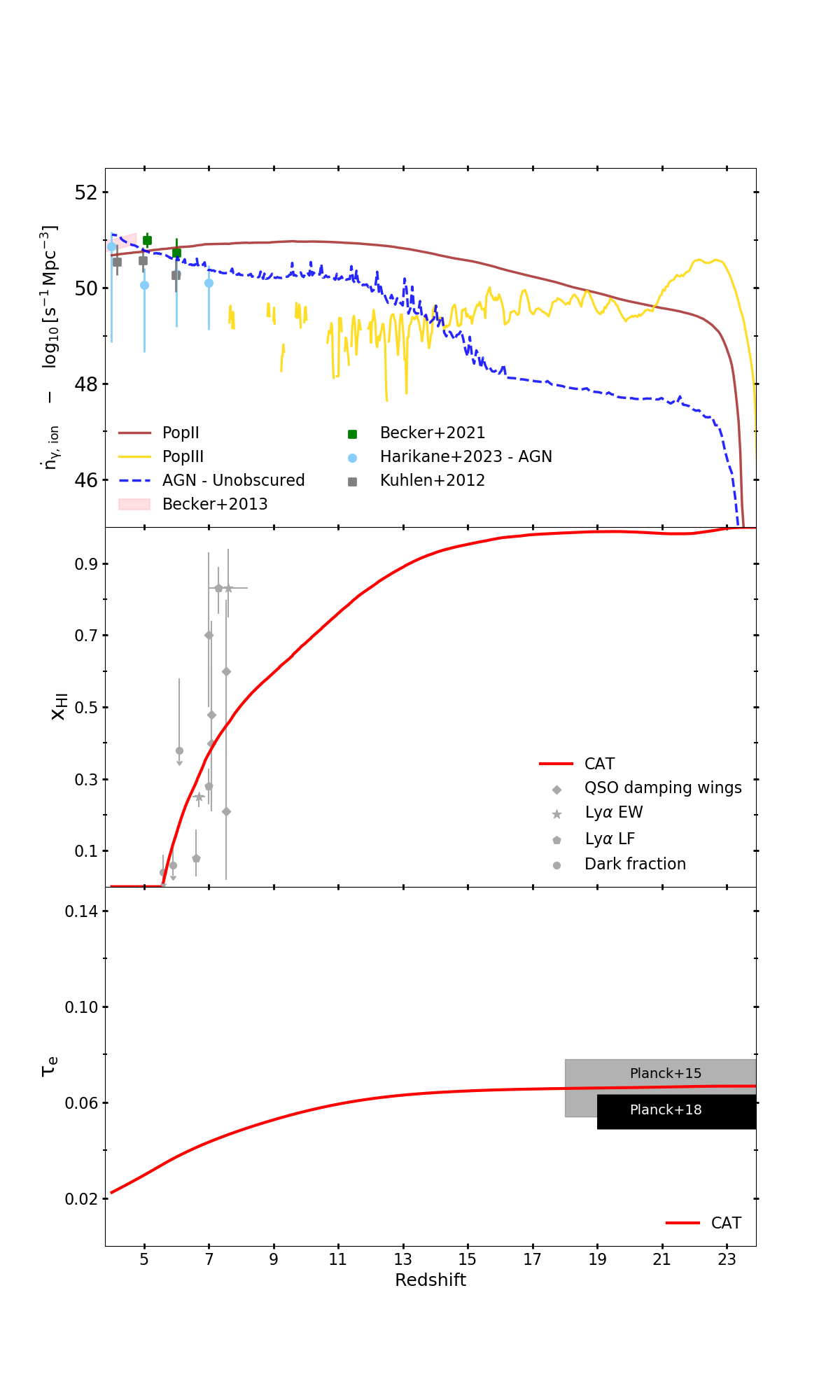}
\caption{\textit{Upper panel:} Global photo-ionizing emissivity as a function of redshift. Brown solid, yellow solid and blue dashed lines represent the contribution of PopII stars, PopIII stars and unobscured AGN, respectively, to the global emissivity. We show as a reference the empirical constraints proposed by \citet[grey data points]{Kuhlen2012}, \citet[pink shaded region]{becker2013}, \citet[green data points]{becker2021} and the recent estimates of the AGN contribution obtained by \citet[light blue circles]{Harikane2023bh}. \textit{Central panel}: evolution of the IGM neutral hydrogen fraction as a function of redshift. CAT predictions are compared to a compilation of observational constraints presented in \citet{bolan2022}, and obtained through different tracers, such as the evolution of Lyman-$\alpha$ equivalent width \citep[][stars]{bolan2022}, Lyman-$\alpha$ luminosity function \citep[][pentagons]{morales2021}, the dark pixel fraction in the Ly-$\alpha$ and Ly-$\beta$ forest \citep[][circles]{McGreer2015}, and quasar damping wings \citep[][diamonds]{davies2018,greig2019,wang2020}. \textit{Bottom panel}: Predicted evolution of the Thomson scattering optical depth as a function of redshift, compared with constraints obtained by the Planck cosmological survey \citep{Planck2015,planck2018}.}
\label{fig:global_constraints}
\end{figure*}

\subsection{Cosmic Reionization}
\label{sec:reionization}

In Figure \ref{fig:global_constraints}, we illustrate the reionization history predicted by CAT, with the relative role played by different sources of ionizing photons.
In the upper panel, we show the evolution in redshift of the global photo-ionizing emissivity, i.e. the rate of ionizing photons injected into the IGM by the entire galaxy population. Different colored lines represent the contribution of Pop II stars (solid brown), Pop III stars (solid yellow), and AGNs (blue dashed), which we assume to include the entire population of early accreting nuclear BHs. 
For all sources, the intrinsic photon rates are corrected for the adopted escape fractions (Eq. \ref{eq:fesc} for Pop II and Pop III stars and Eq. \ref{eq:fobsUeda} for AGNs).

 The emissivity of Pop III stars initially rises quickly and peaks at $z \sim 20$, to remain then almost constant with $\dot{n}_{\rm \gamma, PopIII} \sim 10^{49} - 10^{50} \, \rm s^{-1} \, Mpc^{-1}$, closely tracing the Pop III SFRD evolution, down to the last few episodes of Pop III star formation at $z \sim 10$. The contribution of Pop II stars shows a smooth increase from $z \sim 23$ to $z \sim 12$, below which it slowly declines, despite the increasing trend of the Pop II SFRD, as a consequence of the decrease in the escape fraction $f_{\rm esc}(z)$. The parametric evolution of $f_{\rm esc}(z)$ described by Eq. \ref{eq:fesc}, translates into values of $f_{\rm esc} \simeq 0.04, 0.10, 0.25$ at $z = 5, 10, 20$, respectively. 
 Despite the large uncertainties that still affect observational constraints, these values are consistent with recent results \citep{finkelstein2019,naidu2020,schaerer2002,mascia2023}, which suggest a global averaged escape fraction between $5 - 10 \%$ at $z<10$, and a rising emissivity with increasing redshift throughout the epoch of reionization. 
It is interesting to note that the Pop II emissivity starts to significantly dominate over the Pop III contribution only at $z \lesssim 16$, when the SFRD of 
Pop III stars is already $\sim 2 \rm \, dex$ below the Pop II one (see Figure \ref{fig:SFR_PopII_PopIII}). This is a result of the different stellar IMF and metallicity of the two populations, with massive and very massive Pop III stars having much harder spectra and higher ionizing photon emissivities per unit stellar mass. 

In our reference model, stars are the primary source of ionizing photons in the IGM. At $z \gtrsim 15$, the BHs hosted in galaxy nuclei are descendants of light seeds, with typical initial masses of $M_{\rm BH} \lesssim 10^3 \, M_\odot$. Their mass growth proceeds in the Bondi-Hoyle gas accretion regime, and is highly inefficient, as shown in \citet{trinca2022}. As a consequence, the first accreting BHs at $z>15$ are intrinsically faint sources of photo-ionizing radiation. At $z \lesssim 15$, heavy BH seeds form, with masses $M_{\rm BH} = 10^5 \, M_\odot$, and their gas accretion is more efficient. As a result, the AGN ionizing emissivity starts to increase, but remains subdominant with respect to the Pop II stellar emission down to $z \sim 5$. 

In the redshift range $4< z < 6.1$ the photo-ionizing emissivity predicted by CAT is in good agreement with the empirical constraints on the ionizing ultraviolet background obtained by \citet{becker2013} and \citet{becker2021}, with unobscured AGNs (blue dashed line) providing the dominant contribution only at $z<5$. The predicted AGN emissivity appears to be consistent with the recent estimates from \citet[][]{Harikane2023bh}, based on the first census of 10 faint broad-line AGNs at $z \sim 4-7$ detected in early JWST observations.

In the central panel of Figure \ref{fig:global_constraints}, we show the IGM fraction of neutral hydrogen, $x_{\rm HI} = 1 - Q_{\rm HII}$, predicted by CAT (red solid curve), compared with empirical constraints presented by \citet[][grey data points]{bolan2022} and obtained with different inference methods, based on Lyman-break galaxy (LBG) samples \citep{morales2021,bolan2022}, constraints from the dark pixel fraction in the Ly-$\alpha$ and Ly-$\beta$ forest \citep{McGreer2015}, and quasar damping wings \citep{davies2018,greig2019,wang2020}. We find that the ionization process is complete around $z \sim 5.5$, in agreement with the end of the reionization epoch expected from observational constraints. In addition, we find a good agreement with observations at $z \lesssim 8$, 
although it is important to notice that empirical estimates at $z>7$ are inferred from different tracers and show a large scatter.
In particular, analysis based on damping wings of bright quasars \citep{davies2018} suggest a lower neutral fraction at $z \sim 7.5$ with respect to estimates obtained from LBG samples, and are in closer agreement with CAT predictions. These uncertainties in the higher redshift range will improve with forthcoming deep JWST surveys at $z = 7-10$, which will put tighter constraints on the evolution of the IGM neutral fraction with combined photometric and spectroscopic data.

Finally, the lower panel of Figure \ref{fig:global_constraints} shows the evolution of the electron scattering optical depth $\tau_{\rm e}$, compared to recent constrains obtained by the Planck cosmological survey and, in particular, with the results from \citet[][]{planck2018}. We find a value $\tau_{\rm e, \rm CAT} = 0.067$, which is consistent within $2\sigma$ with the Planck estimates $\tau_{\rm e, \rm Planck} = 0.054 \pm 0.007$.\\

\section{Predicting galaxy UV emission} 
\label{sec:UVlum}

In the previous sections, we showed that CAT model predictions appear to be consistent with current constraints on the redshift evolution of the SFRD and the history of cosmic reionization. 
The natural extension of this analysis is therefore to characterize the luminosity distribution of the population of high-redshift galaxies, and follow its evolution in time. In what follows, we present CAT model predictions for the galaxy UV luminosity function at $4 \leq z \leq 16$, considering the contribution of stellar populations - including Pop III stars -,  BH accretion, and the effects of a gradual change in the stellar IMF with redshift and metallicity, as suggested by recent hydrodynamical simulations \citep{chon2021, chon2022}.

\begin{figure}
\centering
  \includegraphics[width=1\linewidth]{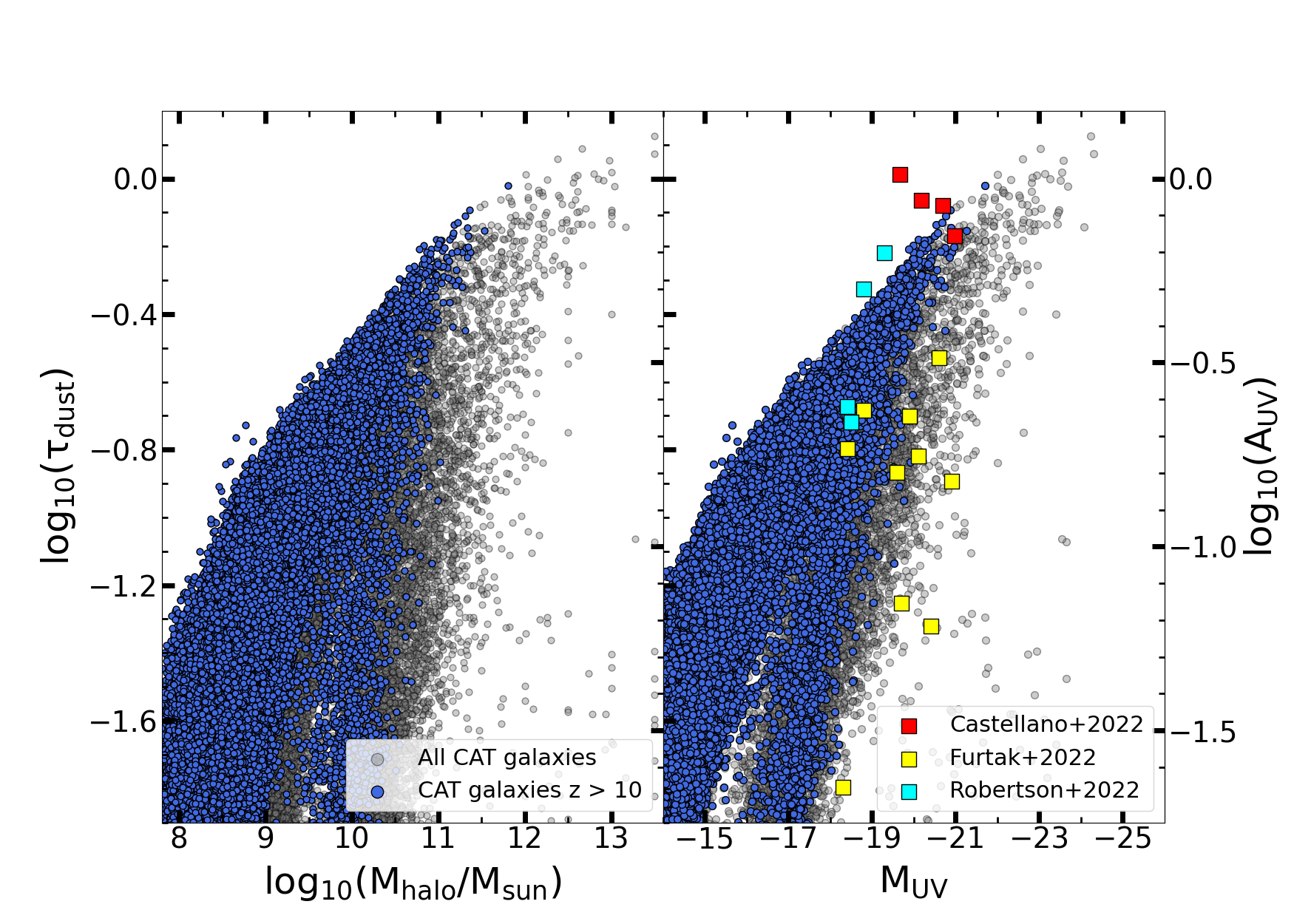}
    \caption{Predicted values of $\rm \tau_{\rm dust}$ (left axis) and $\rm A_{UV}$ (right axis) as a function of the halo mass (left panel) and UV magnitude (right panel). The properties of the galaxy population predicted by CAT at $z>10$ (blue circles) are compared with the results obtained for high-z galaxies observed with JWST by \citet{Castellano2023}, \citet{furtak2023} and \citet{robertson2023}. As a reference, we also show with grey data points the values predicted by CAT for the galaxy population at all redshifts.}
    \label{fig:Tau_dust}
\end{figure}

\subsection{Galaxy UV luminosity function at $4 \leq z \leq 16$ }
\label{sec:galaxyUVlum}
For the present study, we improve our modeling of the total UV emission arising from each galaxy with respect to what has been presented in \citet[][]{trinca2022}, where the galaxy intrinsic UV luminosity was obtained from the SFR adopting a standard conversion factor \citep[][]{madau2014b}. 

\begin{figure*}
\centering
  \includegraphics[width=18.0cm]{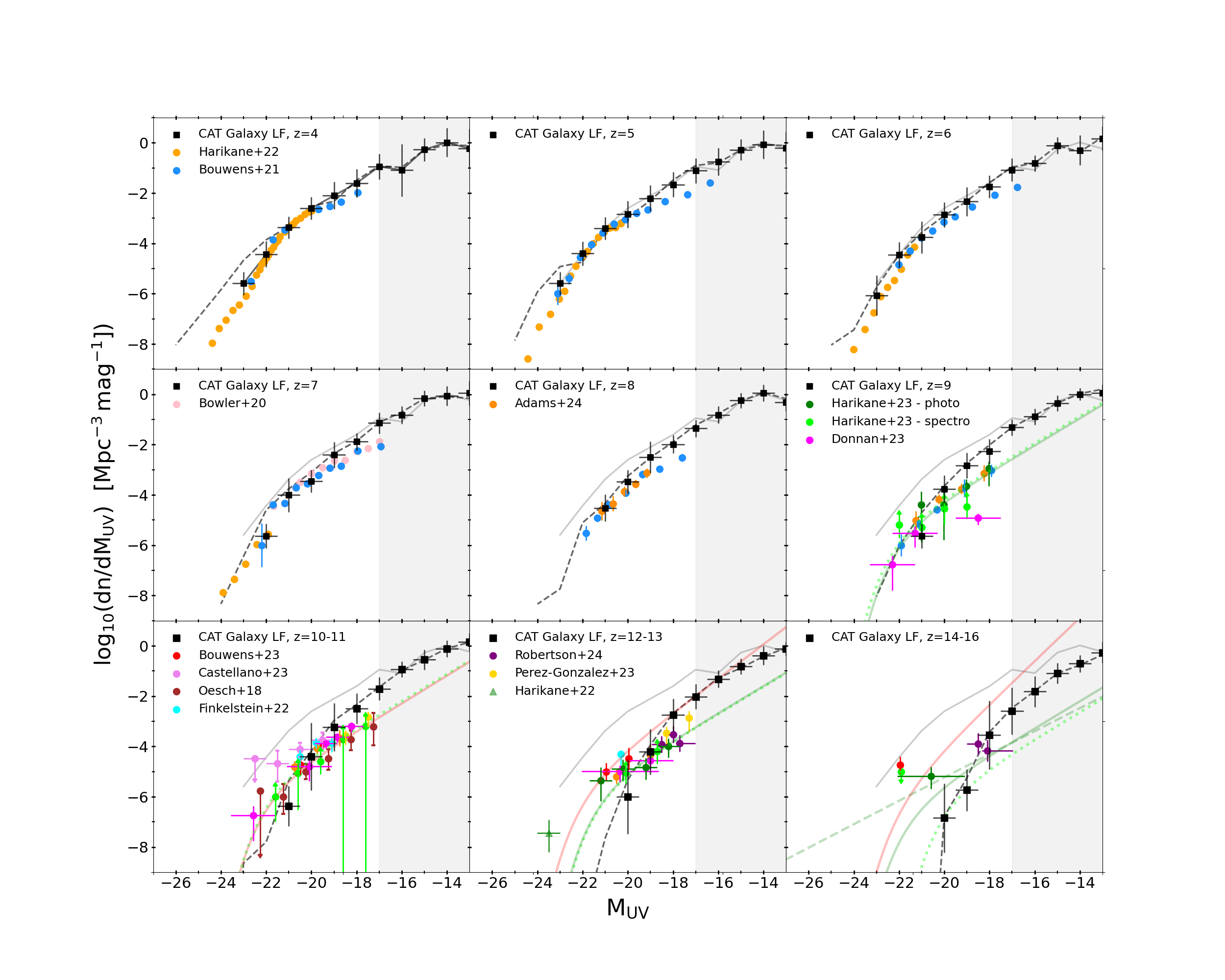}
    \caption{Observable (obscuration-corrected) galaxy UV LF between $z=4$ and $z=16$. CAT predictions for the dust-corrected (black data points) and intrinsic (black dashed lines) distributions are compared with observational data from \citet[brown]{Oesch2018} \citet[pink]{Bowler2020}, \citet[blue]{Bouwens2021}, \citet[red]{Bouwens2023}, \citet[dark green]{harikane2022a,Harikane2023photo},\citet[light green]{Harikane2023spectro}, \citet[cyan]{Finkelstein2022,Finkelstein2022b},  \citet[orange]{harikane2022b}, \citet[magenta]{Donnan2023}, \citet[dark orange]{Adams2023b}, \citet[violet]{Castellano2023}, \citet[dark violet]{robertson2023} and \citet[yellow]{perezgonzalez2023}. The figure shows that stellar emission can account for the observed UV luminosity evolution from $z \sim 4$ to $z \sim 10$. At higher redshifts, similar to other standard galaxy formation models, CAT predictions fail to account for the UV bright-end of the luminosity function sampled by JWST observations. The grey shaded area highlights the population of sources with $M_{\rm UV} > -17$, which contribute to the unresolved SFRD shown in Figure \ref{fig:SFR_PopII_PopIII}. In all panels, we show as a reference the UV LF predicted by CAT at $z=4$ (solid grey lines) for an easier comparison with the distribution at higher redshift. In the lower panels, we also show the best-fit distributions obtained by \citet[$z \sim 10,13,17$, red lines]{Bouwens2023} and \citet[$z \sim 9,12,16$, green lines]{Harikane2023photo}, assuming a Schechter function (solid) or a double power-law (dashed) distribution. The light green-dotted lines report instead the best-fit of the LF at $z \sim 9,10,12$ and $16$ obtained by \citet{Harikane2023spectro} considering only galaxies with a spectroscopic confirmation.}
    \label{fig:GLF_staronly}
\end{figure*}

We compute the UV luminosity of each galaxy, $L_{\rm UV, *}$, by summing over the emission of its active stellar populations,  adopting age and metallicity dependent SEDs, as explained in section \ref{sec:ModelReionization}.  
Following \citet{mancini2016}, we account for dust obscuration by correcting the galaxy UV luminosity as:
\begin{equation}
L_{\rm UV, obs} = L_{\rm UV} \, \exp[-\Sigma_{\rm gas} \, {\cal D} \, k_{\rm UV}]
\label{eq:DustObs}
\end{equation}
where $\Sigma_{\rm gas} = M_{\rm gas}/\pi r_{\rm d}^2$ is the gas surface density within a radius $r_{\rm d} = 0.18 \, r_{\rm vir}$ \citep{mo1998}, ${\cal D}$ the dust-to-gas mass ratio, and $k_{\rm UV}$ is the extinction coefficient per unit mass in the energy band of interest. 
The value of $k_{\rm UV}$ has been inferred considering the extinction curve of the Small Magellanic Cloud (SMC, \citealt{weingartner2001}).
Since CAT is able to track the fraction of the ISM that resides in the warm/hot diffuse medium (see section \ref{sec:metal}), we assume here a simple screen model, where the optical depth is computed considering the contribution of the diffuse gas and dust mass inside the galaxy. 
However, this modeling may appear oversimplified if compared to more sophisticated two-phase dust extinction models \citep[see e.g.][]{mancini2016}, and may underestimate the impact of dust obscuration. Indeed, the increased compactness of high-redshift galaxies would result in lower values of $r_{\rm d}$, leading to a larger column density and dust optical depth. At the same time, while here we assume every galaxy to form a disk, these early systems are expected to show more complex morphologies and dust distributions, with UV dark and bright regions within the same system. For these reasons, the dust obscuration provided by our model has to be considered as an average over systems with comparable halo masses, but very likely different relative stellar/dust geometries and dust optical depths.
In Figure \ref{fig:Tau_dust} we show the attenuation predicted by CAT (grey points) as a function of the host halo mass (left panel) and UV magnitude (right panel), highlighting the systems at redshift $z>10$ (blue points). When compared with the extinction values derived by \citet{Castellano2023, furtak2023, robertson2023} for a sample of high-redshift galaxies observed with JWST at $z \gtrsim 9$, we see that - despite representing a lower limit to the potential dust obscuration - our simplified approach predicts a range of values consistent with the observations. \\ 

In Figure \ref{fig:GLF_staronly} we show the predicted evolution of the galaxy UV luminosity function in the redshift range $4 \leq z \leq 16$. Here we only consider the UV luminosity coming from stellar emission, $L_{\rm UV,*}$ (i.e. we do not consider the additional contribution to the UV emission from accreting BHs). At $z > 10$, given the restricted number of sources currently observed and the potential uncertainties in their redshift determination, we decided to show the galactic LF predicted by CAT averaged over three redshift ranges $10 < z < 11$, $12 < z < 13$, $14 < z < 16$.

CAT results are compared to several observational data (see \citealt{Harikane2023photo} and references therein), including results coming from JWST observations \citep{Castellano2023,Harikane2023photo, Harikane2023spectro,Finkelstein2022b,Donnan2023,Adams2023b,robertson2023,perezgonzalez2023}. 
The model predictions are in good agreement with observational data in the redshift range $z \simeq 4 - 9$.

At $z \sim 10 - 13$, the mild evolution observed in the galaxy LF is well reproduced by CAT, except for the bins of highest luminosity, at $M_{\rm UV} < -19$, where, despite the large statistical uncertainties, the galaxy number density predicted by the model is smaller than the value inferred by some observational studies. These, however, show significant variations, depending on the surveyed area and JWST program considered, hinting to the potential presence of galaxy overdensities in some of these fields \citep{Castellano2023}, to the effect of cosmic variance \citep{yung2023, Adams2023}, as well as to the possible contamination of low-redshift systems in the photometric samples \citep{Naidu2022, Zavala2023, ArrabalHaro2023}. 

Finally, at $z \sim 14-16$, CAT predictions are compared with the recent constraints from \citet{Harikane2023photo} and \citet{Bouwens2023}, based on two candidate galaxies with estimated redshift $z_1 = 16.25^{+0.24}_{-0.46}$ and $z_2 = 16.41^{+0.66}_{-0.55}$. At this very early epoch, the model predicts a number density of bright galaxies that appears to be significantly lower with respect to what is suggested by JWST observations, with a number density of sources with $M_{\rm UV} \sim -20$, which is $\sim 1.2 \, \rm dex$ lower than the best-fit distribution obtained by \citet{Harikane2023spectro} assuming a Schechter function (see also Figure \ref{fig:NumbDensity_comparison} for a quantitative comparison). It has to be noted, though, that if the spectroscopic confirmation were to favour redshift values on the lower bound of the uncertainty range, CAT predictions would stand at $\sim 1 \sigma$ from the constraint proposed by \citet{Harikane2023spectro}. 
In addition, we also show the best-fit distribution obtained by \citet{Harikane2023photo} at $z \sim 16$, where they perform the analysis considering only sources with a spectroscopic confirmation. Given the lack of any spectroscopically confirmed galaxy at $z > 14$, they extrapolated the best-fit Schechter function obtained for $z \sim 9-12$ toward higher redshift. In this case CAT predictions for $M_{\rm UV} \leq - 19$ are consistent with the observed LF, which,  however, needs to be interpreted as a lower limit. Hence, spectroscopic identification of galaxy candidates observed at $z \simeq 14 - 16$ will be crucial to confirm the excess of UV bright sources compared to current model predictions. 
CAT predictions are also compared with recent observational constraints on the galaxy LF at $\rm M_{\rm UV} \sim -18$ by \citet{robertson2023}, based on the photometric detection of three candidate galaxies at $z > 13.5$. Interestingly, the number density predicted by CAT is in good agreement with the data at these fainter luminosities.

To evaluate the potential impact of dust obscuration in the comparison between CAT predictions and observational constraints, we overplot in Figure \ref{fig:GLF_staronly} the intrinsic galaxy UVLF at different redshifts as black-dashed lines. It is  clear how dust attenuation mostly affects the range of luminosity covered by the galaxy population, while it does not affect the overall normalization of the distribution. Indeed, only the bright-end of the LF in the lower redshift range, $z \lesssim 7$, is significantly lowered when accounting for obscuration, and at $z \gtrsim 10$ even the intrinsic distribution cannot reproduce the high number density of bright galaxies observed by JWST.
In the following section, we will explore whether the contribution to the UV luminosity from accreting BHs can partially relieve the current tension between CAT predictions and the bright end of the UV LF at $z \sim 14-16$ derived from photometric candidates \citep{Harikane2023spectro, Bouwens2023}.

\subsubsection{Evolution of the galaxy faint-end}

\begin{figure}
\centering
  \includegraphics[width=1\linewidth]{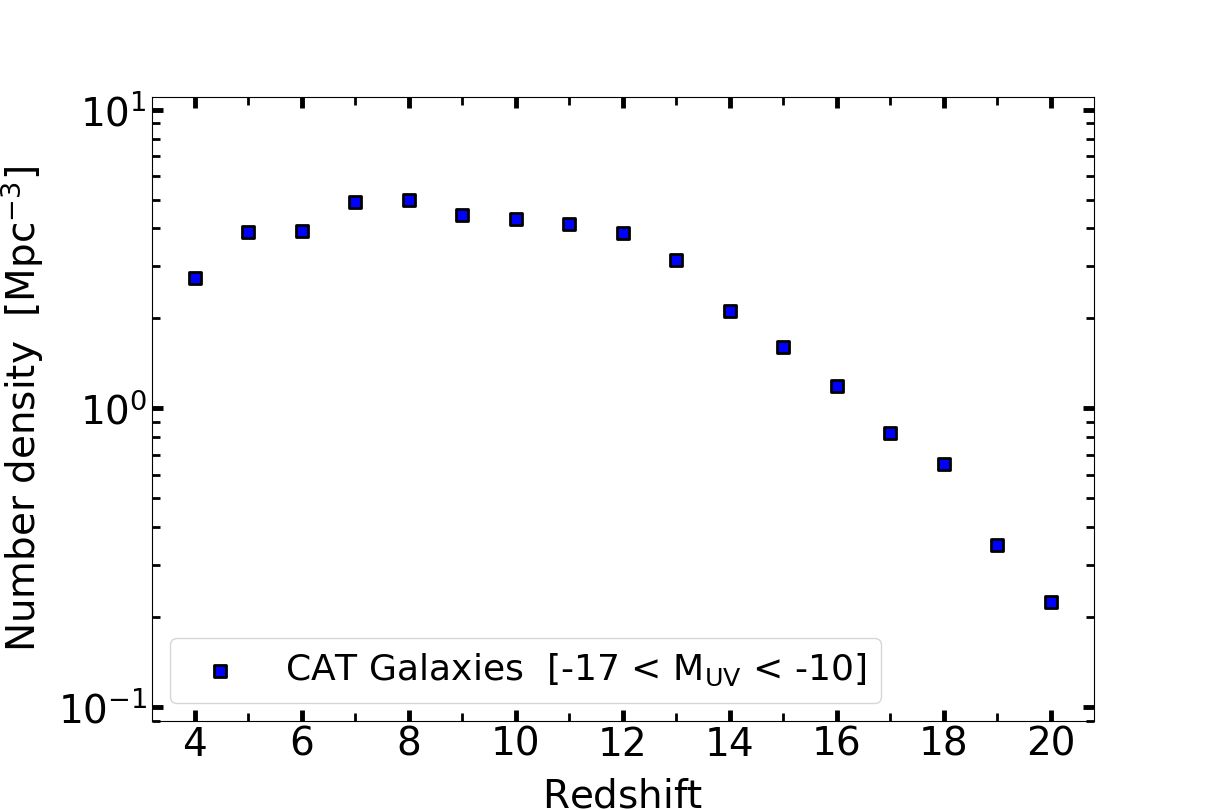}
    \caption{Redshift evolution of the cumulative number density of galaxies with $\rm -17 < M_{\rm UV} \ -10$ predicted by CAT.}
    \label{fig:faintGalND}
\end{figure}

An interesting prediction of the CAT model is the number density of sources fainter than $M_{\rm UV} = - 17$ at redshift $z \gtrsim 9$, which appears to be larger than the value obtained by extrapolating the faint end of the observationally estimated LF. This might be due to the incompleteness of the observed samples at these faint magnitudes, or to a higher impact of dust attenuation \citep[see e.g.][]{barrufet2023}. At the same time, it may point to a too efficient rate of star formation or to a too inefficient feedback in the galaxies populating the faint end of the galaxy UV LF at these epochs in our model. 
We also note that, at these faint magnitudes, the UV LF predicted by CAT shows a very mild evolution between $z \sim 14-16$ and $z=4$, which can be appreciated in Figure \ref{fig:GLF_staronly} by comparing the predicted LF at $z>4$ with the light grey solid line, which represents the LF predicted at $z = 4$. This is an interesting prediction of the model, reflecting the interplay of different physical processes during the evolution of fainter star forming systems. To investigate more closely this low luminosity regime, in Figure \ref{fig:faintGalND} we show the predicted evolution in redshift of the cumulative number density of galaxies with $\rm -17 < M_{\rm UV} \ -10$. 
At early times, a progressive increase in the number density of faint galaxies is observed down to $z \simeq 12$. This reflects the widespread availability of gas in the early phases of galaxy evolution, and the increasing fraction of faint galaxies hosting active Pop III stellar populations (as it will be further discussed in Section \ref{sec:PopIII_LF}), which can boost the UV luminosity of less massive galaxies. 
At later times, the number density of faint galaxies remains almost constant, tracing a plateau which extends between $8 \lesssim z < 12$. This behaviour reflects the increasing importance of mechanical feedback due to SNe in regulating star formation of less massive systems. In addition, the gradual build up of the Lyman-Werner background leads to a decrease in the star formation efficiency of mini-halos and, consequently, of the galaxy number density at the faint end. At $z \lesssim 8$, the number density starts to mildly decline, 
leading to a reduction in the number density of faint galaxies of ~50\% between $z = 8$ and $z = 4$. This effect is due to photo-heating of reionized regions, which prevents gas accretion onto mini-halos. The net result of this evolution is a faint-end slope of the UV LF which increases by $\lesssim 1 \rm \, dex$ from $z \sim 16$ to $z=4$. A similar mild evolution has been recently obtained also by \citet{williams2023}. Performing detailed hydrodynamical simulations, they found that streaming velocities, i.e. the supersonic relative motions between dark matter and baryonic overdensities, have a strong impact in enhancing the faint end of the LF at $z \sim 12$, since the suppression of star formation inside small halos at very early times leads to brighter dwarf galaxies at later times. While this is an intriguing result on its own that needs additional analysis, in the following sections we primarily focus on the model predictions for the bright-end of the galaxy LF, at $M_{\rm UV} < -18$.

\begin{figure}
\centering
  \includegraphics[width=1\linewidth]{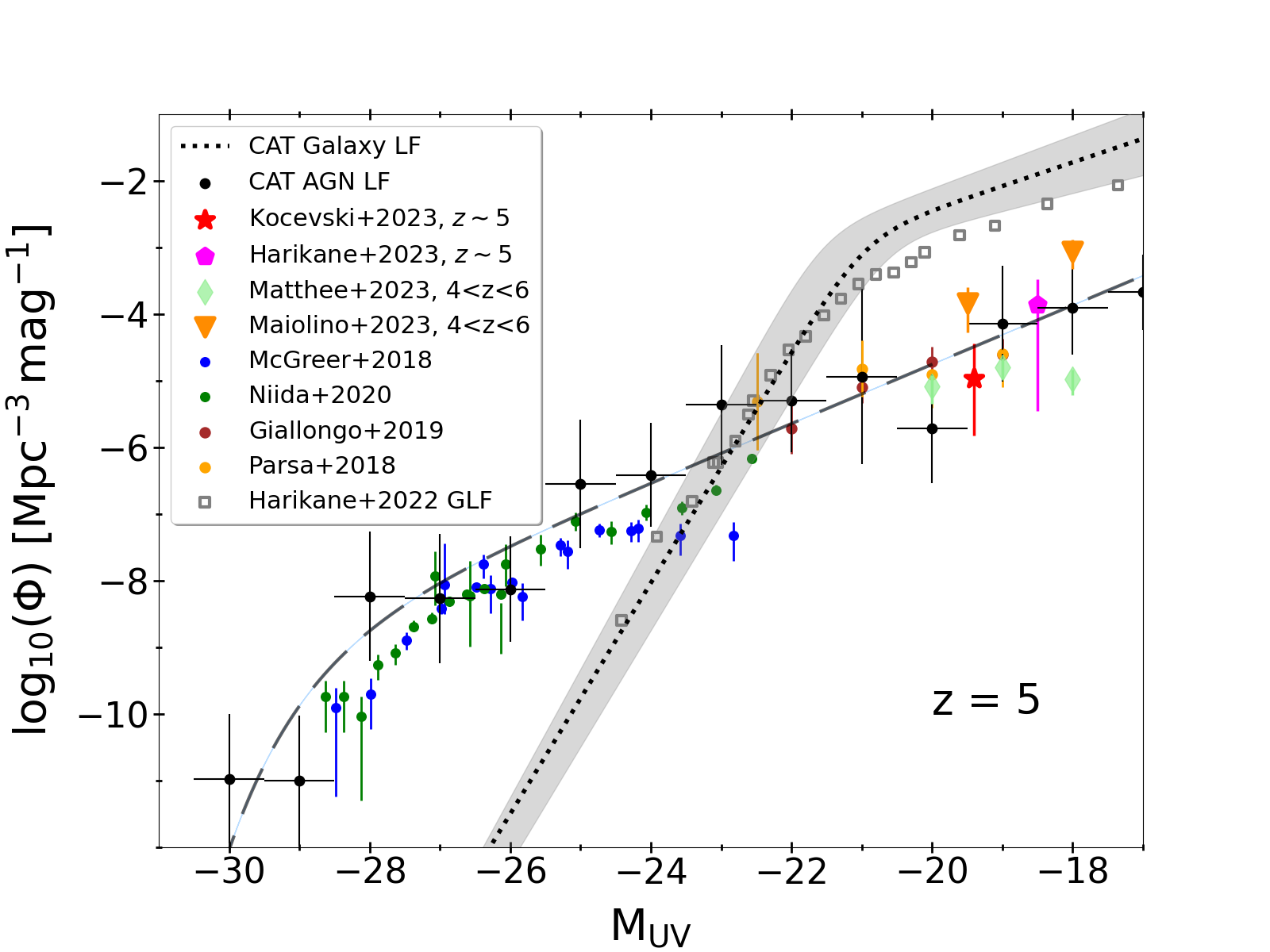}
    \caption{UVLF predicted by CAT for the AGN and galaxy population at redshift $z = 5$. Similarly to Figure \ref{fig:GLF_staronly}, the galaxy luminosity function is computed considering only the emission from stellar populations (the dotted line and gray shaded region represent, respectively, the best-fit and $1 \sigma$ spread of the distribution), and it is compared with observations from \citet{harikane2022a} (empty gray squares). The AGN luminosity function is shown by the black filled dots, and fitted with the black dashed line. This is compared to observations by \citet{mcgreer2018, parsa2018, giallongo2019, niida2020} and to the recent estimates based on JWST data by \citet{Kocevski2023}, \citet{Harikane2023bh}, \citet{Matthee2023} and \citet{Maiolino2023AGN}. CAT model predictions are in good agreement with available constraints at $z = 5$.}
    \label{fig:AGN_LF5}
\end{figure}

\subsection{Can BH accretion explain the UV luminosity function at $z > 10$?}
\label{sec:AGNcont}
In addition to stellar emission, we also account for the possible contribution of the UV luminosity emitted by the nuclear BH, $L_{\rm UV, AGN}$. In fact, for high redshift systems, AGN contamination in the UV rest-frame emission might be significant \citep{Pacucci2022}, as we will analyze more in detail below.
\begin{figure*}
\centering
  \includegraphics[width=1.0\linewidth]{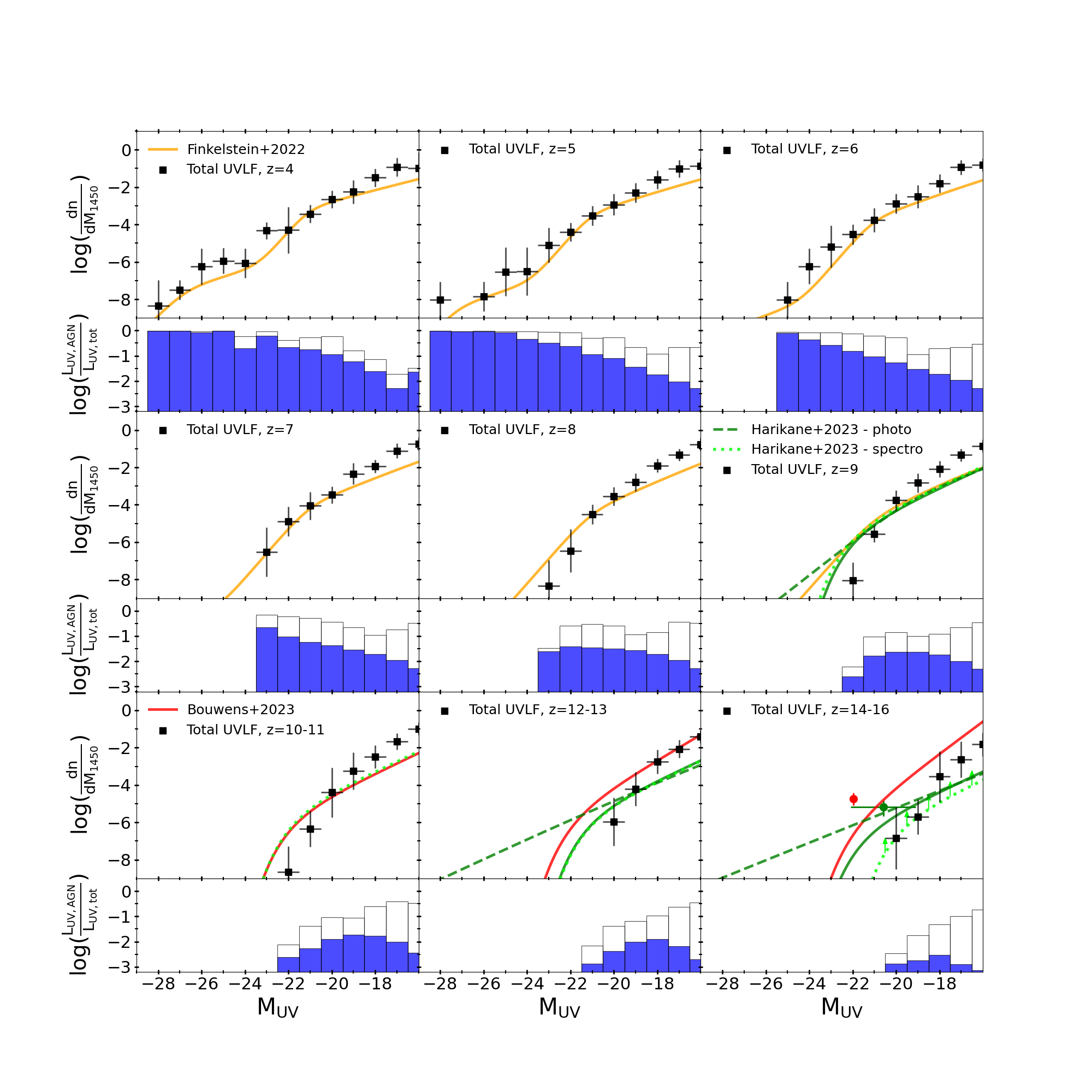}
    \caption{Observable (obscuration-corrected) global UV luminosity function (galaxy + AGN) predicted by CAT at $z = 4,5,6,7,8,9$ and in the redshift ranges $10 \leq z \leq 11, 12 \leq z \leq 13$ and $14 \leq z \leq 16$ (from top left to bottom right). For each panel, we also show the mean (filled histograms) and maximum (empty histograms) AGN contribution to the total UV luminosity (galaxy + AGN) in each magnitude bin. In the upper panels, CAT predictions are compared with the compilation of observational data presented by \citet[][]{finkelstein2022d} for the global UV LF. In the lower panels, we show instead the best-fit distributions obtained by \citet[][]{Bouwens2023} ($z \sim 10,13,17$, red lines) and \citet[][]{Harikane2023spectro,Harikane2023photo} ($z \sim 9,10,12,16$, green lines) based on observational constraints on the UV LF at $z \gtrsim 9$ coming from recent JWST data.}
    \label{fig:GLF_AGN}
\end{figure*}
From the BH accretion rate (see Eq. \ref{eq:BHLacc}), we estimate the BH bolometric luminosity as:
\begin{equation}
    L_{\rm bol} = \epsilon_{\rm r} \, \dot{M}_{\rm accr} \, c^2,
\end{equation}
and then convert this into the UV luminosity relying on the bolometric correction proposed by \citet{duras2020}, and assuming $L_{\nu} \propto \nu^{-0.44}$ \citep[as described in detail in ][]{trinca2022}. 

A thorough comparison between the AGN LF predicted by CAT in the UV and X-ray bands has been presented by \citet{trinca2022}, finding a good agreement between our reference model prediction and the available observational constraints. To provide an example, here we show in Figure \ref{fig:AGN_LF5} the predicted AGN UV luminosity function at $z = 5$ (filled black data points and black dashed line). This is compared to observational constraints from \citet{mcgreer2018} and \citet{niida2020} at the bright-end ($M_{1500} < -22$), and from \citet{parsa2018, giallongo2019, Kocevski2023,Harikane2023bh,Maiolino2023AGN} and \citet{Matthee2023} at the faint-end. We also show the galaxy UV LF predicted by CAT (the black dotted line and grey shaded region represent, respectively, the double power-law best fit and $1\sigma$ spread of the distribution) and observed (empty grey squares, \citealt{harikane2022a}). The comparison confirms that CAT model predictions are in good agreement with the observed galaxy and AGN UV luminosity function at $z = 5$, reproducing remarkably the surprisingly high number density of faint AGNs suggested by the most recent estimates based on JWST data \citep{Matthee2023,Maiolino2023AGN,Greene2023}. 

We then recompute the UV LF considering both emission from stars and accreting BHs, $L_{\rm UV,*} + L_{\rm UV, AGN}$ at $z = 4 - 16$, and estimate the mean and maximum contribution of the AGNs to the total galaxy emission in different ranges of magnitude. This is shown in Figure \ref{fig:GLF_AGN}.

\begin{figure*}
\centering
  \includegraphics[width=17cm]{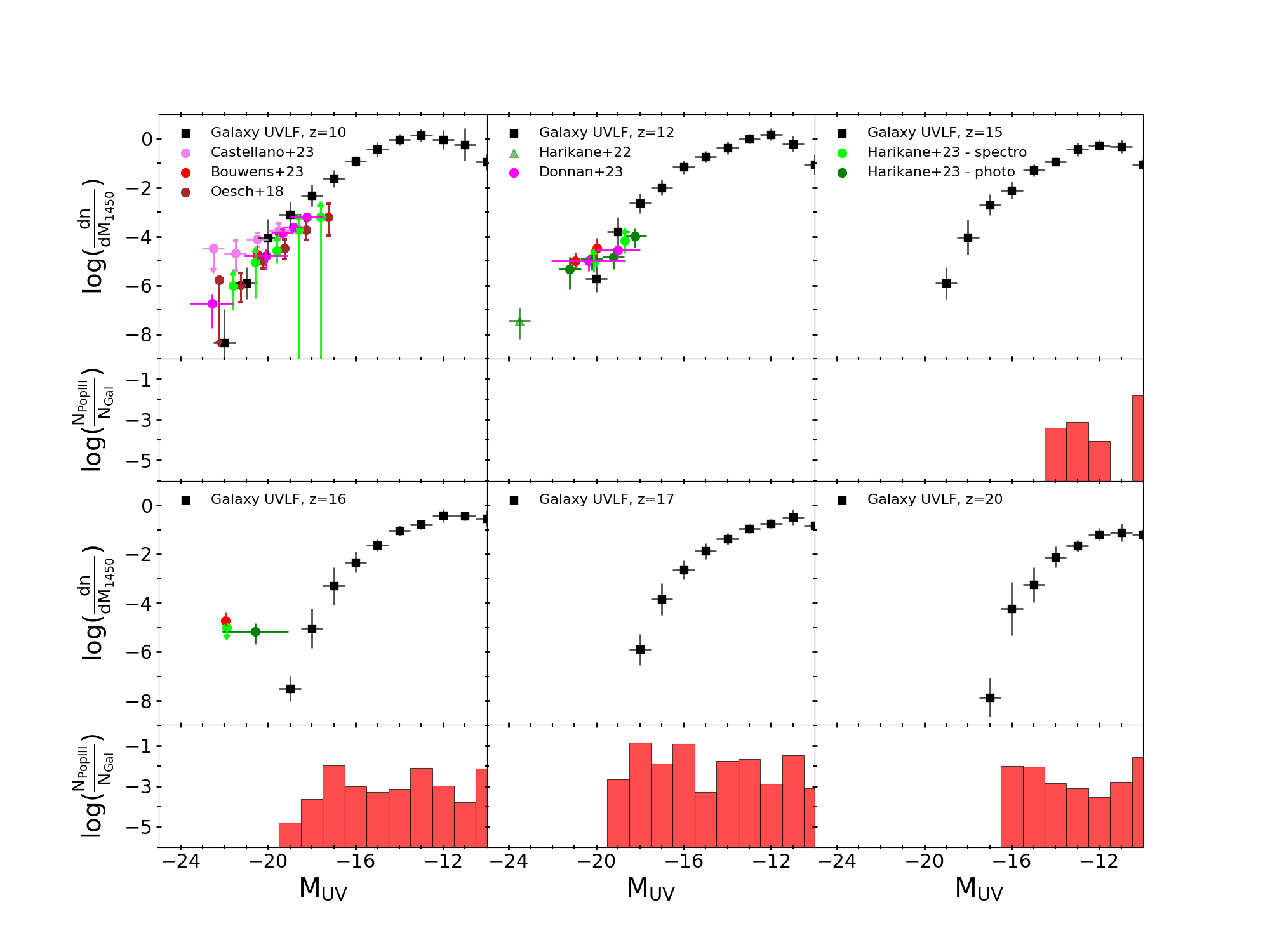}
    \caption{Galaxy UV LF predicted by CAT at $z =10, 13, 15, 16, 17$ and $20$ (from top left to bottom right). Below each panel, the red histograms show the fraction of galaxies hosting active Pop III stellar populations in each bin of magnitude. Observational data follow the same color-coding adopted in Figure \ref{fig:GLF_staronly}.}
    \label{fig:GLF_PopIII}
\end{figure*}
At $z = 4 - 9$, the total (galaxy + AGN) UV LF predicted by CAT is compared to the fit of the combined LF for both AGN and star forming galaxies derived by \citet{finkelstein2022d}, finding a good agreement. At $z \ge 10$, we compare our predictions with the best-fit distributions proposed by \citet[][]{Bouwens2023, Harikane2023spectro}, also shown in Figure \ref{fig:GLF_staronly}. On average, we find that the contribution of accreting BHs to the galaxy UV luminosity is negligible at these redshifts (see the filled histograms). At $10 \lesssim z \lesssim 15$, during the formation epoch of heavy BH seeds, the AGN emission contributes on average to $\sim 1 - 3 \%$ of the total UV emission, with the largest contribution being at most $\lesssim 10 \%$ in the brightest bins of magnitude, at $\rm M_{\rm UV} \lesssim -19$ (see the empty histograms). At even higher redshifts, $z \gtrsim 15$, the AGN emission is even smaller, due to the stunted growth of light seeds predicted by our reference model \citep[see][]{trinca2022}. Considering the joint UV emission of stars and accreting BHs, we still find that the number density of systems with $M_{\rm UV} \sim -20$ predicted by CAT is $\sim 1.2 \, \rm dex$ lower than the best-fit distribution obtained by \citet{Harikane2023spectro} assuming a Schechter function, similarly to what was found when only stellar emission was considered. At the same time, model predictions closely match the distribution obtained by \citet{Harikane2023photo} considering only spectroscopically confirmed galaxies (see Figure \ref{fig:NumbDensity_comparison}).

At $7 \lesssim z \lesssim 10$, we observe a growing impact of AGNs to the total UV emission, with an average contribution between $1 - 20 \% $ of the galaxy luminosity, and peaks of maximum contribution reaching $> 50 \%$ of the total emission, especially in the brightest luminosity bins.
Finally, at $z < 7$, we predict the AGN population to dominate the bright-end of the total UV LF, accounting on average for $20 - 100\%$ of the emission coming from systems brighter than $M_{\rm UV} ~ -22$. The average AGN contribution decreases for fainter galaxies, where it appears to be subdominant, though it might still reach $> 50 \%$ of the total UV emission for specific systems. 

We thus find that the AGN contribution does not have a significant impact on the emission of $z \gtrsim 10$ galaxies, and it cannot resolve the discrepancy between model predictions and observations on the number density of bright sources with $M_{\rm UV} < -19$.
However, while the AGN UV luminosity appears to be - on average - subdominant in fainter high-redshift galaxies, detailed selection criteria applied to deep JWST surveys \citep{Trinca2023} might be able to identify systems hosting the brightest and more luminous BHs from the general population \citep{Nakajima2022,Goulding2022, volonteri2022}, enabling to constrain their contribution to the total luminosity function.

Recent works advocated for super-Eddington accretion to explain the observed deviation from the local scaling relation $\rm M_{\rm BH}/M_{\rm *}$ of the population of high-redshift AGNs. This could be relevant to interpret the bright excess observed in the galaxy UV LF, since the presence of BHs accreting very efficiently already at early epochs might in principle provide a higher contribution to the global emission of the host galaxy with respect to what is shown in Figure \ref{fig:GLF_AGN}. 
In a recent work \citep{schneider2023} we compared our model predictions for the population of accreting BHs with recent JWST observations of $z>5$ AGNs for both an Eddington-limited and a super-Eddington BH growth scenario, where brief periods of enhanced accretion are assumed to be triggered during major galaxy mergers. 
Both models are able to account for current observations of extreme BH candidates at $z>8$, despite a super-Eddington accretion history predicts a significantly higher number density of massive systems at these high redshifts, and they both provide similar predictions for the bright end of the AGN luminosity function \citep[see][]{trinca2022}. However, given the typical short timescales of the super-Eddington accretion phase, on the order of a few Myr, even in this alternative growth scenario it is unlikely that accreting BHs will provide a significant boost to the galaxy LF at early times. A more thorough study is however required to support this conclusion and hence we defer a deeper analysis to a future work, where we will implement a more refined treatment of the super-Eddington accretion mechanism to investigate its potential impact on the overall galaxy evolution through cosmic time.

\subsection{Can a top-heavy IMF explain the UV luminosity function at $z > 10$?}
\label{sec:PopIII_LF}

The UV emissivity of stellar populations is very sensitive to their IMF, metallicity and ages. \citet{Mason2023} argue that current $z \gtrsim 10$ galaxies observed by JWST are dominated by systems with young ages ($\lesssim 10$ Myr) and high star formation rates. \citet{Harikane2023spectro} show that, for a given star formation rate, the UV luminosity from Pop III stars characterized by a top-heavy IMF is $\sim 3  - 4$ times larger than that of Pop II stars with a normal Salpeter-like IMF. Here we first discuss CAT model predictions regarding the UV emission from Pop III stars, and then we estimate the potential effect on the galaxy UV LF of a more gradual transition from a top-heavy IMF for Pop III stars to a standard, Salpeter-like IMF for Pop II stars, modulated by metallicity and redshift, as suggested by recent numerical simulations of star cluster formation in low-metallicity environments resolving individual forming stars \citep{chon2020, chon2022}.

\subsubsection{The role of Pop III stars: CAT model prediction}

CAT models stellar populations in high redshift galaxies depending on whether their initial metallicity is smaller or larger than the critical metallicity $Z_{\rm cr}$. We therefore have an abrupt transition in the IMF, from a top-heavy IMF for Pop III stars ($Z < Z_{\rm cr}$) to a normal, Salpeter-like IMF for Pop II/I stars ($Z \geq Z_{\rm cr}$). 
We have shown in Figure \ref{fig:GLF_staronly} that, according to CAT, stellar emission, including Pop III stars, cannot account for the overabundance of bright UV galaxies observed by JWST. 
To further clarify this point, in Figure \ref{fig:GLF_PopIII}, we report the fraction of galaxies hosting active Pop III populations predicted by CAT at $z = 10, 13, 15, 16, 17$ and 20 (red histograms), together with the corresponding galaxy UV LF at the same redshift. 
At $15 \lesssim z \lesssim 20$, the fraction of galaxies hosting Pop III stars is always relatively small, and progressively shifted to fainter luminosity bins. Interestingly, however, we find that 10\% of some of the brightest systems at $z \sim 17$, with $\rm M_{UV} \sim -18$ host active Pop III stars. This may be ascribed to a less efficient chemical feedback at these very early times, where subsequent generations of Pop III stars form before the medium is enriched above the critical metallicity, leading to a transition in the stellar IMF. The occurrence of active Pop III stars inside bright and more massive galaxies might also be favoured by mergers with smaller metal-free DM halos, which are supposed to occur with higher frequency at very high redshift. 
At $z \lesssim 16$, chemical feedback leads to a drop in the occupation fraction of active Pop III stars, especially at the bright-end of the distribution. At $z \sim 15$ Pop III populations survive only in fainter and less evolved galaxies, where the gas still maintains a pristine composition. In these smaller halos, Pop III star formation is less efficient and leads preferentially to the formation of stars at the lower mass end of the Pop III IMF, which evolve on longer timescales (see Figure \ref{fig:PopIII_lt}).
Finally, below $z \sim 15$ CAT predicts a sharp transition in the brightest systems towards Pop II star formation. Thereafter, Pop III stars continue to form only inside more pristine and fainter galaxies, with $M_{\rm UV} \gtrsim -10$, despite a sustained Pop III star formation survives down to $z \sim 10$, where the rapid enrichment leads to a complete suppression of metal-poor star formation, as shown in Figure \ref{fig:SFR_PopII_PopIII}.

An important caveat of the current model is represented by the assumption of homogeneous metal enrichment and radiative feedback. Various techniques based on statistical descriptions of the DM halo distribution \citep{dijkstra2014feedback, salvadori2014, inayoshi2014, sassano2021} or on N-body simulations to reconstruct their spatial distributions and redshift evolution (see e.g. \citealt{Visbal2020, spinoso2022, hartwig2022, Ventura2024} for recent studies) 
have been developed to allow semi-analytical models to account for these spatial inhomogeneities. Our model lacks this information, although the DM halo merger trees generated with \textsc{galform} \citep[][]{parkinson2008,trinca2022} enables us to explore different overdensities, sampling at the same time less- and more-evolved environments. 

In addition, small and large scale gas dynamics and turbulence have been shown to be very important to allow the formation of Pop III and Pop II stellar populations in different regions of the same galaxy \citep{tornatore2007population, johnson2013, pallottini2015, xu2016, sarmento2019, LiuBromm2020, sarmento2022}. Recent cosmological simulations by \citet{Venditti2023} show that Pop III SF might extend down to $z \sim 6 - 8$ in the outskirts of more metal-enriched galaxies. In particular, they predict that active Pop III stars might survive in $\geq 10\%$ of massive galaxies with $M_* \geq 3 \times 10^9 M_\odot$ at $z \simeq 6.7$, although with a Pop III/Pop II mass fraction $\leq 0.1$ \%. This suggests that Pop III stars formed in particularly unpolluted regions of standard main sequence galaxies might survive below the redshift range predicted by CAT, and in a larger fraction of galaxies. 
However, even in this scenario, it is very unlikely that Pop III stellar systems will provide a significant contribution to the observed UV luminosity function. The same numerical simulations show that the fraction in mass of Pop III stars inside the galaxies which mostly contribute to the $z > 10$ LF ($ \rm M_{*} \sim 10^8 - 10^{11} M_{\odot}$) is between $\rm 10^{-2} \, -\, 10^{-4}$. Pop III dominated systems are expected to have fainter luminosities ($\rm M_{\rm UV} \lesssim 11$) and lower stellar masses (see e.g. the recent observations by \citealt{Vanzella2023} and \citealt{Maiolino2023popIII}).
\begin{figure*}
\centering
  \includegraphics[width=18.0cm]{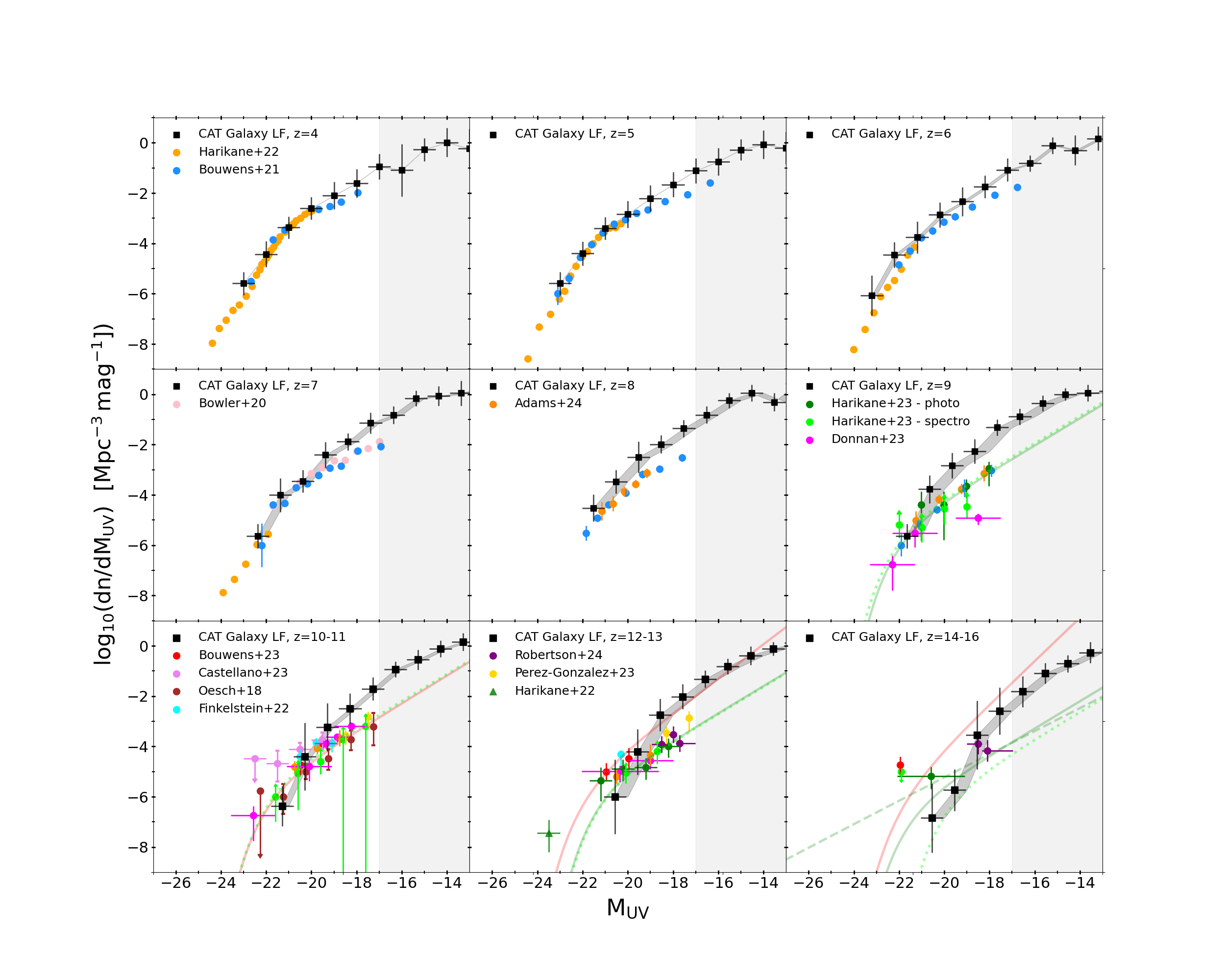}
    \caption{Galaxy UV LF, as in Figure \ref{fig:GLF_staronly}, but assuming the $\cal{K}_{\rm FUV}$ expected for a stellar population of $10 \rm \, Myr$ with a transitional IMF, as described in the text. Black shaded regions show how the predicted LFs are shifted towards higher luminosities with respect to the previous distribution. Observational data follow the same color-coding adopted in Figure \ref{fig:GLF_staronly}.}
    \label{fig:GLF_mod_kUV}
\end{figure*}

\subsubsection{A more gradual transition in the stellar IMF}
\label{sec:CompIMF}

Recent high-resolution 3D hydrodynamical simulations of low-metallicity star forming regions show that the metallicity-driven transition in the stellar IMF might be smoother than predicted by the critical metallicity scenario, and that a larger fraction of massive stars than predicted by a standard Salpeter-like IMF persists up to a metallicity of $Z \sim 10^{-2} \, Z_\odot$ \citep{chon2021}. The main effect is due to the interplay between the cooling timescale and the timescale of turbulence decay. When $Z \lesssim 10^{-2} \, Z_\odot$, star formation begins after the turbulent motion decays, and a single massive cloud core monolithically collapses to form a central massive stellar cluster; despite dust-induced fragmentation occurs at $Z \geq Z_{\rm cr}$, promoting the formation of low-mass stars $m_* \lesssim 0.1 \, M_\odot$, the large gas accretion rate from the circumstellar disc preferentially feeds the central massive stars, making the mass distribution top-heavy. When $Z \geq 0.1 \, Z_\odot$, efficient metal-line cooling and collisions of the turbulent flows promote the onset of star formation in a highly filamentary
gas structure. In this case, the mass supply to the massive stars is limited by the local gas reservoir and shared among the stars, leading to a standard Salpeter-like IMF. 

In addition, the larger temperature of the CMB radiation at $z \gtrsim 10$ suppresses cloud fragmentation and reduces the number of low-mass stars in star-forming regions with metallicities $Z \gtrsim 10^{-2} Z_\odot$ \citep{schneider2010, chon2022}. As a result, stellar populations with metallicity of $Z \leq 10^{-2} Z_\odot$ or forming at $z \gtrsim 10$ are expected to be characterised by a mass spectrum consisting of a low-mass Salpeter-like component, peaking at $\sim 0.1 M_\odot$, and a top-heavy component at  $\gtrsim 10\, M_\odot$, with the mass fraction in the latter increasing with redshift, and decreasing with metallicity. 

While these results rely so far only on sophisticated theoretical studies, it is tempting to investigate their potential impact on the high-$z$ galaxy UV LF, particularly given that galaxies are expected to be more metal-poor at high redshift. Indeed, galaxies observed with JWST at $4 < z < 9$ have been found to be relatively young, with estimated ages $t_* < 30$ Myr, and with metallicities in the range $\sim 0.04  - 0.7 \, Z_\odot$ (see \citealt{Nakajima2022} and references therein), and similar properties have been derived for galaxies confirmed to be at $10 \lesssim z \lesssim 13$ \citep{Curtislake2022, Tacchella2022, Bunker2023, Tacchella2023}. 

To quantify the effect of a redshift-modulated stellar IMF on the galaxy UV LF, here we take a simple approach and we assume that galaxies populating the bright-end of the galaxy UV LF at $z \ge 10$ have metallicities $Z_* \sim 0.1 Z_\odot$ and stellar ages $t_* \sim 10$ Myr. Following \citet{Tanikawa2022}, we model the {\it transitional} stellar IMF as a composition of a Kroupa IMF in the mass range $0.08 M_\odot \leq m_* \leq 300 M_\odot$ \citep{Kroupa2001},
\begin{equation*}
\qquad \Phi(m_*) dm_* \propto \left \{ \begin{aligned}
                                        m_*^{-1.3} \, \, & \text{for} \,\, 0.08 M_\odot \leq m_* < 0.5 M_\odot \\
                                        m_*^{-2.3} \, \, & \text{for} \,\, 0.5 M_\odot \leq m_* \leq 300 M_\odot
                                        \end{aligned} \right.
\end{equation*}
\noindent
and of a Log-flat IMF for $10 M_\odot \leq m_* \leq 300 M_\odot$,
\[
\Phi(m_*) dm_* \propto m_*^{-1},
\]
\noindent
with a relative mass-weight that depends on $z$ and $Z_*$ and that have been obtained by fitting the simulations results of \citet{chon2022}. For a metallicity of $Z_* = 0.1 Z_\odot$, the weight of the Log-flat IMF can be expressed as 
$w_{\rm LF} = 0.04 \times (z - 5)$ for $z > 5$ and it is $w_{\rm LF} = 0$ at $z \leq 5$, meaning that below this redshift all the stars follow a Kroupa IMF. Following \citet{madau2014b}, we use the flexible stellar population synthesis code (FSPS, \citealt{Conroy2009, Conroy2010}) to compute the conversion factor between the intrinsic specific luminosity at 1500 \AA, $L_\nu (\rm FUV)$ (expressed in units of erg s$^{-1}$ Hz$^{-1}$) and the SFR (in units of $M_\odot$ yr$^{-1}$), 
\begin{equation}
    {\cal K}_{\rm FUV} = \frac{\rm SFR}{L_\nu (\rm FUV)}
\end{equation}
\noindent
assuming a constant SFR and our adopted transitional stellar IMF for $Z_* = 0.1 \, Z_\odot$. At redshift $z = (0 - 5, 10, 15, 20)$, we find that ${\cal K}_{\rm FUV} = (1.46, 1.21, 1.04, 0.91) \times 10^{-28}$ for a stellar age of $t_* = 10$ Myr, and ${\cal K}_{\rm FUV} = (1.12, 0.96, 0.90, 0.81) \times 10^{-28}$ for $t_* = 20$ Myr. Hence, a stellar population with the same metallicity and age is predicted to emit up to $1.4 - 1.8$ times more FUV radiation per unit SFR at $z \sim 20$ than at $z \lesssim 10$. 

\begin{figure*}
\centering
  \includegraphics[width=18.0cm]{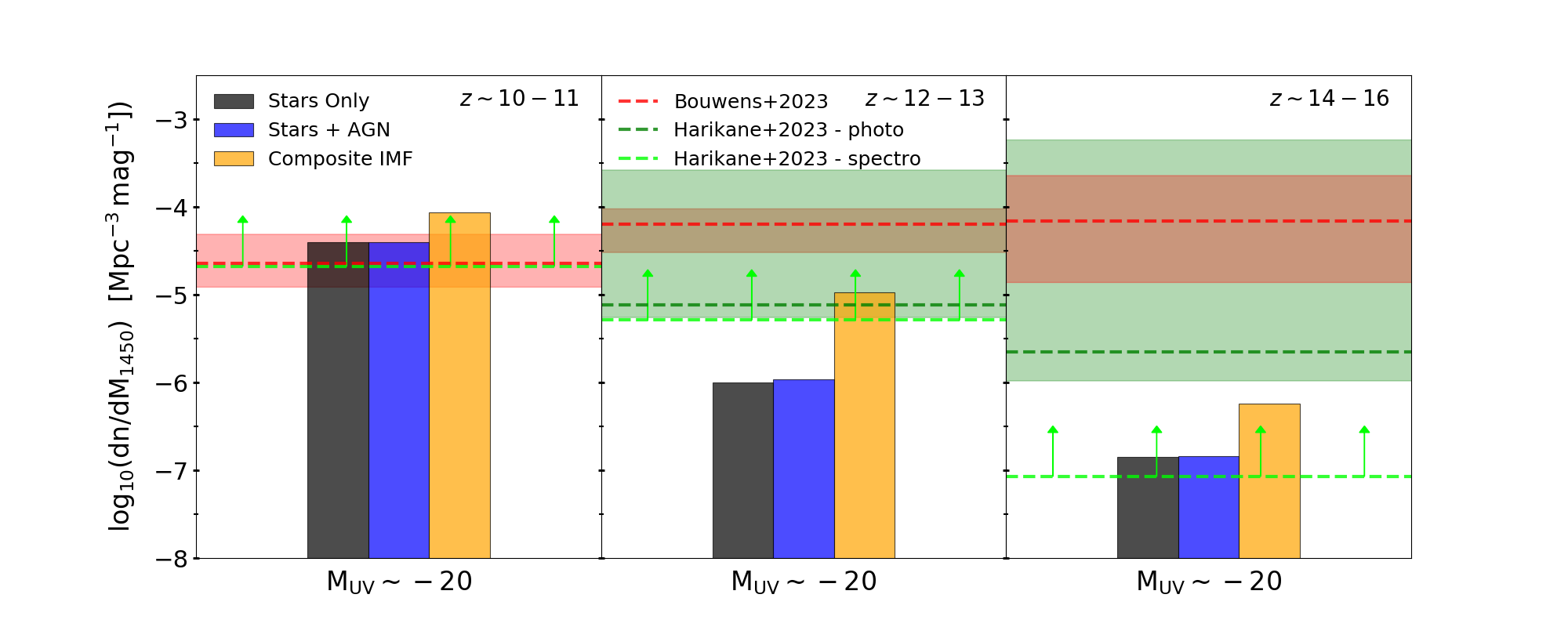}
    \caption{CAT prediction for the number density of galaxies with $M_{\rm UV} \sim -20$ in the redshift range $z \sim 10-11$, $z \sim 12-13$ and $z \sim 14-16$. The different colored histograms show the obtained values for the stellar-only galaxy LF (black bars), the total UVLF (stars + AGN, blue bars) and the galaxy LF corrected for a composite stellar IMF (orange bars).
    The horizontal lines represent the empirical estimates obtained from the best-fit LFs of \citet[dark green]{Harikane2023photo} and \citet[red]{Bouwens2023}, with the shaded regions showing the $1 \, \sigma$ error. The light green horizontal lines show instead the estimates based only on spectroscopically confirmed galaxies obtained by \citet{Harikane2023spectro}, which are represented here as lower limit of the galaxy number density.}
    \label{fig:NumbDensity_comparison}
\end{figure*}

In Figure \ref{fig:GLF_mod_kUV} we show how the predicted galaxy UV LF changes when applying to the Pop II stellar emission a correction for ${\cal K}_{\rm FUV}$ consistent with a stellar population characterized by a composite IMF. We assume a boost in the luminosity of each galaxy of a factor ${\cal K}_{\rm FUV} (z) / {\cal K}_{\rm FUV, Kroupa}$, where ${\cal K}_{\rm FUV, Kroupa} = 1.46 \times 10^{-28}$ and ${\cal K}_{\rm FUV} (z)$ are the expected correction factors for, respectively, the Kroupa and the composite IMFs, assuming a stellar population of $10 \, \rm Myr$ with $Z_* = 0.1 \, Z_\odot$. We see how this correction impacts significantly on the UV LF, especially at $z \gtrsim 9$. In particular, CAT predictions are now consistent with observational constraints in the redshift ranges $z \sim 10-11$ and $z \sim 12-13$. At $z \sim 14 - 16$, the model predicts number density of UV bright sources that is still smaller than estimated from photometric candidates \citep[][]{Harikane2023spectro} and consistent with the lower limit inferred from the spectroscopic analysis \citep[][]{Harikane2023spectro} but the difference from their best-fit model is now reduced to $0.5 \rm \, dex$ and $0.8 \rm \, dex$ at, respectively, $M_{\rm UV} \simeq -19.5$ and $M_{\rm UV} \simeq -20.5$ (see Figure \ref{fig:NumbDensity_comparison}). 

Hence, by considering a gradual transition in the stellar IMF modulated by metallicity and redshift, we are able to recover a better match with the observational estimates of the number density of bright sources at $z \gtrsim 10$. A minor discrepancy persists in the highest redshift range, $z > 14$, where the model predictions are also affected by large statistical uncertainties.
In a future study, we plan to incorporate these new theoretical findings into a more sophisticated modeling of stellar populations in CAT,
as a transitional IMF that depends on metallicity and redshift not only affect the emissivity of the first galaxies, but it also changes
the rate of supernova explosion, hence the effects of mechanical and chemical feedback. In addition, it modifies the mass function of
BH remnants, with interesting consequences on the BH seed populations.

\subsection{Quantifying the tension with JWST data}
\label{sec:ObsComparison}
 We quantify the tension between the predictions of the model that we have explored with current observations in Figure \ref{fig:NumbDensity_comparison}, where we compare the number density of galaxies with $M_{\rm UV} \sim -20$ predicted by CAT (coloured histograms) in the redshift range $z \sim 10-11$ (left panel), $z \sim 12-13$ (middle panel) and $z \sim 14-16$ (right panel), with the observational estimates of \citet[horizontal red line]{Bouwens2023} and \citet[horizontal green line]{Harikane2023photo}, based on photometric candidates, and with analysis of \citet[horizontal light green]{Harikane2023spectro}, based on spectroscopically confirmed galaxies. The values obtained from the photometric samples have been estimated from the Schechter fits\footnote{In general, the LF are best-fit assuming both a double-power-law and a Schechter function, but due to the uncertainties in the measurements, the difference between the two is small at $z \sim 9$, and negligible at $z \sim 12$ and 16 \citet{Harikane2023photo}.} of the LF at $M_{\rm UV} \simeq -20$ reported in the original papers, with their associated errors. The value estimated from the spectroscopic sample is reported as a lower limit. The three coloured histograms represent the number density of sources predicted by CAT when the UV luminosity is computed according to CAT reference model considering only the emission from stars (black) and from stars and accreting BHs (blue), while the orange histograms show the result of computing the stellar emission according to a transitional IMF. As anticipated, considering the AGN contribution in addition to the stellar emission does not significantly increase the number density of sources with $M_{\rm UV} \simeq -20$ at $z \gtrsim 10$. CAT reference model predictions are consistent with observations at $z \simeq 10 - 11$, but fall short by 0.8 dex at $z \simeq 12 - 13$, and 1.2 dex at $z \sim 14 - 16$, compared to the values estimated by \citet{Harikane2023photo}, and by 1.8 dex at $z \simeq 12 - 13$, and 2.7 dex at $z \sim 14 - 16$, compared to the values estimated by \citet{Bouwens2023}, while being consistent with the lower limit estimated by \citet{Harikane2023spectro} at $z \sim 14 - 16$. Interestingly, the increase in the UV luminosity predicted by a transitional IMF brings the model in agreement with the observational estimate at $z \sim 12 - 13$ by \citet{Harikane2023photo}, and reduces the disagreement with their best-fit value to 0.6 dex at $z \sim 14 - 16$, while the estimated values from \citet{Bouwens2023} are still 0.8 dex (2.1 dex) higher than the model predictions at $z \sim 12 - 13$ ($z \sim 14 - 16$). 
 
\section{Conclusions}
\label{sec:conclusion}

In this study we have explored the contribution of Pop III/II stars and accreting BH seeds
to the UV luminosity evolution at $4 \leq z \leq 20$ using the Cosmic Archaeology Tool (CAT) model.
We first presented the predictions of CAT regarding the cosmic star formation history and the contributions of the first galaxies and their nuclear BHs to cosmic reionization. We then compared the galaxy UV luminosity function at $4 \leq z \leq 20$ with available observations, from deep HST and JWST data, discussing the contribution of emission from stars, including Pop III stars, and accreting BHs. We find that:
\begin{itemize}
    \item The model predicts a cosmic star formation history dominated by UV faint 
($M_{\rm UV} > - 18$) galaxies populated by Pop II/I stars. In the redshift range $4 < z < 10$, CAT predictions for the total SFRD are in very good agreement with data inferred from gamma-ray burst observations, which are sensitive to both obscured and unobscured star formation \citep{kistler2009, robertson2012}. Conversely, the SFRD derived from the rest-frame UV luminosity are better reproduced by CAT when only the contribution of the sources brightest than $M_{\rm UV} \simeq -18$ is considered \citep{trinca2022}. 

\item At $10\lesssim z \lesssim 20$, the formation of Pop III stars is strongly self-regulated, their SFRD remains almost constant, and their contribution to the total SFRD ranges from $\lesssim 10\%$ down to $\lesssim 0.5\%$. Below $z \sim 10$, chemical feedback prevents further episodes of Pop III star formation. 

\item CAT predicts a cosmic reionization process consistent with current observational constraints. Stars are the primary sources of cosmic reionization, with $5 - 10 \%$ of ionizing photons escaping into the IGM at $5 \leq z \leq 10$, in good agreement with recent results \citep{finkelstein2019,naidu2020,schaerer2002,mascia2023}. Due to their top-heavy IMF and lower metallicity, Pop III stars dominate the emissivity down to $z \simeq 16$. The AGN ionizing emissivity remains subdominant with respect to the Pop II stellar emission down to $z \sim 5$. 

\end{itemize}

Having satisfied these global constraints, we have investigated the redshift evolution of the galaxy UV luminosity function predicted by CAT, comparing it with deep HST data and with recent JWST observations. We find that:
\begin{itemize}
    \item The stellar and AGN UV luminosity functions predicted by CAT are in good agreement with observations at $5 \lesssim z \lesssim 9 - 10$. At higher redshift, CAT predicts a steeper faint-end slope than the observed best-fit luminosity functions extrapolated at $M_{\rm UV} > - 18$, suggesting that stars may form too efficiently, feedback may be too inefficient, or that current observational samples may be still incomplete at these faint luminosities. 
    
    \item When considering only the emission from stars, the UV LF predicted by CAT at the bright end shows a mild evolution at $10 \lesssim z \lesssim 13$, consistent with observations, except for the highest luminosity bins, at $M_{\rm UV} < -19$, where the model seems to underestimate the number of bright objects, despite the large statistical uncertainties. We quantified this tension by comparing the number density of $M_{\rm UV} \sim -20$ galaxies with recent JWST data, finding that the model predictions are consistent with observations at $z \simeq 10 - 11$, but fall short by 0.8 dex at $z \simeq 12 - 13$, and 1.2 dex at $z \sim 14 - 16$, compared to the values estimated by \citet{Harikane2023photo}, and by 1.8 dex at $z \simeq 12 - 13$, and 2.7 dex at $z \sim 14 - 16$, compared to the values estimated by \citet{Bouwens2023}, while being consistent with the lower limit estimated by \citet{Harikane2023spectro} at $z \sim 14 - 16$. 
    
    \item Including the emission by AGNs does not affect these findings. In fact, at $z \gtrsim 15$, the AGN emission appears negligible, due to the stunted growth of light seeds predicted by our reference model \citep[see][]{trinca2022}. At $10 \lesssim z \lesssim 15$, during the formation epoch of heavy BH seeds, the AGN emission contributes on average to $\sim 1 - 3 \%$ of the total UV emission, and their largest contamination reaches $\lesssim 10 \%$ in the brightest bins of magnitude, at $\rm M_{\rm UV} \lesssim -19$. However, the AGN emission becomes progressively more important at lower redshift, with an average (maximum) contribution of $\lesssim 20 \%$ ($> 50\%$) at  $7 \lesssim z \lesssim 10$, and of $20 - 100\%$ at $z < 7$ in systems with $M_{\rm UV} \lesssim -22$.

    \item Metal-free and very metal-poor stellar populations might also increase the UV luminosity of galaxies at $z \gtrsim 10$. Our results suggest that Pop III stars, with their harder emission spectra, contribute to the UV luminosity in up to $\sim 10 \%$ of high-redshift systems at $z \gtrsim 16$, while their contribution becomes significantly smaller with decreasing redshift, due to the progression of metal enrichment. 

    \item Finally, we have explored the effects on the UV luminosities of a more gradual transition in the stellar IMF, as suggested by recent high-resolution numerical simulations \citep{chon2021, chon2022}. We model this {\it transitional} IMF as a superposition of a Kroupa and a Log-flat IMF, with a relative weight that depends both on redshift and on the initial stellar metallicity, $Z_*$. Assuming a fixed value of $Z_* = 0.1 \, Z_\odot$ and  a constant stellar age of $t_* = 10$ Myr, we find that galaxies emit $30 \%$ more UV photons per unit SFR at $z \simeq 10$ and $60 \%$ at $z \simeq 20$ than at $z \lesssim 5$, where a standard Kroupa IMF applies. When accounting for this effect, the number density of $M_{\rm UV} \sim -20$ galaxies predicted by the model is in agreement with the observational estimate at $z \sim 12 - 13$ by \citet{Harikane2023photo}, and reduces the disagreement with their best-fit value to 0.6 dex at $z \sim 14 - 16$, while the estimated values from \citet{Bouwens2023} are still 0.8 dex (2.1 dex) higher than the model predictions at $z \sim 12 - 13$ ($z \sim 14 - 16$). 
\end{itemize}

If the current tension between theoretical models and JWST observations will consolidate on the basis of a larger sample of spectroscopically confirmed galaxies at $z \gtrsim 10$, this opens up the prospective of exploring the nature of the first sources that inhabit early galaxies, and to improve our understanding of the physical processes that shape the assembly of cosmic structures at these remote cosmic epochs.
    
\section*{Acknowledgements}
We acknowledge support from the PRIN 2022 MUR project 2022CB3PJ3 - First Light And Galaxy aSsembly (FLAGS) funded by the European Union – Next Generation EU, from the Amaldi Research
Center funded by the MIUR program “Dipartimento di Eccellenza ”(CUP:B81I18001170001), and from the INFN TEONGRAV specific initiative.
This research is supported in part by Grants-in-Aid for Scientific Research (KO: 17H06360, 17H01102, 17H02869, 22H00149) from the Japan Society for the Promotion of Science.

\section*{Data Availability}
The simulated data underlying this article will be shared on reasonable request to the corresponding author.

\bibliographystyle{mnras}
\bibliography{Bibliography} 

\begin{thebibliography}{}
\makeatletter
\relax
\def\mn@urlcharsother{\let\do\@makeother \do\$\do\&\do\#\do\^\do\_\do\%\do\~}
\def\mn@doi{\begingroup\mn@urlcharsother \@ifnextchar [ {\mn@doi@}
  {\mn@doi@[]}}
\def\mn@doi@[#1]#2{\def\@tempa{#1}\ifx\@tempa\@empty \href
  {http://dx.doi.org/#2} {doi:#2}\else \href {http://dx.doi.org/#2} {#1}\fi
  \endgroup}
\def\mn@eprint#1#2{\mn@eprint@#1:#2::\@nil}
\def\mn@eprint@arXiv#1{\href {http://arxiv.org/abs/#1} {{\tt arXiv:#1}}}
\def\mn@eprint@dblp#1{\href {http://dblp.uni-trier.de/rec/bibtex/#1.xml}
  {dblp:#1}}
\def\mn@eprint@#1:#2:#3:#4\@nil{\def\@tempa {#1}\def\@tempb {#2}\def\@tempc
  {#3}\ifx \@tempc \@empty \let \@tempc \@tempb \let \@tempb \@tempa \fi \ifx
  \@tempb \@empty \def\@tempb {arXiv}\fi \@ifundefined
  {mn@eprint@\@tempb}{\@tempb:\@tempc}{\expandafter \expandafter \csname
  mn@eprint@\@tempb\endcsname \expandafter{\@tempc}}}

\bibitem[\protect\citeauthoryear{{Adams} et~al.,}{{Adams}
  et~al.}{2023a}]{Adams2023b}
{Adams} N.~J.,  et~al., 2023a, \mn@doi [arXiv e-prints]
  {10.48550/arXiv.2304.13721}, \href
  {https://ui.adsabs.harvard.edu/abs/2023arXiv230413721A} {p. arXiv:2304.13721}

\bibitem[\protect\citeauthoryear{{Adams} et~al.,}{{Adams}
  et~al.}{2023b}]{Adams2023}
{Adams} N.~J.,  et~al., 2023b, \mn@doi [\mnras]
  {10.1093/mnras/stac334710.48550/arXiv.2207.11217}, \href
  {https://ui.adsabs.harvard.edu/abs/2023MNRAS.518.4755A} {518, 4755}

\bibitem[\protect\citeauthoryear{{Aguado} et~al.,}{{Aguado}
  et~al.}{2023}]{Aguado_2023a}
{Aguado} D.~S.,  et~al., 2023, \mn@doi [\aap] {10.1051/0004-6361/202245392},
  \href {https://ui.adsabs.harvard.edu/abs/2023A&A...669L...4A} {669, L4}

\bibitem[\protect\citeauthoryear{{Algera} et~al.,}{{Algera}
  et~al.}{2023}]{algera2023}
{Algera} H. S.~B.,  et~al., 2023, \mn@doi [\mnras] {10.1093/mnras/stac3195},
  \href {https://ui.adsabs.harvard.edu/abs/2023MNRAS.518.6142A} {518, 6142}

\bibitem[\protect\citeauthoryear{{Arrabal Haro} et~al.,}{{Arrabal Haro}
  et~al.}{2023}]{ArrabalHaro2023}
{Arrabal Haro} P.,  et~al., 2023, \mn@doi [arXiv e-prints]
  {10.48550/arXiv.2303.15431}, \href
  {https://ui.adsabs.harvard.edu/abs/2023arXiv230315431A} {p. arXiv:2303.15431}

\bibitem[\protect\citeauthoryear{{Atek} et~al.,}{{Atek}
  et~al.}{2023}]{Atek2023}
{Atek} H.,  et~al., 2023, \mn@doi [\mnras]
  {10.1093/mnras/stac314410.48550/arXiv.2207.12338}, \href
  {https://ui.adsabs.harvard.edu/abs/2023MNRAS.519.1201A} {519, 1201}

\bibitem[\protect\citeauthoryear{{Backhaus} et~al.,}{{Backhaus}
  et~al.}{2022}]{backhaus2022}
{Backhaus} B.~E.,  et~al., 2022, \mn@doi [\apj] {10.3847/1538-4357/ac3919},
  \href {https://ui.adsabs.harvard.edu/abs/2022ApJ...926..161B} {926, 161}

\bibitem[\protect\citeauthoryear{{Baldwin}, {Phillips}  \&
  {Terlevich}}{{Baldwin} et~al.}{1981}]{baldwin1981}
{Baldwin} J.~A.,  {Phillips} M.~M.,   {Terlevich} R.,  1981, \mn@doi [\pasp]
  {10.1086/130766}, \href
  {https://ui.adsabs.harvard.edu/abs/1981PASP...93....5B} {93, 5}

\bibitem[\protect\citeauthoryear{{Barkana} \& {Loeb}}{{Barkana} \&
  {Loeb}}{2001}]{Barkana2001}
{Barkana} R.,  {Loeb} A.,  2001, \mn@doi [\physrep]
  {10.1016/S0370-1573(01)00019-9}, \href
  {https://ui.adsabs.harvard.edu/abs/2001PhR...349..125B} {349, 125}

\bibitem[\protect\citeauthoryear{{Barrufet} et~al.,}{{Barrufet}
  et~al.}{2023}]{barrufet2023}
{Barrufet} L.,  et~al., 2023, \mn@doi [\mnras] {10.1093/mnras/stad947}, \href
  {https://ui.adsabs.harvard.edu/abs/2023MNRAS.tmp..908B} {}

\bibitem[\protect\citeauthoryear{{Becerra}, {Greif}, {Springel}  \&
  {Hernquist}}{{Becerra} et~al.}{2015}]{becerra2015}
{Becerra} F.,  {Greif} T.~H.,  {Springel} V.,   {Hernquist} L.~E.,  2015,
  \mn@doi [\mnras] {10.1093/mnras/stu2284}, \href
  {https://ui.adsabs.harvard.edu/abs/2015MNRAS.446.2380B} {446, 2380}

\bibitem[\protect\citeauthoryear{{Becerra}, {Marinacci}, {Bromm}  \&
  {Hernquist}}{{Becerra} et~al.}{2018}]{becerra2018}
{Becerra} F.,  {Marinacci} F.,  {Bromm} V.,   {Hernquist} L.~E.,  2018, \mn@doi
  [\mnras] {10.1093/mnras/sty2210}, \href
  {https://ui.adsabs.harvard.edu/abs/2018MNRAS.480.5029B} {480, 5029}

\bibitem[\protect\citeauthoryear{{Becker} \& {Bolton}}{{Becker} \&
  {Bolton}}{2013}]{becker2013}
{Becker} G.~D.,  {Bolton} J.~S.,  2013, \mn@doi [\mnras]
  {10.1093/mnras/stt1610}, \href
  {https://ui.adsabs.harvard.edu/abs/2013MNRAS.436.1023B} {436, 1023}

\bibitem[\protect\citeauthoryear{Becker, D'Aloisio, Christenson, Zhu, Worseck
  \& Bolton}{Becker et~al.}{2021}]{becker2021}
Becker G.~D.,  D'Aloisio A.,  Christenson H.~M.,  Zhu Y.,  Worseck G.,   Bolton
  J.~S.,  2021, \mn@doi [\mnras] {10.1093/mnras/stab2696}, 508, 1853

\bibitem[\protect\citeauthoryear{{Behroozi}, {Wechsler}, {Hearin}  \&
  {Conroy}}{{Behroozi} et~al.}{2019}]{Behroozi2019}
{Behroozi} P.,  {Wechsler} R.~H.,  {Hearin} A.~P.,   {Conroy} C.,  2019,
  \mn@doi [\mnras] {10.1093/mnras/stz1182}, \href
  {https://ui.adsabs.harvard.edu/abs/2019MNRAS.488.3143B} {488, 3143}

\bibitem[\protect\citeauthoryear{{Bolan} et~al.,}{{Bolan}
  et~al.}{2022}]{bolan2022}
{Bolan} P.,  et~al., 2022, \mn@doi [\mnras] {10.1093/mnras/stac1963}, \href
  {https://ui.adsabs.harvard.edu/abs/2022MNRAS.517.3263B} {517, 3263}

\bibitem[\protect\citeauthoryear{{Bondi}}{{Bondi}}{1952}]{bondi1952}
{Bondi} H.,  1952, \mn@doi [\mnras] {10.1093/mnras/112.2.195}, \href
  {https://ui.adsabs.harvard.edu/abs/1952MNRAS.112..195B} {112, 195}

\bibitem[\protect\citeauthoryear{{Bouwens} et~al.,}{{Bouwens}
  et~al.}{2012}]{bouwens2012}
{Bouwens} R.~J.,  et~al., 2012, \mn@doi [\apj] {10.1088/0004-637X/754/2/83},
  \href {https://ui.adsabs.harvard.edu/abs/2012ApJ...754...83B} {754, 83}

\bibitem[\protect\citeauthoryear{{Bouwens} et~al.,}{{Bouwens}
  et~al.}{2014}]{bouwens2014}
{Bouwens} R.~J.,  et~al., 2014, \mn@doi [\apj] {10.1088/0004-637X/795/2/126},
  \href {https://ui.adsabs.harvard.edu/abs/2014ApJ...795..126B} {795, 126}

\bibitem[\protect\citeauthoryear{{Bouwens} et~al.,}{{Bouwens}
  et~al.}{2021}]{Bouwens2021}
{Bouwens} R.~J.,  et~al., 2021, \mn@doi [\aj] {10.3847/1538-3881/abf83e}, \href
  {https://ui.adsabs.harvard.edu/abs/2021AJ....162...47B} {162, 47}

\bibitem[\protect\citeauthoryear{{Bouwens}, {Illingworth}, {Oesch}, {Stefanon},
  {Naidu}, {van Leeuwen}  \& {Magee}}{{Bouwens} et~al.}{2023}]{Bouwens2023}
{Bouwens} R.,  {Illingworth} G.,  {Oesch} P.,  {Stefanon} M.,  {Naidu} R.,
  {van Leeuwen} I.,   {Magee} D.,  2023, \mn@doi [\mnras]
  {10.1093/mnras/stad1014}, \href
  {https://ui.adsabs.harvard.edu/abs/2023MNRAS.tmp.1019B} {}

\bibitem[\protect\citeauthoryear{{Bowler}, {Jarvis}, {Dunlop}, {McLure},
  {McLeod}, {Adams}, {Milvang-Jensen}  \& {McCracken}}{{Bowler}
  et~al.}{2020}]{Bowler2020}
{Bowler} R.~A.~A.,  {Jarvis} M.~J.,  {Dunlop} J.~S.,  {McLure} R.~J.,  {McLeod}
  D.~J.,  {Adams} N.~J.,  {Milvang-Jensen} B.,   {McCracken} H.~J.,  2020,
  \mn@doi [\mnras] {10.1093/mnras/staa313}, \href
  {https://ui.adsabs.harvard.edu/abs/2020MNRAS.493.2059B} {493, 2059}

\bibitem[\protect\citeauthoryear{{Bradley} et~al.,}{{Bradley}
  et~al.}{2022}]{Bradley2022}
{Bradley} L.~D.,  et~al., 2022, \mn@doi [arXiv e-prints]
  {10.48550/arXiv.2210.01777}, \href
  {https://ui.adsabs.harvard.edu/abs/2022arXiv221001777B} {p. arXiv:2210.01777}

\bibitem[\protect\citeauthoryear{{Bromm}, {Kudritzki}  \& {Loeb}}{{Bromm}
  et~al.}{2001}]{bromm2001}
{Bromm} V.,  {Kudritzki} R.~P.,   {Loeb} A.,  2001, \mn@doi [\apj]
  {10.1086/32054910.48550/arXiv.astro-ph/0007248}, \href
  {https://ui.adsabs.harvard.edu/abs/2001ApJ...552..464B} {552, 464}

\bibitem[\protect\citeauthoryear{{Bruzual} \& {Charlot}}{{Bruzual} \&
  {Charlot}}{2003}]{bruzual2003}
{Bruzual} G.,  {Charlot} S.,  2003, \mn@doi [\mnras]
  {10.1046/j.1365-8711.2003.06897.x}, \href
  {https://ui.adsabs.harvard.edu/abs/2003MNRAS.344.1000B} {344, 1000}

\bibitem[\protect\citeauthoryear{{Bunker} et~al.,}{{Bunker}
  et~al.}{2023}]{Bunker2023}
{Bunker} A.~J.,  et~al., 2023, \mn@doi [arXiv e-prints]
  {10.48550/arXiv.2302.07256}, \href
  {https://ui.adsabs.harvard.edu/abs/2023arXiv230207256B} {p. arXiv:2302.07256}

\bibitem[\protect\citeauthoryear{{Castellano} et~al.,}{{Castellano}
  et~al.}{2022}]{Castellano2022}
{Castellano} M.,  et~al., 2022, \mn@doi [\apjl] {10.3847/2041-8213/ac94d0},
  \href {https://ui.adsabs.harvard.edu/abs/2022ApJ...938L..15C} {938, L15}

\bibitem[\protect\citeauthoryear{{Castellano} et~al.,}{{Castellano}
  et~al.}{2023}]{Castellano2023}
{Castellano} M.,  et~al., 2023, \mn@doi [\apjl] {10.3847/2041-8213/accea5},
  \href {https://ui.adsabs.harvard.edu/abs/2023ApJ...948L..14C} {948, L14}

\bibitem[\protect\citeauthoryear{{Chon} \& {Omukai}}{{Chon} \&
  {Omukai}}{2020}]{chon2020}
{Chon} S.,  {Omukai} K.,  2020, \mn@doi [\mnras] {10.1093/mnras/staa863}, \href
  {https://ui.adsabs.harvard.edu/abs/2020MNRAS.494.2851C} {494, 2851}

\bibitem[\protect\citeauthoryear{{Chon}, {Hosokawa}  \& {Omukai}}{{Chon}
  et~al.}{2021}]{chon2021}
{Chon} S.,  {Hosokawa} T.,   {Omukai} K.,  2021, \mn@doi [\mnras]
  {10.1093/mnras/stab061}, \href
  {https://ui.adsabs.harvard.edu/abs/2021MNRAS.502..700C} {502, 700}

\bibitem[\protect\citeauthoryear{{Chon}, {Ono}, {Omukai}  \&
  {Schneider}}{{Chon} et~al.}{2022}]{chon2022}
{Chon} S.,  {Ono} H.,  {Omukai} K.,   {Schneider} R.,  2022, \mn@doi [\mnras]
  {10.1093/mnras/stac1549}, \href
  {https://ui.adsabs.harvard.edu/abs/2022MNRAS.514.4639C} {514, 4639}

\bibitem[\protect\citeauthoryear{{Cleri} et~al.,}{{Cleri}
  et~al.}{2023}]{cleri2023}
{Cleri} N.~J.,  et~al., 2023, \mn@doi [arXiv e-prints]
  {10.48550/arXiv.2301.07745}, \href
  {https://ui.adsabs.harvard.edu/abs/2023arXiv230107745C} {p. arXiv:2301.07745}

\bibitem[\protect\citeauthoryear{{Conroy} \& {Gunn}}{{Conroy} \&
  {Gunn}}{2010}]{Conroy2010}
{Conroy} C.,  {Gunn} J.~E.,  2010, \mn@doi [\apj]
  {10.1088/0004-637X/712/2/833}, \href
  {https://ui.adsabs.harvard.edu/abs/2010ApJ...712..833C} {712, 833}

\bibitem[\protect\citeauthoryear{{Conroy}, {Gunn}  \& {White}}{{Conroy}
  et~al.}{2009}]{Conroy2009}
{Conroy} C.,  {Gunn} J.~E.,   {White} M.,  2009, \mn@doi [\apj]
  {10.1088/0004-637X/699/1/486}, \href
  {https://ui.adsabs.harvard.edu/abs/2009ApJ...699..486C} {699, 486}

\bibitem[\protect\citeauthoryear{{Curti} et~al.,}{{Curti}
  et~al.}{2023}]{curti2023}
{Curti} M.,  et~al., 2023, \mn@doi [\mnras] {10.1093/mnras/stac2737}, \href
  {https://ui.adsabs.harvard.edu/abs/2023MNRAS.518..425C} {518, 425}

\bibitem[\protect\citeauthoryear{{Curtis-Lake} et~al.,}{{Curtis-Lake}
  et~al.}{2022}]{Curtislake2022}
{Curtis-Lake} E.,  et~al., 2022, \mn@doi [arXiv e-prints]
  {10.48550/arXiv.2212.04568}, \href
  {https://ui.adsabs.harvard.edu/abs/2022arXiv221204568C} {p. arXiv:2212.04568}

\bibitem[\protect\citeauthoryear{{Davies} et~al.,}{{Davies}
  et~al.}{2018}]{davies2018}
{Davies} F.~B.,  et~al., 2018, \mn@doi [\apj] {10.3847/1538-4357/aad6dc}, \href
  {https://ui.adsabs.harvard.edu/abs/2018ApJ...864..142D} {864, 142}

\bibitem[\protect\citeauthoryear{{Dayal} et~al.,}{{Dayal}
  et~al.}{2020}]{dayal2020}
{Dayal} P.,  et~al., 2020, \mn@doi [\mnras] {10.1093/mnras/staa1138}, \href
  {https://ui.adsabs.harvard.edu/abs/2020MNRAS.495.3065D} {495, 3065}

\bibitem[\protect\citeauthoryear{{Dekel}, {Sarkar}, {Birnboim}, {Mandelker}  \&
  {Li}}{{Dekel} et~al.}{2023}]{Dekel2023}
{Dekel} A.,  {Sarkar} K.~S.,  {Birnboim} Y.,  {Mandelker} N.,   {Li} Z.,  2023,
  \mn@doi [arXiv e-prints] {10.48550/arXiv.2303.04827}, \href
  {https://ui.adsabs.harvard.edu/abs/2023arXiv230304827D} {p. arXiv:2303.04827}

\bibitem[\protect\citeauthoryear{Dijkstra, Ferrara  \& Mesinger}{Dijkstra
  et~al.}{2014}]{dijkstra2014feedback}
Dijkstra M.,  Ferrara A.,   Mesinger A.,  2014, Monthly Notices of the Royal
  Astronomical Society, 442, 2036

\bibitem[\protect\citeauthoryear{{Donnan} et~al.,}{{Donnan}
  et~al.}{2023}]{Donnan2023}
{Donnan} C.~T.,  et~al., 2023, \mn@doi [\mnras]
  {10.1093/mnras/stac347210.48550/arXiv.2207.12356}, \href
  {https://ui.adsabs.harvard.edu/abs/2023MNRAS.518.6011D} {518, 6011}

\bibitem[\protect\citeauthoryear{{Dubois}, {Volonteri}  \& {Silk}}{{Dubois}
  et~al.}{2014}]{dubois2014}
{Dubois} Y.,  {Volonteri} M.,   {Silk} J.,  2014, \mn@doi [\mnras]
  {10.1093/mnras/stu373}, \href
  {https://ui.adsabs.harvard.edu/abs/2014MNRAS.440.1590D} {440, 1590}

\bibitem[\protect\citeauthoryear{{Duras} et~al.,}{{Duras}
  et~al.}{2020}]{duras2020}
{Duras} F.,  et~al., 2020, \mn@doi [\aap] {10.1051/0004-6361/201936817}, \href
  {https://ui.adsabs.harvard.edu/abs/2020A&A...636A..73D} {636, A73}

\bibitem[\protect\citeauthoryear{{Ellis} et~al.,}{{Ellis}
  et~al.}{2013}]{ellis2013}
{Ellis} R.~S.,  et~al., 2013, \mn@doi [\apjl] {10.1088/2041-8205/763/1/L7},
  \href {https://ui.adsabs.harvard.edu/abs/2013ApJ...763L...7E} {763, L7}

\bibitem[\protect\citeauthoryear{{Ferrara}, {Salvadori}, {Yue}  \&
  {Schleicher}}{{Ferrara} et~al.}{2014}]{ferrara2014}
{Ferrara} A.,  {Salvadori} S.,  {Yue} B.,   {Schleicher} D.,  2014, \mn@doi
  [\mnras] {10.1093/mnras/stu1280}, \href
  {https://ui.adsabs.harvard.edu/abs/2014MNRAS.443.2410F} {443, 2410}

\bibitem[\protect\citeauthoryear{{Ferrara}, {Pallottini}  \& {Dayal}}{{Ferrara}
  et~al.}{2022}]{ferrara2022}
{Ferrara} A.,  {Pallottini} A.,   {Dayal} P.,  2022, \mn@doi [arXiv e-prints]
  {10.48550/arXiv.2208.00720}, \href
  {https://ui.adsabs.harvard.edu/abs/2022arXiv220800720F} {p. arXiv:2208.00720}

\bibitem[\protect\citeauthoryear{{Fialkov}, {Barkana}, {Visbal},
  {Tseliakhovich}  \& {Hirata}}{{Fialkov} et~al.}{2013}]{Fialkov2013}
{Fialkov} A.,  {Barkana} R.,  {Visbal} E.,  {Tseliakhovich} D.,   {Hirata}
  C.~M.,  2013, \mn@doi [\mnras] {10.1093/mnras/stt650}, \href
  {https://ui.adsabs.harvard.edu/abs/2013MNRAS.432.2909F} {432, 2909}

\bibitem[\protect\citeauthoryear{{Finkelstein} \& {Bagley}}{{Finkelstein} \&
  {Bagley}}{2022}]{finkelstein2022d}
{Finkelstein} S.~L.,  {Bagley} M.~B.,  2022, \mn@doi [\apj]
  {10.3847/1538-4357/ac89eb}, \href
  {https://ui.adsabs.harvard.edu/abs/2022ApJ...938...25F} {938, 25}

\bibitem[\protect\citeauthoryear{{Finkelstein} et~al.,}{{Finkelstein}
  et~al.}{2019}]{finkelstein2019}
{Finkelstein} S.~L.,  et~al., 2019, \mn@doi [\apj] {10.3847/1538-4357/ab1ea8},
  \href {https://ui.adsabs.harvard.edu/abs/2019ApJ...879...36F} {879, 36}

\bibitem[\protect\citeauthoryear{{Finkelstein} et~al.,}{{Finkelstein}
  et~al.}{2022a}]{Finkelstein2022}
{Finkelstein} S.~L.,  et~al., 2022a, \mn@doi [arXiv e-prints]
  {10.48550/arXiv.2211.05792}, \href
  {https://ui.adsabs.harvard.edu/abs/2022arXiv221105792F} {p. arXiv:2211.05792}

\bibitem[\protect\citeauthoryear{{Finkelstein} et~al.,}{{Finkelstein}
  et~al.}{2022b}]{Finkelstein2022b}
{Finkelstein} S.~L.,  et~al., 2022b, \mn@doi [\apjl]
  {10.3847/2041-8213/ac966e10.48550/arXiv.2207.12474}, \href
  {https://ui.adsabs.harvard.edu/abs/2022ApJ...940L..55F} {940, L55}

\bibitem[\protect\citeauthoryear{{Fiore}, {Ferrara}, {Bischetti}, {Feruglio}
  \& {Travascio}}{{Fiore} et~al.}{2022}]{fiore2022}
{Fiore} F.,  {Ferrara} A.,  {Bischetti} M.,  {Feruglio} C.,   {Travascio} A.,
  2022, \mn@doi [arXiv e-prints] {10.48550/arXiv.2211.08937}, \href
  {https://ui.adsabs.harvard.edu/abs/2022arXiv221108937F} {p. arXiv:2211.08937}

\bibitem[\protect\citeauthoryear{{Fraser}, {Casey}, {Gilmore}, {Heger}  \&
  {Chan}}{{Fraser} et~al.}{2017}]{Fraser_2017}
{Fraser} M.,  {Casey} A.~R.,  {Gilmore} G.,  {Heger} A.,   {Chan} C.,  2017,
  \mn@doi [\mnras] {10.1093/mnras/stx480}, \href
  {https://ui.adsabs.harvard.edu/abs/2017MNRAS.468..418F} {468, 418}

\bibitem[\protect\citeauthoryear{{Furtak}, {Shuntov}, {Atek}, {Zitrin},
  {Richard}, {Lehnert}  \& {Chevallard}}{{Furtak} et~al.}{2023}]{furtak2023}
{Furtak} L.~J.,  {Shuntov} M.,  {Atek} H.,  {Zitrin} A.,  {Richard} J.,
  {Lehnert} M.~D.,   {Chevallard} J.,  2023, \mn@doi [\mnras]
  {10.1093/mnras/stac3717}, \href
  {https://ui.adsabs.harvard.edu/abs/2023MNRAS.519.3064F} {519, 3064}

\bibitem[\protect\citeauthoryear{{Ganguly}, {Brotherton}, {Cales}, {Scoggins},
  {Shang}  \& {Vestergaard}}{{Ganguly} et~al.}{2007}]{Ganguly2007}
{Ganguly} R.,  {Brotherton} M.~S.,  {Cales} S.,  {Scoggins} B.,  {Shang} Z.,
  {Vestergaard} M.,  2007, \mn@doi [\apj] {10.1086/519759}, \href
  {https://ui.adsabs.harvard.edu/abs/2007ApJ...665..990G} {665, 990}

\bibitem[\protect\citeauthoryear{{Giallongo} et~al.,}{{Giallongo}
  et~al.}{2019}]{giallongo2019}
{Giallongo} E.,  et~al., 2019, \mn@doi [\apj] {10.3847/1538-4357/ab39e1}, \href
  {https://ui.adsabs.harvard.edu/abs/2019ApJ...884...19G} {884, 19}

\bibitem[\protect\citeauthoryear{{Goulding} \& {Greene}}{{Goulding} \&
  {Greene}}{2022}]{Goulding2022}
{Goulding} A.~D.,  {Greene} J.~E.,  2022, arXiv e-prints, \href
  {https://ui.adsabs.harvard.edu/abs/2022arXiv220802822G} {p. arXiv:2208.02822}

\bibitem[\protect\citeauthoryear{{Greene} et~al.,}{{Greene}
  et~al.}{2023}]{Greene2023}
{Greene} J.~E.,  et~al., 2023, \mn@doi [arXiv e-prints]
  {10.48550/arXiv.2309.05714}, \href
  {https://ui.adsabs.harvard.edu/abs/2023arXiv230905714G} {p. arXiv:2309.05714}

\bibitem[\protect\citeauthoryear{{Greig}, {Mesinger}  \& {Ba{\~n}ados}}{{Greig}
  et~al.}{2019}]{greig2019}
{Greig} B.,  {Mesinger} A.,   {Ba{\~n}ados} E.,  2019, \mn@doi [\mnras]
  {10.1093/mnras/stz230}, \href
  {https://ui.adsabs.harvard.edu/abs/2019MNRAS.484.5094G} {484, 5094}

\bibitem[\protect\citeauthoryear{{Gruppioni} et~al.,}{{Gruppioni}
  et~al.}{2020}]{gruppioni2020}
{Gruppioni} C.,  et~al., 2020, \mn@doi [\aap] {10.1051/0004-6361/202038487},
  \href {https://ui.adsabs.harvard.edu/abs/2020A&A...643A...8G} {643, A8}

\bibitem[\protect\citeauthoryear{{Harikane} et~al.,}{{Harikane}
  et~al.}{2022a}]{harikane2022a}
{Harikane} Y.,  et~al., 2022a, \mn@doi [\apjs]
  {10.3847/1538-4365/ac3dfc10.48550/arXiv.2108.01090}, \href
  {https://ui.adsabs.harvard.edu/abs/2022ApJS..259...20H} {259, 20}

\bibitem[\protect\citeauthoryear{{Harikane} et~al.,}{{Harikane}
  et~al.}{2022b}]{harikane2022b}
{Harikane} Y.,  et~al., 2022b, \mn@doi [\apj] {10.3847/1538-4357/ac53a9}, \href
  {https://ui.adsabs.harvard.edu/abs/2022ApJ...929....1H} {929, 1}

\bibitem[\protect\citeauthoryear{{Harikane} et~al.,}{{Harikane}
  et~al.}{2023a}]{Harikane2023bh}
{Harikane} Y.,  et~al., 2023a, arXiv e-prints, \href
  {https://ui.adsabs.harvard.edu/abs/2023arXiv230311946H} {p. arXiv:2303.11946}

\bibitem[\protect\citeauthoryear{{Harikane}, {Nakajima}, {Ouchi}, {Umeda},
  {Isobe}, {Ono}, {Xu}  \& {Zhang}}{{Harikane}
  et~al.}{2023b}]{Harikane2023spectro}
{Harikane} Y.,  {Nakajima} K.,  {Ouchi} M.,  {Umeda} H.,  {Isobe} Y.,  {Ono}
  Y.,  {Xu} Y.,   {Zhang} Y.,  2023b, \mn@doi [arXiv e-prints]
  {10.48550/arXiv.2304.06658}, \href
  {https://ui.adsabs.harvard.edu/abs/2023arXiv230406658H} {p. arXiv:2304.06658}

\bibitem[\protect\citeauthoryear{{Harikane} et~al.,}{{Harikane}
  et~al.}{2023c}]{Harikane2023photo}
{Harikane} Y.,  et~al., 2023c, \mn@doi [\apjs] {10.3847/1538-4365/acaaa9},
  \href {https://ui.adsabs.harvard.edu/abs/2023ApJS..265....5H} {265, 5}

\bibitem[\protect\citeauthoryear{{Hartwig} et~al.,}{{Hartwig}
  et~al.}{2022}]{hartwig2022}
{Hartwig} T.,  et~al., 2022, \mn@doi [\apj] {10.3847/1538-4357/ac7150}, \href
  {https://ui.adsabs.harvard.edu/abs/2022ApJ...936...45H} {936, 45}

\bibitem[\protect\citeauthoryear{{Heintz} et~al.,}{{Heintz}
  et~al.}{2022a}]{Heintz2022b}
{Heintz} K.~E.,  et~al., 2022a, \mn@doi [arXiv e-prints]
  {10.48550/arXiv.2212.02890}, \href
  {https://ui.adsabs.harvard.edu/abs/2022arXiv221202890H} {p. arXiv:2212.02890}

\bibitem[\protect\citeauthoryear{{Heintz} et~al.,}{{Heintz}
  et~al.}{2022b}]{Heintz2022a}
{Heintz} K.~E.,  et~al., 2022b, \mn@doi [arXiv e-prints]
  {10.48550/arXiv.2212.06877}, \href
  {https://ui.adsabs.harvard.edu/abs/2022arXiv221206877H} {p. arXiv:2212.06877}

\bibitem[\protect\citeauthoryear{Hirano, Hosokawa, Yoshida, Umeda, Omukai,
  Chiaki  \& Yorke}{Hirano et~al.}{2014}]{hirano2014}
Hirano S.,  Hosokawa T.,  Yoshida N.,  Umeda H.,  Omukai K.,  Chiaki G.,
  Yorke H.~W.,  2014, The Astrophysical Journal, 781, 60

\bibitem[\protect\citeauthoryear{{Hirano}, {Zhu}, {Yoshida}, {Spergel}  \&
  {Yorke}}{{Hirano} et~al.}{2015}]{hirano2015}
{Hirano} S.,  {Zhu} N.,  {Yoshida} N.,  {Spergel} D.,   {Yorke} H.~W.,  2015,
  \mn@doi [\apj] {10.1088/0004-637X/814/1/18}, \href
  {https://ui.adsabs.harvard.edu/abs/2015ApJ...814...18H} {814, 18}

\bibitem[\protect\citeauthoryear{Hosokawa, Omukai, Yoshida  \& Yorke}{Hosokawa
  et~al.}{2011}]{hosokawa2011protostellar}
Hosokawa T.,  Omukai K.,  Yoshida N.,   Yorke H.~W.,  2011, Science, 334, 1250

\bibitem[\protect\citeauthoryear{{Hosokawa}, {Omukai}  \& {Yorke}}{{Hosokawa}
  et~al.}{2012}]{hosokawa2012}
{Hosokawa} T.,  {Omukai} K.,   {Yorke} H.~W.,  2012, \mn@doi [\apj]
  {10.1088/0004-637X/756/1/93}, \href
  {https://ui.adsabs.harvard.edu/abs/2012ApJ...756...93H} {756, 93}

\bibitem[\protect\citeauthoryear{{Hosokawa}, {Hirano}, {Kuiper}, {Yorke},
  {Omukai}  \& {Yoshida}}{{Hosokawa} et~al.}{2016}]{hosokawa2016}
{Hosokawa} T.,  {Hirano} S.,  {Kuiper} R.,  {Yorke} H.~W.,  {Omukai} K.,
  {Yoshida} N.,  2016, \mn@doi [\apj] {10.3847/0004-637X/824/2/119}, \href
  {https://ui.adsabs.harvard.edu/abs/2016ApJ...824..119H} {824, 119}

\bibitem[\protect\citeauthoryear{{Hoyle} \& {Lyttleton}}{{Hoyle} \&
  {Lyttleton}}{1941}]{hoyle1941}
{Hoyle} F.,  {Lyttleton} R.~A.,  1941, \mn@doi [\mnras]
  {10.1093/mnras/101.4.227}, \href
  {https://ui.adsabs.harvard.edu/abs/1941MNRAS.101..227H} {101, 227}

\bibitem[\protect\citeauthoryear{{Iliev}, {Scannapieco}  \& {Shapiro}}{{Iliev}
  et~al.}{2005}]{Iliev2005}
{Iliev} I.~T.,  {Scannapieco} E.,   {Shapiro} P.~R.,  2005, \mn@doi [\apj]
  {10.1086/429083}, \href
  {https://ui.adsabs.harvard.edu/abs/2005ApJ...624..491I} {624, 491}

\bibitem[\protect\citeauthoryear{{Inayoshi}, {Omukai}  \& {Tasker}}{{Inayoshi}
  et~al.}{2014}]{inayoshi2014}
{Inayoshi} K.,  {Omukai} K.,   {Tasker} E.,  2014, \mn@doi [\mnras]
  {10.1093/mnrasl/slu151}, \href
  {https://ui.adsabs.harvard.edu/abs/2014MNRAS.445L.109I} {445, L109}

\bibitem[\protect\citeauthoryear{{Inayoshi}, {Visbal}  \& {Haiman}}{{Inayoshi}
  et~al.}{2020}]{inayoshi2020}
{Inayoshi} K.,  {Visbal} E.,   {Haiman} Z.,  2020, \mn@doi [\araa]
  {10.1146/annurev-astro-120419-014455}, \href
  {https://ui.adsabs.harvard.edu/abs/2020ARA&A..58...27I} {58, 27}

\bibitem[\protect\citeauthoryear{{Inayoshi}, {Onoue}, {Sugahara}, {Inoue}  \&
  {Ho}}{{Inayoshi} et~al.}{2022}]{inayoshi2022}
{Inayoshi} K.,  {Onoue} M.,  {Sugahara} Y.,  {Inoue} A.~K.,   {Ho} L.~C.,
  2022, \mn@doi [\apjl] {10.3847/2041-8213/ac6f01}, \href
  {https://ui.adsabs.harvard.edu/abs/2022ApJ...931L..25I} {931, L25}

\bibitem[\protect\citeauthoryear{{Ishiyama} et~al.,}{{Ishiyama}
  et~al.}{2021}]{Ishiyama2021}
{Ishiyama} T.,  et~al., 2021, \mn@doi [\mnras] {10.1093/mnras/stab1755}, \href
  {https://ui.adsabs.harvard.edu/abs/2021MNRAS.506.4210I} {506, 4210}

\bibitem[\protect\citeauthoryear{{Jaacks}, {Finkelstein}  \& {Bromm}}{{Jaacks}
  et~al.}{2019}]{Jaacks2019}
{Jaacks} J.,  {Finkelstein} S.~L.,   {Bromm} V.,  2019, \mn@doi [\mnras]
  {10.1093/mnras/stz1529}, \href
  {https://ui.adsabs.harvard.edu/abs/2019MNRAS.488.2202J} {488, 2202}

\bibitem[\protect\citeauthoryear{{Johnson}, {Dalla Vecchia}  \&
  {Khochfar}}{{Johnson} et~al.}{2013}]{johnson2013}
{Johnson} J.~L.,  {Dalla Vecchia} C.,   {Khochfar} S.,  2013, \mn@doi [\mnras]
  {10.1093/mnras/sts011}, \href
  {https://ui.adsabs.harvard.edu/abs/2013MNRAS.428.1857J} {428, 1857}

\bibitem[\protect\citeauthoryear{{Katz}, {Kimm}, {Ellis}, {Devriendt}  \&
  {Slyz}}{{Katz} et~al.}{2022}]{Katz2022}
{Katz} H.,  {Kimm} T.,  {Ellis} R.~S.,  {Devriendt} J.,   {Slyz} A.,  2022,
  \mn@doi [arXiv e-prints] {10.48550/arXiv.2207.04751}, \href
  {https://ui.adsabs.harvard.edu/abs/2022arXiv220704751K} {p. arXiv:2207.04751}

\bibitem[\protect\citeauthoryear{{Keller}, {Munshi}, {Trebitsch}  \&
  {Tremmel}}{{Keller} et~al.}{2023}]{keller2023}
{Keller} B.~W.,  {Munshi} F.,  {Trebitsch} M.,   {Tremmel} M.,  2023, \mn@doi
  [\apjl] {10.3847/2041-8213/acb148}, \href
  {https://ui.adsabs.harvard.edu/abs/2023ApJ...943L..28K} {943, L28}

\bibitem[\protect\citeauthoryear{{Khusanova} et~al.,}{{Khusanova}
  et~al.}{2021}]{khusanova2020}
{Khusanova} Y.,  et~al., 2021, \mn@doi [\aap] {10.1051/0004-6361/202038944},
  \href {https://ui.adsabs.harvard.edu/abs/2021A&A...649A.152K} {649, A152}

\bibitem[\protect\citeauthoryear{{Kistler}, {Y{\"u}ksel}, {Beacom}, {Hopkins}
  \& {Wyithe}}{{Kistler} et~al.}{2009}]{kistler2009}
{Kistler} M.~D.,  {Y{\"u}ksel} H.,  {Beacom} J.~F.,  {Hopkins} A.~M.,
  {Wyithe} J. S.~B.,  2009, \mn@doi [\apjl] {10.1088/0004-637X/705/2/L104},
  \href {https://ui.adsabs.harvard.edu/abs/2009ApJ...705L.104K} {705, L104}

\bibitem[\protect\citeauthoryear{{Kocevski} et~al.,}{{Kocevski}
  et~al.}{2023}]{Kocevski2023}
{Kocevski} D.~D.,  et~al., 2023, \mn@doi [arXiv e-prints]
  {10.48550/arXiv.2302.00012}, \href
  {https://ui.adsabs.harvard.edu/abs/2023arXiv230200012K} {p. arXiv:2302.00012}

\bibitem[\protect\citeauthoryear{{Kroupa}}{{Kroupa}}{2001}]{Kroupa2001}
{Kroupa} P.,  2001, \mn@doi [\mnras] {10.1046/j.1365-8711.2001.04022.x}, \href
  {https://ui.adsabs.harvard.edu/abs/2001MNRAS.322..231K} {322, 231}

\bibitem[\protect\citeauthoryear{{Kuhlen} \& {Faucher-Gigu{\`e}re}}{{Kuhlen} \&
  {Faucher-Gigu{\`e}re}}{2012}]{Kuhlen2012}
{Kuhlen} M.,  {Faucher-Gigu{\`e}re} C.-A.,  2012, \mn@doi [\mnras]
  {10.1111/j.1365-2966.2012.20924.x}, \href
  {https://ui.adsabs.harvard.edu/abs/2012MNRAS.423..862K} {423, 862}

\bibitem[\protect\citeauthoryear{{Labbe} et~al.,}{{Labbe}
  et~al.}{2022}]{Labbe2022}
{Labbe} I.,  et~al., 2022, \mn@doi [arXiv e-prints]
  {10.48550/arXiv.2207.12446}, \href
  {https://ui.adsabs.harvard.edu/abs/2022arXiv220712446L} {p. arXiv:2207.12446}

\bibitem[\protect\citeauthoryear{{Larson} et~al.,}{{Larson}
  et~al.}{2023}]{Larson2023}
{Larson} R.~L.,  et~al., 2023, \mn@doi [arXiv e-prints]
  {10.48550/arXiv.2303.08918}, \href
  {https://ui.adsabs.harvard.edu/abs/2023arXiv230308918L} {p. arXiv:2303.08918}

\bibitem[\protect\citeauthoryear{{Latif} \& {Ferrara}}{{Latif} \&
  {Ferrara}}{2016}]{latif2016a}
{Latif} M.~A.,  {Ferrara} A.,  2016, \mn@doi [\pasa] {10.1017/pasa.2016.41},
  \href {https://ui.adsabs.harvard.edu/abs/2016PASA...33...51L} {33, e051}

\bibitem[\protect\citeauthoryear{{Latif}, {Schleicher}, {Schmidt}  \&
  {Niemeyer}}{{Latif} et~al.}{2013}]{latif2013}
{Latif} M.~A.,  {Schleicher} D.~R.~G.,  {Schmidt} W.,   {Niemeyer} J.~C.,
  2013, \mn@doi [\mnras] {10.1093/mnras/stt1786}, \href
  {https://ui.adsabs.harvard.edu/abs/2013MNRAS.436.2989L} {436, 2989}

\bibitem[\protect\citeauthoryear{{Liu} \& {Bromm}}{{Liu} \&
  {Bromm}}{2020}]{LiuBromm2020}
{Liu} B.,  {Bromm} V.,  2020, \mn@doi [\mnras] {10.1093/mnras/staa2143}, \href
  {https://ui.adsabs.harvard.edu/abs/2020MNRAS.497.2839L} {497, 2839}

\bibitem[\protect\citeauthoryear{{Madau} \& {Dickinson}}{{Madau} \&
  {Dickinson}}{2014}]{madau2014b}
{Madau} P.,  {Dickinson} M.,  2014, \mn@doi [\araa]
  {10.1146/annurev-astro-081811-125615}, \href
  {https://ui.adsabs.harvard.edu/abs/2014ARA&A..52..415M} {52, 415}

\bibitem[\protect\citeauthoryear{{Madau}, {Haardt}  \& {Dotti}}{{Madau}
  et~al.}{2014}]{madau2014a}
{Madau} P.,  {Haardt} F.,   {Dotti} M.,  2014, \mn@doi [\apjl]
  {10.1088/2041-8205/784/2/L38}, \href
  {https://ui.adsabs.harvard.edu/abs/2014ApJ...784L..38M} {784, L38}

\bibitem[\protect\citeauthoryear{{Magg}, {Schauer}, {Klessen}, {Glover},
  {Tress}  \& {Jaura}}{{Magg} et~al.}{2022}]{Magg_2022}
{Magg} M.,  {Schauer} A. T.~P.,  {Klessen} R.~S.,  {Glover} S. C.~O.,  {Tress}
  R.~G.,   {Jaura} O.,  2022, \mn@doi [\apj] {10.3847/1538-4357/ac5aac}, \href
  {https://ui.adsabs.harvard.edu/abs/2022ApJ...929..119M} {929, 119}

\bibitem[\protect\citeauthoryear{{Maio}, {Petkova}, {De Lucia}  \&
  {Borgani}}{{Maio} et~al.}{2016}]{Maio2016}
{Maio} U.,  {Petkova} M.,  {De Lucia} G.,   {Borgani} S.,  2016, \mn@doi
  [\mnras] {10.1093/mnras/stw1196}, \href
  {https://ui.adsabs.harvard.edu/abs/2016MNRAS.460.3733M} {460, 3733}

\bibitem[\protect\citeauthoryear{{Maiolino} et~al.,}{{Maiolino}
  et~al.}{2023a}]{Maiolino2023BHGNz11}
{Maiolino} R.,  et~al., 2023a, \mn@doi [arXiv e-prints]
  {10.48550/arXiv.2305.12492}, \href
  {https://ui.adsabs.harvard.edu/abs/2023arXiv230512492M} {p. arXiv:2305.12492}

\bibitem[\protect\citeauthoryear{{Maiolino} et~al.,}{{Maiolino}
  et~al.}{2023b}]{Maiolino2023popIII}
{Maiolino} R.,  et~al., 2023b, \mn@doi [arXiv e-prints]
  {10.48550/arXiv.2306.00953}, \href
  {https://ui.adsabs.harvard.edu/abs/2023arXiv230600953M} {p. arXiv:2306.00953}

\bibitem[\protect\citeauthoryear{{Maiolino} et~al.,}{{Maiolino}
  et~al.}{2023c}]{Maiolino2023AGN}
{Maiolino} R.,  et~al., 2023c, \mn@doi [arXiv e-prints]
  {10.48550/arXiv.2308.01230}, \href
  {https://ui.adsabs.harvard.edu/abs/2023arXiv230801230M} {p. arXiv:2308.01230}

\bibitem[\protect\citeauthoryear{{Mancini}, {Schneider}, {Graziani},
  {Valiante}, {Dayal}, {Maio}  \& {Ciardi}}{{Mancini}
  et~al.}{2016}]{mancini2016}
{Mancini} M.,  {Schneider} R.,  {Graziani} L.,  {Valiante} R.,  {Dayal} P.,
  {Maio} U.,   {Ciardi} B.,  2016, \mn@doi [\mnras] {10.1093/mnras/stw1783},
  \href {https://ui.adsabs.harvard.edu/abs/2016MNRAS.462.3130M} {462, 3130}

\bibitem[\protect\citeauthoryear{{Mascia} et~al.,}{{Mascia}
  et~al.}{2023}]{mascia2023}
{Mascia} S.,  et~al., 2023, \mn@doi [arXiv e-prints]
  {10.48550/arXiv.2301.02816}, \href
  {https://ui.adsabs.harvard.edu/abs/2023arXiv230102816M} {p. arXiv:2301.02816}

\bibitem[\protect\citeauthoryear{{Mason}, {Trenti}  \& {Treu}}{{Mason}
  et~al.}{2023}]{Mason2023}
{Mason} C.~A.,  {Trenti} M.,   {Treu} T.,  2023, \mn@doi [\mnras]
  {10.1093/mnras/stad035}, \href
  {https://ui.adsabs.harvard.edu/abs/2023MNRAS.tmp...65M} {}

\bibitem[\protect\citeauthoryear{{Matthee} et~al.,}{{Matthee}
  et~al.}{2023}]{Matthee2023}
{Matthee} J.,  et~al., 2023, \mn@doi [arXiv e-prints]
  {10.48550/arXiv.2306.05448}, \href
  {https://ui.adsabs.harvard.edu/abs/2023arXiv230605448M} {p. arXiv:2306.05448}

\bibitem[\protect\citeauthoryear{{McCaffrey}, {Hardin}, {Wise}  \&
  {Regan}}{{McCaffrey} et~al.}{2023}]{mccaffrey2023}
{McCaffrey} J.,  {Hardin} S.,  {Wise} J.,   {Regan} J.,  2023, \mn@doi [arXiv
  e-prints] {10.48550/arXiv.2304.13755}, \href
  {https://ui.adsabs.harvard.edu/abs/2023arXiv230413755M} {p. arXiv:2304.13755}

\bibitem[\protect\citeauthoryear{{McGreer}, {Mesinger}  \&
  {D'Odorico}}{{McGreer} et~al.}{2015}]{McGreer2015}
{McGreer} I.~D.,  {Mesinger} A.,   {D'Odorico} V.,  2015, \mn@doi [\mnras]
  {10.1093/mnras/stu2449}, \href
  {https://ui.adsabs.harvard.edu/abs/2015MNRAS.447..499M} {447, 499}

\bibitem[\protect\citeauthoryear{{McGreer}, {Fan}, {Jiang}  \& {Cai}}{{McGreer}
  et~al.}{2018}]{mcgreer2018}
{McGreer} I.~D.,  {Fan} X.,  {Jiang} L.,   {Cai} Z.,  2018, \mn@doi [\aj]
  {10.3847/1538-3881/aaaab4}, \href
  {https://ui.adsabs.harvard.edu/abs/2018AJ....155..131M} {155, 131}

\bibitem[\protect\citeauthoryear{{McLeod} et~al.,}{{McLeod}
  et~al.}{2023}]{McLeod2023}
{McLeod} D.~J.,  et~al., 2023, \mn@doi [arXiv e-prints]
  {10.48550/arXiv.2304.14469}, \href
  {https://ui.adsabs.harvard.edu/abs/2023arXiv230414469M} {p. arXiv:2304.14469}

\bibitem[\protect\citeauthoryear{{Menci}, {Castellano}, {Santini}, {Merlin},
  {Fontana}  \& {Shankar}}{{Menci} et~al.}{2022}]{Menci2022}
{Menci} N.,  {Castellano} M.,  {Santini} P.,  {Merlin} E.,  {Fontana} A.,
  {Shankar} F.,  2022, \mn@doi [\apjl] {10.3847/2041-8213/ac96e9}, \href
  {https://ui.adsabs.harvard.edu/abs/2022ApJ...938L...5M} {938, L5}

\bibitem[\protect\citeauthoryear{{Merlin} et~al.,}{{Merlin}
  et~al.}{2019}]{merlin2019}
{Merlin} E.,  et~al., 2019, \mn@doi [\mnras] {10.1093/mnras/stz2615}, \href
  {https://ui.adsabs.harvard.edu/abs/2019MNRAS.490.3309M} {490, 3309}

\bibitem[\protect\citeauthoryear{{Mirocha} \& {Furlanetto}}{{Mirocha} \&
  {Furlanetto}}{2023}]{Mirocha2023}
{Mirocha} J.,  {Furlanetto} S.~R.,  2023, \mn@doi [\mnras]
  {10.1093/mnras/stac3578}, \href
  {https://ui.adsabs.harvard.edu/abs/2023MNRAS.519..843M} {519, 843}

\bibitem[\protect\citeauthoryear{{Mo}, {Mao}  \& {White}}{{Mo}
  et~al.}{1998}]{mo1998}
{Mo} H.~J.,  {Mao} S.,   {White} S. D.~M.,  1998, \mn@doi [\mnras]
  {10.1046/j.1365-8711.1998.01227.x}, \href
  {https://ui.adsabs.harvard.edu/abs/1998MNRAS.295..319M} {295, 319}

\bibitem[\protect\citeauthoryear{{Morales}, {Mason}, {Bruton}, {Gronke},
  {Haardt}  \& {Scarlata}}{{Morales} et~al.}{2021}]{morales2021}
{Morales} A.~M.,  {Mason} C.~A.,  {Bruton} S.,  {Gronke} M.,  {Haardt} F.,
  {Scarlata} C.,  2021, \mn@doi [\apj] {10.3847/1538-4357/ac1104}, \href
  {https://ui.adsabs.harvard.edu/abs/2021ApJ...919..120M} {919, 120}

\bibitem[\protect\citeauthoryear{{Naidu}, {Tacchella}, {Mason}, {Bose}, {Oesch}
   \& {Conroy}}{{Naidu} et~al.}{2020}]{naidu2020}
{Naidu} R.~P.,  {Tacchella} S.,  {Mason} C.~A.,  {Bose} S.,  {Oesch} P.~A.,
  {Conroy} C.,  2020, \mn@doi [\apj] {10.3847/1538-4357/ab7cc9}, \href
  {https://ui.adsabs.harvard.edu/abs/2020ApJ...892..109N} {892, 109}

\bibitem[\protect\citeauthoryear{{Naidu} et~al.,}{{Naidu}
  et~al.}{2022}]{Naidu2022}
{Naidu} R.~P.,  et~al., 2022, \mn@doi [\apjl]
  {10.3847/2041-8213/ac9b2210.48550/arXiv.2207.09434}, \href
  {https://ui.adsabs.harvard.edu/abs/2022ApJ...940L..14N} {940, L14}

\bibitem[\protect\citeauthoryear{{Nakajima} \& {Maiolino}}{{Nakajima} \&
  {Maiolino}}{2022}]{Nakajima2022}
{Nakajima} K.,  {Maiolino} R.,  2022, \mn@doi [\mnras]
  {10.1093/mnras/stac1242}, \href
  {https://ui.adsabs.harvard.edu/abs/2022MNRAS.513.5134N} {513, 5134}

\bibitem[\protect\citeauthoryear{Natarajan, Pacucci, Ferrara, Agarwal, Ricarte,
  Zackrisson  \& Cappelluti}{Natarajan et~al.}{2017}]{natarajan2017unveiling}
Natarajan P.,  Pacucci F.,  Ferrara A.,  Agarwal B.,  Ricarte A.,  Zackrisson
  E.,   Cappelluti N.,  2017, The Astrophysical Journal, 838, 117

\bibitem[\protect\citeauthoryear{{Negri} \& {Volonteri}}{{Negri} \&
  {Volonteri}}{2017}]{negri2017}
{Negri} A.,  {Volonteri} M.,  2017, \mn@doi [\mnras] {10.1093/mnras/stx362},
  \href {https://ui.adsabs.harvard.edu/abs/2017MNRAS.467.3475N} {467, 3475}

\bibitem[\protect\citeauthoryear{{Niida} et~al.,}{{Niida}
  et~al.}{2020}]{niida2020}
{Niida} M.,  et~al., 2020, \mn@doi [\apj] {10.3847/1538-4357/abbe11}, \href
  {https://ui.adsabs.harvard.edu/abs/2020ApJ...904...89N} {904, 89}

\bibitem[\protect\citeauthoryear{{Oesch}, {Bouwens}, {Illingworth}, {Labb{\'e}}
   \& {Stefanon}}{{Oesch} et~al.}{2018}]{Oesch2018}
{Oesch} P.~A.,  {Bouwens} R.~J.,  {Illingworth} G.~D.,  {Labb{\'e}} I.,
  {Stefanon} M.,  2018, \mn@doi [\apj] {10.3847/1538-4357/aab03f}, \href
  {https://ui.adsabs.harvard.edu/abs/2018ApJ...855..105O} {855, 105}

\bibitem[\protect\citeauthoryear{Omukai \& Nishi}{Omukai \&
  Nishi}{1998}]{omukai1998formation}
Omukai K.,  Nishi R.,  1998, The Astrophysical Journal, 508, 141

\bibitem[\protect\citeauthoryear{Omukai \& Palla}{Omukai \&
  Palla}{2003}]{omukai2003formation}
Omukai K.,  Palla F.,  2003, The Astrophysical Journal, 589, 677

\bibitem[\protect\citeauthoryear{Omukai, Tsuribe, Schneider  \& Ferrara}{Omukai
  et~al.}{2005}]{omukai2005}
Omukai K.,  Tsuribe T.,  Schneider R.,   Ferrara A.,  2005, The Astrophysical
  Journal, 626, 627

\bibitem[\protect\citeauthoryear{{Pacucci} \& {Loeb}}{{Pacucci} \&
  {Loeb}}{2020}]{pacucci2020}
{Pacucci} F.,  {Loeb} A.,  2020, \mn@doi [\apj] {10.3847/1538-4357/ab886e},
  \href {https://ui.adsabs.harvard.edu/abs/2020ApJ...895...95P} {895, 95}

\bibitem[\protect\citeauthoryear{{Pacucci}, {Dayal}, {Harikane}, {Inoue}  \&
  {Loeb}}{{Pacucci} et~al.}{2022}]{Pacucci2022}
{Pacucci} F.,  {Dayal} P.,  {Harikane} Y.,  {Inoue} A.~K.,   {Loeb} A.,  2022,
  \mn@doi [\mnras] {10.1093/mnrasl/slac03510.48550/arXiv.2201.00823}, \href
  {https://ui.adsabs.harvard.edu/abs/2022MNRAS.514L...6P} {514, L6}

\bibitem[\protect\citeauthoryear{{Pallottini} et~al.,}{{Pallottini}
  et~al.}{2015}]{pallottini2015}
{Pallottini} A.,  et~al., 2015, \mn@doi [\mnras] {10.1093/mnras/stv1795}, \href
  {https://ui.adsabs.harvard.edu/abs/2015MNRAS.453.2465P} {453, 2465}

\bibitem[\protect\citeauthoryear{{Parkinson}, {Cole}  \& {Helly}}{{Parkinson}
  et~al.}{2008}]{parkinson2008}
{Parkinson} H.,  {Cole} S.,   {Helly} J.,  2008, \mn@doi [\mnras]
  {10.1111/j.1365-2966.2007.12517.x}, \href
  {https://ui.adsabs.harvard.edu/abs/2008MNRAS.383..557P} {383, 557}

\bibitem[\protect\citeauthoryear{{Parsa}, {Dunlop}  \& {McLure}}{{Parsa}
  et~al.}{2018}]{parsa2018}
{Parsa} S.,  {Dunlop} J.~S.,   {McLure} R.~J.,  2018, \mn@doi [\mnras]
  {10.1093/mnras/stx2887}, \href
  {https://ui.adsabs.harvard.edu/abs/2018MNRAS.474.2904P} {474, 2904}

\bibitem[\protect\citeauthoryear{{P{\'e}rez-Gonz{\'a}lez}
  et~al.,}{{P{\'e}rez-Gonz{\'a}lez} et~al.}{2023}]{perezgonzalez2023}
{P{\'e}rez-Gonz{\'a}lez} P.~G.,  et~al., 2023, \mn@doi [arXiv e-prints]
  {10.48550/arXiv.2302.02429}, \href
  {https://ui.adsabs.harvard.edu/abs/2023arXiv230202429P} {p. arXiv:2302.02429}

\bibitem[\protect\citeauthoryear{{Pfister}, {Volonteri}, {Dubois}, {Dotti}  \&
  {Colpi}}{{Pfister} et~al.}{2019}]{pfister2019}
{Pfister} H.,  {Volonteri} M.,  {Dubois} Y.,  {Dotti} M.,   {Colpi} M.,  2019,
  \mn@doi [\mnras] {10.1093/mnras/stz822}, \href
  {https://ui.adsabs.harvard.edu/abs/2019MNRAS.486..101P} {486, 101}

\bibitem[\protect\citeauthoryear{{Piana}, {Dayal}  \& {Choudhury}}{{Piana}
  et~al.}{2022}]{piana2022}
{Piana} O.,  {Dayal} P.,   {Choudhury} T.~R.,  2022, \mn@doi [\mnras]
  {10.1093/mnras/stab3757}, \href
  {https://ui.adsabs.harvard.edu/abs/2022MNRAS.510.5661P} {510, 5661}

\bibitem[\protect\citeauthoryear{{Planck Collaboration} et~al.,}{{Planck
  Collaboration} et~al.}{2016}]{Planck2015}
{Planck Collaboration} et~al., 2016, \mn@doi [\aap]
  {10.1051/0004-6361/201525830}, \href
  {https://ui.adsabs.harvard.edu/abs/2016A&A...594A..13P} {594, A13}

\bibitem[\protect\citeauthoryear{{Planck Collaboration} et~al.,}{{Planck
  Collaboration} et~al.}{2018}]{planck2018}
{Planck Collaboration} et~al., 2018, arXiv e-prints, \href
  {https://ui.adsabs.harvard.edu/abs/2018arXiv180706209P} {p. arXiv:1807.06209}

\bibitem[\protect\citeauthoryear{{Prada}, {Behroozi}, {Ishiyama}, {Klypin}  \&
  {P{\'e}rez}}{{Prada} et~al.}{2023}]{Prada2023}
{Prada} F.,  {Behroozi} P.,  {Ishiyama} T.,  {Klypin} A.,   {P{\'e}rez} E.,
  2023, \mn@doi [arXiv e-prints] {10.48550/arXiv.2304.11911}, \href
  {https://ui.adsabs.harvard.edu/abs/2023arXiv230411911P} {p. arXiv:2304.11911}

\bibitem[\protect\citeauthoryear{{Reines} \& {Volonteri}}{{Reines} \&
  {Volonteri}}{2015}]{reines2015}
{Reines} A.~E.,  {Volonteri} M.,  2015, \mn@doi [\apj]
  {10.1088/0004-637X/813/2/82}, \href
  {https://ui.adsabs.harvard.edu/abs/2015ApJ...813...82R} {813, 82}

\bibitem[\protect\citeauthoryear{{Ricci}, {Marchesi}, {Shankar}, {La Franca}
  \& {Civano}}{{Ricci} et~al.}{2017}]{ricci2017}
{Ricci} F.,  {Marchesi} S.,  {Shankar} F.,  {La Franca} F.,   {Civano} F.,
  2017, \mn@doi [\mnras] {10.1093/mnras/stw2909}, \href
  {https://ui.adsabs.harvard.edu/abs/2017MNRAS.465.1915R} {465, 1915}

\bibitem[\protect\citeauthoryear{{Robertson} \& {Ellis}}{{Robertson} \&
  {Ellis}}{2012}]{robertson2012}
{Robertson} B.~E.,  {Ellis} R.~S.,  2012, \mn@doi [\apj]
  {10.1088/0004-637X/744/2/95}, \href
  {https://ui.adsabs.harvard.edu/abs/2012ApJ...744...95R} {744, 95}

\bibitem[\protect\citeauthoryear{{Robertson} et~al.,}{{Robertson}
  et~al.}{2023}]{robertson2023}
{Robertson} B.~E.,  et~al., 2023, \mn@doi [Nature Astronomy]
  {10.1038/s41550-023-01921-1}, \href
  {https://ui.adsabs.harvard.edu/abs/2023NatAs.tmp...67R} {}

\bibitem[\protect\citeauthoryear{{Salvadori}, {Tolstoy}, {Ferrara}  \&
  {Zaroubi}}{{Salvadori} et~al.}{2014}]{salvadori2014}
{Salvadori} S.,  {Tolstoy} E.,  {Ferrara} A.,   {Zaroubi} S.,  2014, \mn@doi
  [\mnras] {10.1093/mnrasl/slt132}, \href
  {https://ui.adsabs.harvard.edu/abs/2014MNRAS.437L..26S} {437, L26}

\bibitem[\protect\citeauthoryear{{Sarmento} \& {Scannapieco}}{{Sarmento} \&
  {Scannapieco}}{2022}]{sarmento2022}
{Sarmento} R.,  {Scannapieco} E.,  2022, \mn@doi [\apj]
  {10.3847/1538-4357/ac815c}, \href
  {https://ui.adsabs.harvard.edu/abs/2022ApJ...935..174S} {935, 174}

\bibitem[\protect\citeauthoryear{{Sarmento}, {Scannapieco}  \&
  {C{\^o}t{\'e}}}{{Sarmento} et~al.}{2019}]{sarmento2019}
{Sarmento} R.,  {Scannapieco} E.,   {C{\^o}t{\'e}} B.,  2019, \mn@doi [\apj]
  {10.3847/1538-4357/aafa1a}, \href
  {https://ui.adsabs.harvard.edu/abs/2019ApJ...871..206S} {871, 206}

\bibitem[\protect\citeauthoryear{{Sassano}, {Schneider}, {Valiante},
  {Inayoshi}, {Chon}, {Omukai}, {Mayer}  \& {Capelo}}{{Sassano}
  et~al.}{2021}]{sassano2021}
{Sassano} F.,  {Schneider} R.,  {Valiante} R.,  {Inayoshi} K.,  {Chon} S.,
  {Omukai} K.,  {Mayer} L.,   {Capelo} P.~R.,  2021, \mn@doi [\mnras]
  {10.1093/mnras/stab1737}, \href
  {https://ui.adsabs.harvard.edu/abs/2021MNRAS.506..613S} {506, 613}

\bibitem[\protect\citeauthoryear{{Sassano}, {Capelo}, {Mayer}, {Schneider}  \&
  {Valiante}}{{Sassano} et~al.}{2023}]{sassano2023}
{Sassano} F.,  {Capelo} P.~R.,  {Mayer} L.,  {Schneider} R.,   {Valiante} R.,
  2023, \mn@doi [\mnras] {10.1093/mnras/stac3608}, \href
  {https://ui.adsabs.harvard.edu/abs/2023MNRAS.519.1837S} {519, 1837}

\bibitem[\protect\citeauthoryear{{Schaerer}}{{Schaerer}}{2002}]{schaerer2002}
{Schaerer} D.,  2002, \mn@doi [\aap] {10.1051/0004-6361:20011619}, \href
  {https://ui.adsabs.harvard.edu/abs/2002A&A...382...28S} {382, 28}

\bibitem[\protect\citeauthoryear{{Schaerer}, {Marques-Chaves}, {Barrufet},
  {Oesch}, {Izotov}, {Naidu}, {Guseva}  \& {Brammer}}{{Schaerer}
  et~al.}{2022}]{Schaerer2022}
{Schaerer} D.,  {Marques-Chaves} R.,  {Barrufet} L.,  {Oesch} P.,  {Izotov}
  Y.~I.,  {Naidu} R.,  {Guseva} N.~G.,   {Brammer} G.,  2022, \mn@doi [\aap]
  {10.1051/0004-6361/20224455610.48550/arXiv.2207.10034}, \href
  {https://ui.adsabs.harvard.edu/abs/2022A&A...665L...4S} {665, L4}

\bibitem[\protect\citeauthoryear{{Schenker} et~al.,}{{Schenker}
  et~al.}{2013}]{schenker2013}
{Schenker} M.~A.,  et~al., 2013, \mn@doi [\apj] {10.1088/0004-637X/768/2/196},
  \href {https://ui.adsabs.harvard.edu/abs/2013ApJ...768..196S} {768, 196}

\bibitem[\protect\citeauthoryear{{Schneider} \& {Omukai}}{{Schneider} \&
  {Omukai}}{2010}]{schneider2010}
{Schneider} R.,  {Omukai} K.,  2010, \mn@doi [\mnras]
  {10.1111/j.1365-2966.2009.15891.x}, \href
  {https://ui.adsabs.harvard.edu/abs/2010MNRAS.402..429S} {402, 429}

\bibitem[\protect\citeauthoryear{{Schneider}, {Valiante}, {Trinca}, {Graziani},
  {Volonteri}  \& {Maiolino}}{{Schneider} et~al.}{2023}]{schneider2023}
{Schneider} R.,  {Valiante} R.,  {Trinca} A.,  {Graziani} L.,  {Volonteri} M.,
   {Maiolino} R.,  2023, \mn@doi [\mnras] {10.1093/mnras/stad2503}, \href
  {https://ui.adsabs.harvard.edu/abs/2023MNRAS.526.3250S} {526, 3250}

\bibitem[\protect\citeauthoryear{{Shakura} \& {Sunyaev}}{{Shakura} \&
  {Sunyaev}}{1973}]{shakura1973}
{Shakura} N.~I.,  {Sunyaev} R.~A.,  1973, \aap, \href
  {https://ui.adsabs.harvard.edu/abs/1973A&A....24..337S} {500, 33}

\bibitem[\protect\citeauthoryear{{Skinner} \& {Wise}}{{Skinner} \&
  {Wise}}{2020}]{SkinnerWise2020}
{Skinner} D.,  {Wise} J.~H.,  2020, \mn@doi [\mnras] {10.1093/mnras/staa139},
  \href {https://ui.adsabs.harvard.edu/abs/2020MNRAS.492.4386S} {492, 4386}

\bibitem[\protect\citeauthoryear{{Spinoso}, {Bonoli}, {Valiante}, {Schneider}
  \& {Izquierdo-Villalba}}{{Spinoso} et~al.}{2022}]{spinoso2022}
{Spinoso} D.,  {Bonoli} S.,  {Valiante} R.,  {Schneider} R.,
  {Izquierdo-Villalba} D.,  2022, arXiv e-prints, \href
  {https://ui.adsabs.harvard.edu/abs/2022arXiv220313846S} {p. arXiv:2203.13846}

\bibitem[\protect\citeauthoryear{Sugimura, Omukai  \& Inoue}{Sugimura
  et~al.}{2014}]{sugimura2014critical}
Sugimura K.,  Omukai K.,   Inoue A.~K.,  2014, Monthly Notices of the Royal
  Astronomical Society, 445, 544

\bibitem[\protect\citeauthoryear{{Sugimura}, {Matsumoto}, {Hosokawa}, {Hirano}
  \& {Omukai}}{{Sugimura} et~al.}{2020}]{sugimura2020}
{Sugimura} K.,  {Matsumoto} T.,  {Hosokawa} T.,  {Hirano} S.,   {Omukai} K.,
  2020, arXiv e-prints, \href
  {https://ui.adsabs.harvard.edu/abs/2020arXiv200200012S} {p. arXiv:2002.00012}

\bibitem[\protect\citeauthoryear{{Susa}, {Hasegawa}  \& {Tominaga}}{{Susa}
  et~al.}{2014}]{susa2014}
{Susa} H.,  {Hasegawa} K.,   {Tominaga} N.,  2014, \mn@doi [\apj]
  {10.1088/0004-637X/792/1/32}, \href
  {https://ui.adsabs.harvard.edu/abs/2014ApJ...792...32S} {792, 32}

\bibitem[\protect\citeauthoryear{{Tacchella} et~al.,}{{Tacchella}
  et~al.}{2022}]{Tacchella2022}
{Tacchella} S.,  et~al., 2022, \mn@doi [arXiv e-prints]
  {10.48550/arXiv.2208.03281}, \href
  {https://ui.adsabs.harvard.edu/abs/2022arXiv220803281T} {p. arXiv:2208.03281}

\bibitem[\protect\citeauthoryear{{Tacchella} et~al.,}{{Tacchella}
  et~al.}{2023}]{Tacchella2023}
{Tacchella} S.,  et~al., 2023, \mn@doi [arXiv e-prints]
  {10.48550/arXiv.2302.07234}, \href
  {https://ui.adsabs.harvard.edu/abs/2023arXiv230207234T} {p. arXiv:2302.07234}

\bibitem[\protect\citeauthoryear{{Tanaka} \& {Haiman}}{{Tanaka} \&
  {Haiman}}{2009}]{tanaka2009}
{Tanaka} T.,  {Haiman} Z.,  2009, \mn@doi [\apj]
  {10.1088/0004-637X/696/2/1798}, \href
  {https://ui.adsabs.harvard.edu/abs/2009ApJ...696.1798T} {696, 1798}

\bibitem[\protect\citeauthoryear{{Tanikawa}, {Yoshida}, {Kinugawa}, {Trani},
  {Hosokawa}, {Susa}  \& {Omukai}}{{Tanikawa} et~al.}{2022}]{Tanikawa2022}
{Tanikawa} A.,  {Yoshida} T.,  {Kinugawa} T.,  {Trani} A.~A.,  {Hosokawa} T.,
  {Susa} H.,   {Omukai} K.,  2022, \mn@doi [\apj] {10.3847/1538-4357/ac4247},
  \href {https://ui.adsabs.harvard.edu/abs/2022ApJ...926...83T} {926, 83}

\bibitem[\protect\citeauthoryear{Tornatore, Ferrara  \& Schneider}{Tornatore
  et~al.}{2007}]{tornatore2007population}
Tornatore L.,  Ferrara A.,   Schneider R.,  2007, Monthly Notices of the Royal
  Astronomical Society, 382, 945

\bibitem[\protect\citeauthoryear{{Tremmel}, {Governato}, {Volonteri}, {Quinn}
  \& {Pontzen}}{{Tremmel} et~al.}{2018}]{tremmel2018}
{Tremmel} M.,  {Governato} F.,  {Volonteri} M.,  {Quinn} T.~R.,   {Pontzen} A.,
   2018, \mn@doi [\mnras] {10.1093/mnras/sty139}, \href
  {https://ui.adsabs.harvard.edu/abs/2018MNRAS.475.4967T} {475, 4967}

\bibitem[\protect\citeauthoryear{{Tremmel} et~al.,}{{Tremmel}
  et~al.}{2019}]{tremmel2019}
{Tremmel} M.,  et~al., 2019, \mn@doi [\mnras] {10.1093/mnras/sty3336}, \href
  {https://ui.adsabs.harvard.edu/abs/2019MNRAS.483.3336T} {483, 3336}

\bibitem[\protect\citeauthoryear{{Trinca}, {Schneider}, {Valiante}, {Graziani},
  {Zappacosta}  \& {Shankar}}{{Trinca} et~al.}{2022}]{trinca2022}
{Trinca} A.,  {Schneider} R.,  {Valiante} R.,  {Graziani} L.,  {Zappacosta} L.,
    {Shankar} F.,  2022, \mn@doi [\mnras] {10.1093/mnras/stac062}, \href
  {https://ui.adsabs.harvard.edu/abs/2022MNRAS.511..616T} {511, 616}

\bibitem[\protect\citeauthoryear{{Trinca}, {Schneider}, {Maiolino}, {Valiante},
  {Graziani}  \& {Volonteri}}{{Trinca} et~al.}{2023}]{Trinca2023}
{Trinca} A.,  {Schneider} R.,  {Maiolino} R.,  {Valiante} R.,  {Graziani} L.,
  {Volonteri} M.,  2023, \mn@doi [\mnras] {10.1093/mnras/stac3768}, \href
  {https://ui.adsabs.harvard.edu/abs/2023MNRAS.519.4753T} {519, 4753}

\bibitem[\protect\citeauthoryear{{Trussler}, {Conselice}, {Adams}, {Maiolino},
  {Nakajima}, {Zackrisson}  \& {Ferreira}}{{Trussler}
  et~al.}{2022}]{Trussler2022}
{Trussler} J. A.~A.,  {Conselice} C.~J.,  {Adams} N.~J.,  {Maiolino} R.,
  {Nakajima} K.,  {Zackrisson} E.,   {Ferreira} L.,  2022, \mn@doi [arXiv
  e-prints] {10.48550/arXiv.2211.02038}, \href
  {https://ui.adsabs.harvard.edu/abs/2022arXiv221102038T} {p. arXiv:2211.02038}

\bibitem[\protect\citeauthoryear{{{\"U}bler} et~al.,}{{{\"U}bler}
  et~al.}{2023}]{Ubler2023}
{{\"U}bler} H.,  et~al., 2023, \mn@doi [arXiv e-prints]
  {10.48550/arXiv.2302.06647}, \href
  {https://ui.adsabs.harvard.edu/abs/2023arXiv230206647U} {p. arXiv:2302.06647}

\bibitem[\protect\citeauthoryear{{Ueda}, {Akiyama}, {Hasinger}, {Miyaji}  \&
  {Watson}}{{Ueda} et~al.}{2014}]{ueda2014}
{Ueda} Y.,  {Akiyama} M.,  {Hasinger} G.,  {Miyaji} T.,   {Watson} M.~G.,
  2014, \mn@doi [\apj] {10.1088/0004-637X/786/2/104}, \href
  {https://ui.adsabs.harvard.edu/abs/2014ApJ...786..104U} {786, 104}

\bibitem[\protect\citeauthoryear{Valiante, Schneider, Salvadori  \&
  Bianchi}{Valiante et~al.}{2011}]{valiante2011}
Valiante R.,  Schneider R.,  Salvadori S.,   Bianchi S.,  2011, Monthly Notices
  of the Royal Astronomical Society, 416, 1916

\bibitem[\protect\citeauthoryear{Valiante, Schneider, Salvadori  \&
  Gallerani}{Valiante et~al.}{2014}]{valiante2014}
Valiante R.,  Schneider R.,  Salvadori S.,   Gallerani S.,  2014, Monthly
  Notices of the Royal Astronomical Society, 444, 2442

\bibitem[\protect\citeauthoryear{Valiante, Schneider, Volonteri  \&
  Omukai}{Valiante et~al.}{2016}]{valiante2016}
Valiante R.,  Schneider R.,  Volonteri M.,   Omukai K.,  2016, Monthly Notices
  of the Royal Astronomical Society, 457, 3356

\bibitem[\protect\citeauthoryear{{Valiante}, {Schneider}, {Zappacosta},
  {Graziani}, {Pezzulli}  \& {Volonteri}}{{Valiante}
  et~al.}{2018}]{valiante2018observability}
{Valiante} R.,  {Schneider} R.,  {Zappacosta} L.,  {Graziani} L.,  {Pezzulli}
  E.,   {Volonteri} M.,  2018, \mn@doi [\mnras] {10.1093/mnras/sty213}, \href
  {https://ui.adsabs.harvard.edu/abs/2018MNRAS.476..407V} {476, 407}

\bibitem[\protect\citeauthoryear{{Vanzella} et~al.,}{{Vanzella}
  et~al.}{2023}]{Vanzella2023}
{Vanzella} E.,  et~al., 2023, \mn@doi [\apj] {10.3847/1538-4357/acb59a}, \href
  {https://ui.adsabs.harvard.edu/abs/2023ApJ...945...53V} {945, 53}

\bibitem[\protect\citeauthoryear{{Venditti}, {Graziani}, {Schneider},
  {Pentericci}, {Di Cesare}, {Maio}  \& {Omukai}}{{Venditti}
  et~al.}{2023}]{Venditti2023}
{Venditti} A.,  {Graziani} L.,  {Schneider} R.,  {Pentericci} L.,  {Di Cesare}
  C.,  {Maio} U.,   {Omukai} K.,  2023, arXiv e-prints, \href
  {https://ui.adsabs.harvard.edu/abs/2023arXiv230110259V} {p. arXiv:2301.10259}

\bibitem[\protect\citeauthoryear{{Ventura}, {Qin}, {Balu}  \&
  {Wyithe}}{{Ventura} et~al.}{2024}]{Ventura2024}
{Ventura} E.~M.,  {Qin} Y.,  {Balu} S.,   {Wyithe} J. S.~B.,  2024, \mn@doi
  [arXiv e-prints] {10.48550/arXiv.2401.07396}, \href
  {https://ui.adsabs.harvard.edu/abs/2024arXiv240107396V} {p. arXiv:2401.07396}

\bibitem[\protect\citeauthoryear{{Visbal}, {Bryan}  \& {Haiman}}{{Visbal}
  et~al.}{2020}]{Visbal2020}
{Visbal} E.,  {Bryan} G.~L.,   {Haiman} Z.,  2020, \mn@doi [\apj]
  {10.3847/1538-4357/ab994e}, \href
  {https://ui.adsabs.harvard.edu/abs/2020ApJ...897...95V} {897, 95}

\bibitem[\protect\citeauthoryear{{Volonteri} et~al.,}{{Volonteri}
  et~al.}{2020}]{volonteri2020}
{Volonteri} M.,  et~al., 2020, \mn@doi [\mnras] {10.1093/mnras/staa2384}, \href
  {https://ui.adsabs.harvard.edu/abs/2020MNRAS.498.2219V} {498, 2219}

\bibitem[\protect\citeauthoryear{{Volonteri}, {Habouzit}  \&
  {Colpi}}{{Volonteri} et~al.}{2022}]{volonteri2022}
{Volonteri} M.,  {Habouzit} M.,   {Colpi} M.,  2022, arXiv e-prints, \href
  {https://ui.adsabs.harvard.edu/abs/2022arXiv221204710V} {p. arXiv:2212.04710}

\bibitem[\protect\citeauthoryear{{Wang} et~al.,}{{Wang}
  et~al.}{2020}]{wang2020}
{Wang} F.,  et~al., 2020, \mn@doi [\apj] {10.3847/1538-4357/ab8c45}, \href
  {https://ui.adsabs.harvard.edu/abs/2020ApJ...896...23W} {896, 23}

\bibitem[\protect\citeauthoryear{{Wang} et~al.,}{{Wang}
  et~al.}{2022}]{Wang2022}
{Wang} X.,  et~al., 2022, \mn@doi [arXiv e-prints] {10.48550/arXiv.2212.04476},
  \href {https://ui.adsabs.harvard.edu/abs/2022arXiv221204476W} {p.
  arXiv:2212.04476}

\bibitem[\protect\citeauthoryear{{Weinberger} et~al.,}{{Weinberger}
  et~al.}{2017}]{weinberger2017}
{Weinberger} R.,  et~al., 2017, \mn@doi [\mnras] {10.1093/mnras/stw2944}, \href
  {https://ui.adsabs.harvard.edu/abs/2017MNRAS.465.3291W} {465, 3291}

\bibitem[\protect\citeauthoryear{{Weingartner} \& {Draine}}{{Weingartner} \&
  {Draine}}{2001}]{weingartner2001}
{Weingartner} J.~C.,  {Draine} B.~T.,  2001, \mn@doi [\apj] {10.1086/318651},
  \href {https://ui.adsabs.harvard.edu/abs/2001ApJ...548..296W} {548, 296}

\bibitem[\protect\citeauthoryear{{Welch} et~al.,}{{Welch}
  et~al.}{2022}]{Welch2022}
{Welch} B.,  et~al., 2022, \mn@doi [\apjl] {10.3847/2041-8213/ac9d39}, \href
  {https://ui.adsabs.harvard.edu/abs/2022ApJ...940L...1W} {940, L1}

\bibitem[\protect\citeauthoryear{{Williams} et~al.,}{{Williams}
  et~al.}{2023}]{williams2023}
{Williams} H.,  et~al., 2023, \mn@doi [Science] {10.1126/science.adf5307},
  \href {https://ui.adsabs.harvard.edu/abs/2023Sci...380..416W} {380, 416}

\bibitem[\protect\citeauthoryear{{Wolcott-Green}, {Haiman}  \&
  {Bryan}}{{Wolcott-Green} et~al.}{2017}]{wolcottgreen2017}
{Wolcott-Green} J.,  {Haiman} Z.,   {Bryan} G.~L.,  2017, \mn@doi [\mnras]
  {10.1093/mnras/stx167}, \href
  {https://ui.adsabs.harvard.edu/abs/2017MNRAS.469.3329W} {469, 3329}

\bibitem[\protect\citeauthoryear{{Woods} et~al.,}{{Woods}
  et~al.}{2019}]{woods2019}
{Woods} T.~E.,  et~al., 2019, \mn@doi [\pasa] {10.1017/pasa.2019.14}, \href
  {https://ui.adsabs.harvard.edu/abs/2019PASA...36...27W} {36, e027}

\bibitem[\protect\citeauthoryear{{Xu}, {Norman}, {O'Shea}  \& {Wise}}{{Xu}
  et~al.}{2016}]{xu2016}
{Xu} H.,  {Norman} M.~L.,  {O'Shea} B.~W.,   {Wise} J.~H.,  2016, \mn@doi
  [\apj] {10.3847/0004-637X/823/2/140}, \href
  {https://ui.adsabs.harvard.edu/abs/2016ApJ...823..140X} {823, 140}

\bibitem[\protect\citeauthoryear{{Yoshida}, {Omukai}  \& {Hernquist}}{{Yoshida}
  et~al.}{2008}]{yoshida2008}
{Yoshida} N.,  {Omukai} K.,   {Hernquist} L.,  2008, \mn@doi [Science]
  {10.1126/science.1160259}, \href
  {https://ui.adsabs.harvard.edu/abs/2008Sci...321..669Y} {321, 669}

\bibitem[\protect\citeauthoryear{{Yung}, {Somerville}, {Finkelstein}, {Wilkins}
   \& {Gardner}}{{Yung} et~al.}{2023}]{yung2023}
{Yung} L.~Y.~A.,  {Somerville} R.~S.,  {Finkelstein} S.~L.,  {Wilkins} S.~M.,
  {Gardner} J.~P.,  2023, \mn@doi [arXiv e-prints] {10.48550/arXiv.2304.04348},
  \href {https://ui.adsabs.harvard.edu/abs/2023arXiv230404348Y} {p.
  arXiv:2304.04348}

\bibitem[\protect\citeauthoryear{{Zavala} et~al.,}{{Zavala}
  et~al.}{2023}]{Zavala2023}
{Zavala} J.~A.,  et~al., 2023, \mn@doi [\apjl] {10.3847/2041-8213/acacfe},
  \href {https://ui.adsabs.harvard.edu/abs/2023ApJ...943L...9Z} {943, L9}

\bibitem[\protect\citeauthoryear{{Zhang}, {Behroozi}, {Volonteri}, {Silk},
  {Fan}, {Hopkins}, {Yang}  \& {Aird}}{{Zhang} et~al.}{2021}]{zhang2021}
{Zhang} H.,  {Behroozi} P.,  {Volonteri} M.,  {Silk} J.,  {Fan} X.,  {Hopkins}
  P.~F.,  {Yang} J.,   {Aird} J.,  2021, arXiv e-prints, \href
  {https://ui.adsabs.harvard.edu/abs/2021arXiv210510474Z} {p. arXiv:2105.10474}

\bibitem[\protect\citeauthoryear{{Ziparo}, {Ferrara}, {Sommovigo}  \&
  {Kohandel}}{{Ziparo} et~al.}{2023}]{ziparo2023}
{Ziparo} F.,  {Ferrara} A.,  {Sommovigo} L.,   {Kohandel} M.,  2023, \mn@doi
  [\mnras] {10.1093/mnras/stad125}, \href
  {https://ui.adsabs.harvard.edu/abs/2023MNRAS.tmp..171Z} {}

\bibitem[\protect\citeauthoryear{{de Bennassuti}, {Schneider}, {Valiante}  \&
  {Salvadori}}{{de Bennassuti} et~al.}{2014}]{debennassuti2014}
{de Bennassuti} M.,  {Schneider} R.,  {Valiante} R.,   {Salvadori} S.,  2014,
  \mn@doi [\mnras] {10.1093/mnras/stu1962}, \href
  {https://ui.adsabs.harvard.edu/abs/2014MNRAS.445.3039D} {445, 3039}

\bibitem[\protect\citeauthoryear{{de Bennassuti}, {Salvadori}, {Schneider},
  {Valiante}  \& {Omukai}}{{de Bennassuti} et~al.}{2017}]{debennassuti2017}
{de Bennassuti} M.,  {Salvadori} S.,  {Schneider} R.,  {Valiante} R.,
  {Omukai} K.,  2017, \mn@doi [\mnras] {10.1093/mnras/stw2687}, \href
  {https://ui.adsabs.harvard.edu/abs/2017MNRAS.465..926D} {465, 926}

\makeatother
\end{thebibliography}

\appendix

\bsp	
\label{lastpage}
\end{document}